 \documentclass[twocolumn]{aastex63}
\DeclareUnicodeCharacter{2212}{-}
\revised{\today}
\shorttitle{{Top-of-the-atmosphere} and Vertical Cloud Structure  of a fast-rotating late T-dwarf}
\shortauthors{Manjavacas et al.}
\graphicspath{{./}{figures/}}
\begin{document}

\title{{Tracing} the {Top-of-the-atmosphere} and Vertical Cloud Structure  of a fast-rotating late T-dwarf}

\author[0000-0003-0192-6887]{Elena Manjavacas}
\affiliation{AURA for the European Space Agency (ESA), ESA Office, Space Telescope Science Institute, 3700 San Martin Drive, Baltimore, MD, 21218 USA}
\affiliation{Department of Physics and Astronomy, Johns Hopkins University, Baltimore, MD 21218, USA}

\author[0000-0001-7356-6652]{Theodora Karalidi}
\affiliation{Department of Physics, University of Central Florida, 4000 Central Florida Blvd., Orlando, FL 32816, USA}

\author[0000-0003-2278-6932]{Xianyu Tan}
\affiliation{Atmospheric Oceanic and Planetary Physics, Department of Physics, University of Oxford, OX1 3PU, UK}

\author[0000-0003-0489-1528]{Johanna M. Vos}
\affiliation{Department of Astrophysics, American Museum of Natural History, Central Park West at 79th Street, New York, NY 10024, USA}

\author[0000-0003-1487-6452]{Ben W. P. Lew}
\affiliation{Bay Area Environmental Research Institute and NASA Ames Research Center, Moffett Field, CA 94035, USA}

\author[0000-0003-4614-7035]{Beth A. Biller}
\affiliation{SUPA, Institute for Astronomy, University of Edinburgh, Blackford Hill, Edinburgh EH9 3HJ, UK}
\affiliation{Centre for Exoplanet Science, University of Edinburgh, Edinburgh, UK}

\author[0000-0001-5254-6740]{Natalia Oliveros-G\'omez}
\affiliation{Departamento de Astronomía, Universidad de Guanajuato. Callej\'on de Jalisco, S/N, 36023, Guanajuato, GTO, M\'exico }

\correspondingauthor{Elena Manjavacas}
\email{emanjavacas@stsci.edu}



\begin{abstract}

Only a handful of late-T brown dwarfs have been monitored for spectro-photometric variability, leaving incomplete the study of the atmospheric cloud structures of the coldest brown dwarfs, that share temperatures with some cold, directly-imaged
exoplanets. 2MASS~J00501994--332240 is a T7.0 {rapidly} rotating, field brown dwarf that showed low-level  photometric variability in data obtained with the  \textit{Spitzer} Space telescope. We monitored 2MASS~J00501994--332240 during $\sim$2.6~hr with MOSFIRE, installed at the Keck\,I telescope, {with the aim of constraining its near-infrared spectro-photometric variability}. {We measured fluctuations with a peak-to-peak  amplitude of} 1.48$\pm$0.75\% in the $J$-band photometric {light curve}, an amplitude of 0.62$\pm$0.18\% in the $J$-band spectro-photometric {light curve}, an amplitude of 1.26$\pm$0.93\% in the $H$-band {light curve}, and an amplitude of {5.33$\pm$2.02\%} in the $\mathrm{CH_{4}-H_{2}O}$ band {light curve}. {Nevertheless, the Bayesian Information Criterion does not detect significant variability in any of the light curves. Thus, given the detection limitations due to the MOSFIRE sensitivity, we can only claim tentative low-level variability for 2M0050--3322 in the best case scenario}. The {amplitudes of the peak-to-peak fluctuations}   measured for 2MASS~J00501994--332240 agree with the variability amplitude predictions of General Circulation Models for a T7.0 brown dwarf {for an edge on object}. {Radiative-transfer models predict that the $\mathrm{Na_{2}S}$  and  KCl clouds condense at pressures lower than that traced by the $\mathrm{CH_{4}-H_{2}O}$ band, which might explain the higher peak-to-peak fluctuations measured for this light curve}. Finally, we provide a {visual} recreation of the map provided by General Circulation Models, and the vertical structure of 2MASS~J00501994--332240 provided by radiative-transfer models.

\end{abstract}


\keywords{stars: brown dwarf}


\section{Introduction} \label{intro}

The search for brown dwarf photometric variability started soon after the discovery of the first brown dwarfs (e.g. \citealt{Bailer-Jones1999, Zapatero_Osorio2003, Caballero2004, Scholz2005}). The discovery of {high-amplitude, periodic photometric variability} in the T2.5 brown dwarf SIMP J013656.5+093347 \citep{Artigau2009} suggested that weather patterns that evolve with the object's rotation was the most likely cause of its variability. {Although other scenarios are possible \citep{Tremblin2015}}, these weather patterns are {most likely} produced by heterogeneous clouds with different varying thickness and temperatures \citep{Apai2013}, forming spots and bands in the atmospheres of brown dwarfs that evolve with time \citep{Apai2017}.

Up to now, dozens of L, L-T and T brown dwarfs have been monitored for photometric and spectro-photometric variability (\citealt{Radigan2014, Metchev2015, Biller2018, Vos2019}, among others) with ground-based and space-based facilities. Using ground-based facilities, \cite{Radigan2014} concluded that $39^{+16}_{-14}$\% of L9-T3.5 dwarfs {show in general higher variability amplitude than other spectral types}. Outside the L/T transition, \cite{Radigan2014} inferred that $60^{+22}_{-18}$\% of their sample vary with amplitudes of 0.5\%-1.6\%, concluding that heterogeneous clouds are common among brown dwarfs.  {Similarly, using the \textit{Spitzer} Space telescope, \cite{Metchev2015} was able to measure flux variations $>$0.2\%, determining that most objects in their sample show low-level variability.}

Until now, {the majority of brown dwarfs monitored for variability have been mostly photometrically monitored in one or several photometric bands}. Although {single-band} photometric variability traces {heterogeneities} in the atmospheres of brown dwarfs, it {only traces} one pressure range of the atmosphere of the object at a time.  {Spectro-photometric variability monitoring is more powerful {than single band monitoring}, since it \textit{simultaneously} traces  the variability at different wavelengths, {including the broad band, and} the different molecular and atomic spectral features in brown dwarf spectra (e.g. $\mathrm{H_{2}O}$ and $\mathrm{CH_{4}}$ bands, or the K\,I, Na\,I alkali lines). Thus, using spectro-photometric variability monitoring we are able to trace the atmosphere of brown dwarfs at different pressure levels simultaneously, providing a 3D map of their atmospheres \citep{Yang2016, Manjavacas2021}}.
Unfortunately, spectro-photometric variability monitoring is challenging to carry out, since we are limited almost only to space-based facilities. In fact,  most of the spectro-photometric data we have in the near-infrared have been acquired with the \textit{Hubble Space Telescope (HST)}, and its \textit{Wide Field Camera 3 (WFC3)} (e.g. \citealt{Apai2013, Buenzli2012, Lew2016,Yang2016, Biller2018, Manjavacas2018, Manjavacas2019b, Zhou2020}). From the ground, spectro-photometric monitoring for any given brown dwarf is possible only if multi-object near-infrared spectrographs are used, which are relatively scarce \citep{Manjavacas2021, Heinze2021}, {or if a calibration star is close enough to the target to perform single-slit simultaneous spectroscopic monitoring} {\citep{Schlawin2017, Kellogg2017}}. 

{To date}, only two late-T brown dwarfs have been monitored for spectro-photometric variability: the T6.5 2MASS~J22282889--431026 \citep{Buenzli2012}, and the T8.0 Ross~458c \citep{Manjavacas2019b}, both with \textit{HST/WFC3}. 2MASS J22282889--431026 shows different variability amplitude inside and outside the water band, and phase shifts between its light curves in the $J$- and $H$-bands, and with the water band at 1.40 $\mu$m \citep{Buenzli2012}. Ross 458c {also shows} tentative phase shifts between the $J$- and $H$-band light curves \citep{Manjavacas2019b}. Phase shifts have not been {commonly observed} in \textit{HST} bands for L and L/T brown dwarfs, so that may point at important differences in the dynamics-cloud coupling for mid and late T-dwarfs with respect to early T-dwarfs. {Thus, a larger sample of late-T dwarfs need to be spectrophotometrically monitored to shed light on the dynamics of these objects}.

In this paper we present Keck\,I/MOSFIRE $J$- and $H$-bands spectro-photometric monitoring for another late-T dwarf, 2MASS J00501994--332240 (T7.0, \citealt{Tinney2005}), with the aim of increasing the sample of late-T dwarfs {with spectroscopic variability monitoring}, and {shedding} light into the atmospheric structure of these objects.  2MASS~J00501994--332240 shares its spectral type with some cold directly-imaged giant exoplanets, like 51~Eridani~b \citep{Macintosh2015, Rajan2017}.  {2MASS J00501994--33224 serves as an analog} to provide insights into the cloud structure of cold giant exoplanets for which high signal-to-noise spectro-photometric data are very limited with the {existing} instrumentation.

This paper is structured as follows: in Section \ref{target} we introduce the properties of 2MASS~J00501994--332240. In Section \ref{observations} we provide the details of the Keck\,I/MOSFIRE monitoring for 2MASS~J00501994--332240. In Section \ref{reduction} we describe the data reduction. In Section \ref{light_curves}, we explain how the light curve for the object was produced. In Section \ref{correlation_analysis_section} we evaluate the correlations between our light curve and effects that might introduce spurious variability {in} our target's light curve. In Section \ref{results} we present our results, and their significance. {In Section \ref{discussion} we present a discussion of our results}. In Section \ref{interpretation} we provide an interpretation of 2MASS~J00501994--332240 light curve according to General Circulation and radiative-transfer models. Finally, in Section \ref{conclusions} we summarize our conclusions.

\section{2MASS J00501994--3322402} \label{target}

2MASS J00501994--3322402 (2M0050--3322) was discovered  by \cite{Tinney2005}, and spectral typed as a T7.0 brown dwarf by \cite{Burgasser2006}. {\cite{Dupuy_Liu2012} measured a trigonometric parallax for 2M0050--3322 of $\pi$ = 94.6$\pm$2.4~mas}. {2M0050--3322} has near-infrared magnitudes of $J$~=~15.928$\pm$0.070, $H$~=~15.838$\pm$0.191, and $K$~=~15.241$\pm$0.185 \citep{Cutri2003}. {\cite{Filippazzo2015} determined the fundamental parameters of our target by fitting its IRTF/SpeX spectrum to atmospheric models, which provided an estimation of its $\mathrm{T_{eff}}$ and log~g. Integrating its absolute flux calibrated Spectral Energy Distribution, \cite{Filippazzo2015} derived  its $\mathrm{L_{bol}/L_{\odot}}$. Finally, using the calculated $\mathrm{L_{bol}/L_{\odot}}$, its radius, and mass were derived from evolutionary models.  2M0050--3322 fundamental parameters are}: $\mathrm{log(L/L_{\odot})}$~=~−5.39$\pm$0.02, age~=~0.500–-10.00~Gyrs, R~=~0.94$\pm$0.16~$\mathrm{R_{Jup}}$,  log(g)~=~4.95$\pm$0.49, $\mathrm{T_{eff}}$~=~836$\pm$71~K, and a {mass of}  40.34$\pm$25.54~$\mathrm{M_{Jup}}$, confirming that 2M0050--3322 has the characteristics of a field T7.0 brown dwarf. {No inclination has been measured for 2M0050--3322.}

{\cite{Radigan2014} found no significant $J$-band photometric variability for 2M0050--3322, but they provided an upper level of 1.1\%.} \cite{Metchev2015} found  variability on the [4.5] \textit{Spitzer} channel, {but not in the [3.6] channel}. \cite{Metchev2015} measured a variability amplitude of 1.07$\pm$0.11~\% for 2M0050--3322, and a rotational period of {1.55$\pm$0.02~hr}, being one of the brown dwarfs with the shortest rotational period {\citep{tannock2021weather}}.

\section{Observations} \label{observations}

Near-infrared multiobject spectrographs like MOSFIRE \citep{McLean2010, McLean2012} installed at the Keck\,I telescope, allow us to perform spectro-photometric monitoring of brown dwarfs from the ground. As explained in \cite{Manjavacas2021}, to carry out spectro-photometric monitoring observations of brown dwarfs from the ground, we need calibration stars in the same field of our targets to account for spectral calibration,  telluric contamination, changes in the airmass, humidity and temperature variations in the atmosphere, etc, that might potentially introduce spurious variability signals.  MOSFIRE performs  simultaneous spectroscopy of up to 46 objects in a 6.1'x 6.1' field of view, using the Configurable Slit Unit (CSU),  a cryogenic robotic slit mask system that is reconfigurable electronically in less than 5~minutes without any thermal cycling of the instrument. A single photometric band is covered in each instrument setting ($Y$, $J$, $H$ or $K$).

The total time of $J$- and $H$-band monitoring is $\sim$2.67~hr, with a gap of $\sim$25 min between the $J$-band imaging and spectroscopic data. We observed $\sim$0.45~hr in $J$-imaging mode, $\sim$0.35~hr in $J$-band spectrophotometric mode, and 1.36~hr in $H$-band spectrophotometric mode

\begin{figure}
    \centering
    \includegraphics[width=0.5\textwidth]{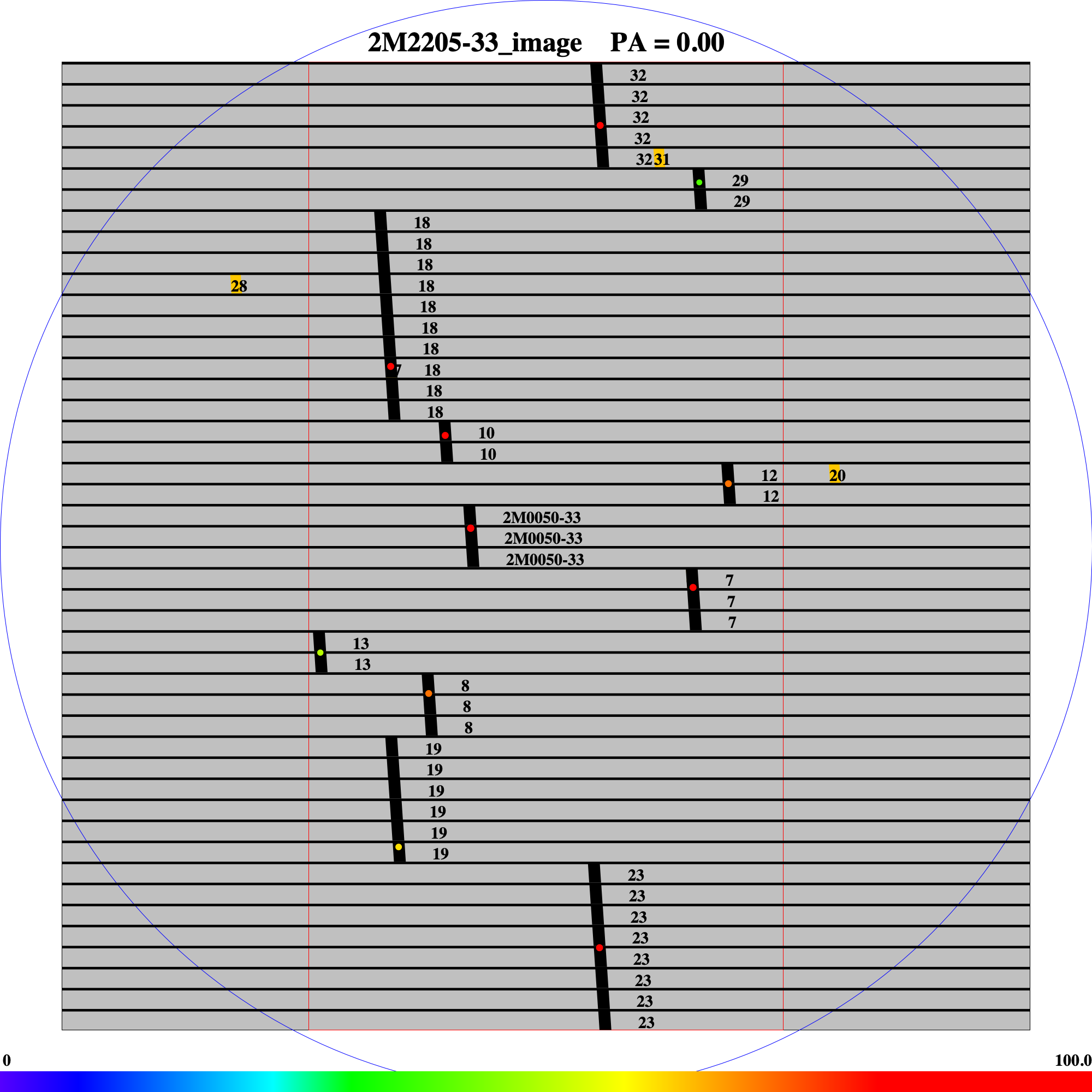}
    \caption{Illustration of the positioning of the CSU bars {(black horizontal lines)} on MOSFIRE to obtain simultaneous multi-object spectroscopy of the field of 2M0050--3322 as produced by MAGMA, the MOSFIRE Automatic GUI-based Mask Application. Our target (named as 2M0050-33) is placed in the center of the field. The position of the comparison stars as shown in Table~\ref{calstars} are also marked. The round colored points show the expected positions of the target and calibration stars. The different colors indicate the weight given to each object for MAGMA to prioritize the desired objects in the field of view. As shown in the color-bar, red is weight 100 and blue weight 0. The yellow squares show the position of the stars used for the alignment of the mask. {The blue circle indicates the field of MOSFIRE. The vertical red lines delimits the area in which the slits are positioned for spectroscopy in any filter}.}
    \label{mosfire_field}
\end{figure}

We observed 2M0050--3322 on UT 2020-10-09  during $\sim$2.6~hr, covering approximately two rotational periods of the object. Since the target {has a} high proper motion, we designed the MOSFIRE mask accordingly with the predicted position of the target at the time of the observation. Nonetheless, we did not find our target in the slit as expected. Thus, we performed $J$-band imaging of the field during the first {$\sim$0.45~hr} of the observation, while we designed a new mask, after having identified our target in the imaging data. We took images of {the} 6.1'$\times$6.1' field of view, with a scale of 0.1798”/pix, performing a 3.0" ABBA pattern. We integrated during 11~s in each ABBA position. After designing a new mask, {we interrupted the observations for $\sim$25 min between the $J$-band imaging and spectroscopic data while installing the new MOSFIRE mask}. Then, we obtained $J$-band spectra of 2M0050--3322 for  {$\sim$0.35~hr}, covering the first rotational period of the target.  We obtained  three  spectra of 2M0050--3322 in the $J$-band (1.153--1.352~$\mu$m) using a 1.5" ABBA pattern. As in \cite{Manjavacas2021}, we used wide slits of 4.5" to avoid slit losses for all  calibration stars and the target, obtaining a spectral resolution of R$\sim$1000. Immediately, we obtained another nine spectra in the $H$-band (1.5280-1.8090~$\mu$m) during $\sim$1.4~hr, covering the second rotational period of 2M0050--3322.  
In Table~\ref{calstars} we show the list of objects used as calibrators, their coordinates, and their $J$-band magnitudes. Those that do not have an object number did not have {sufficient} signal-to-noise (SNR) to be used as calibrator stars (SNR $<$ 5). In Figure \ref{mosfire_field}, we show the configuration of the CSU mask, with the position of the target and the calibration stars. We used exposure times of 120~s for each ABBA nod position in the the first ABBA. We observed over {an} airmass range 1.79 to 2.37.

For data reduction purposes, 11 dome flats of 2~s exposure were obtained in the $J$- and $H$-band. Following the same calibration strategy as in \cite{Manjavacas2021}, we  obtained three $J$- and $H$-band "sky" spectra  using the same configuration for the multiobject mask as for the observations, but using 1.0" slits to obtain higher resolution sky lines during 1~s. The 1.0" slits provided spectra of the skylines with enough resolution to allow the pipeline to produce an accurate wavelength calibration. We used archive imaging flats and darks taken on 2020-09-04.

\begin{table}
	\caption{{Information about the calibration objects in the field of 2M20050--3322.}}  
	\label{calstars}
	\centering
	\begin{center}
		\begin{tabular}{llllll}
        \hline
		\hline 
			
Num. mask & Num. obj. & RA (J2000) & DEC (J2000) & $J$-mag  \\   
\hline

32 & 5 & 00:50:17.976 & -33:19:49.25 & 15.64   \\
29 & --  & 00:50:14.982 & -33:20:10.84 & 18.58   \\
18 & --   & 00:50:24.317 & -33:21:20.61 & 17.48   \\
10 & 3  & 00:50:22.666 & -33:21:46.78 & 16.15   \\
12  & 6   & 00:50:14.095 & -33:22:05.07 & 16.56   \\
2M0050-33 & 2M0050-33 &  00:50:21.900 & -33:22:21.95 & 15.93  \\
7 & -- &  00:50:15.175 & -33:22:44.41 & 17.74  \\
13 & -- &  00:50:26.454 & -33:23:09.12 & 19.54     \\
8 & -- &   00:50:23.165 & -33:23:24.61 & 18.23  \\
19 & 2 & 00:50:24.075 & -33:24:22.80 & 14.93   \\
23 & 1 &  00:50:17.997 & -33:25:01.00 & 15.85  \\

			\hline			
		\end{tabular}
	\end{center}

\end{table}

\section{Data Reduction} \label{reduction}

\subsection{{Imaging data}}

We reduced the $J$-band imaging data independently from the spectroscopic data. All $J$-band raw images were dark subtracted using a median-combined dark. Then we divided them by a median-combined flat field, created using the median-combined and subtracted lamp-on and lamp-off flats. {We created a bad-pixel mask to mask hot and dead pixels. We flagged those pixels by selecting the pixels in the master flat calibration image with values 2$\sigma$ above or below the mean}. We subtracted the A-B pairs and the B-A pairs independently to remove the sky. Finally, the A-B and B-A pairs {were} aligned and coadded.

{To perform the aperture photometry on the reduced imaging data, we used the Python package \textit{DAOStarFinder} from \textit{astropy.photutils}\footnote{\url{https://photutils.readthedocs.io/en/stable/}}, an implementation of the \textit{daofind} algorithm \citep{Stetson1987}, that searches images for local density maximum that have a peak amplitude greater than a specified threshold (the threshold is applied to a convolved image) and have a size and shape similar to a defined 2D Gaussian kernel. Then, we performed aperture photometry using the Python package \textit{aperture\_photometry} from {\textit{astropy.photutils}} with an aperture of 5 pixels, in an annulus between 6 and 10 pixels {to extract the background}. Finally, using the x and y position of the stars, we matched the photometry of all stars in all the $J$-band images to obtain a light curve for each individual star.}

\subsection{{Spectroscopic data}}

\begin{figure*}
    \centering
    \includegraphics[width=0.98\textwidth]{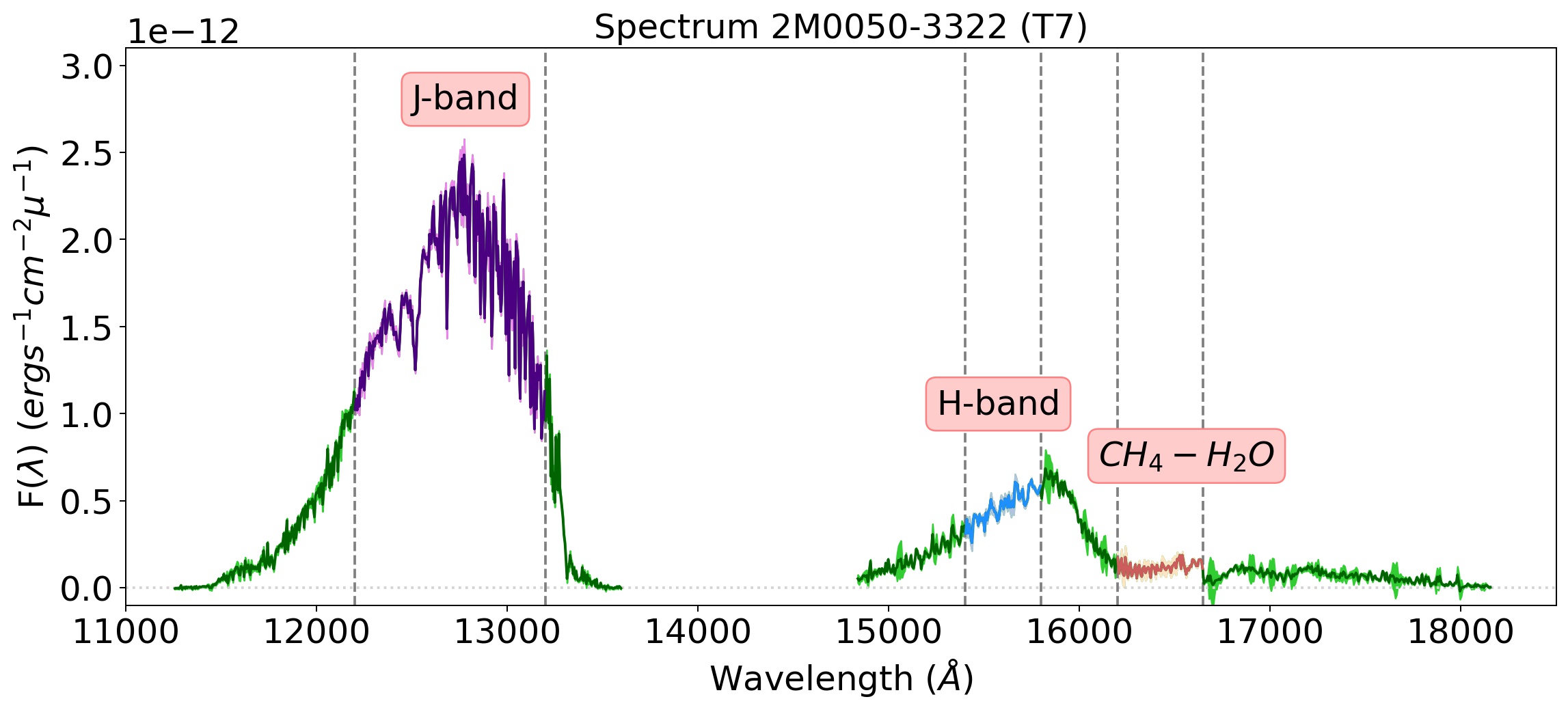}
    \caption{Median-combined, flux-calibrated $J$- and $H$-band MOSFIRE spectrum of 2M0050--3322, {and its uncertainties in lighter colors}. We mark in violet the wavelength range used to obtain the $J$-band spectroscopic light curve, in blue the $H$-band spectroscopic light curve, and in orange the $\mathrm{CH_{4}-H_{2}O}$ light curve. }
    \label{spectrum}
\end{figure*}

We used the  version 1.7.1 of \textit{PypeIt\footnote{\url{https://github.com/pypeit/PypeIt}}} to reduce the multi-object spectro-photometric data acquired with MOSFIRE in the $J$- and $H$-band \citep{Prochaska2019,Prochaska2020}. The pipeline corrected  the raw images from dark current, and a bad pixels mask is generated. The edges of the slits were traced using the dome flats, and a master flat was also created. \textit{PypeIt} produced a wavelength calibration for our data using the sky arc frames taken using the same multiobject mask we employed for our observations, but with narrower slits of 1.0". The wavelength calibration accounted for the spectral tilt across the slit. The calibrations were applied to our science frames, and the sky was subtracted using the A-B or B-A frames following \cite{Kelson2003}. The 1D science spectra were extracted from the 2D sky-corrected frames.  The signal-to-noise achieved for the target and the calibration stars in the $J$-band and $H$-band spectro-photometric data are summarized in Table~\ref{spectra_snr}. No telluric calibration was performed for these spectra. Instead, for the upcoming analysis, we have used the wavelength range between 12200 and 13200~{$\AA$}, avoiding the most prominent telluric contamination in the $J$-band, and the range between 15400 and 15800~$\AA$ for the $H$-band. We show the median-combined, flux-calibrated $J$- and $H$-band MOSFIRE spectrum in Fig. \ref{spectrum}. {The flux calibration was performed using the 2MASS $J$- and $H$-band magnitudes. We convolved our near-infrared spectra with $J$-, and $H$- filter transmission curves of 2MASS. The resulting spectra were integrated. We calculated the flux for our targets corresponding to the $J$, and $H$ bands using 2MASS magnitudes. Finally, we calculated the scaling factor for the J, and H bands and multiplied our near-infrared spectra in J, and H filters to have the same flux as given by 2MASS}.

\begin{table}
	\caption{{Median signal-to-noise of the $J$-band and $H$-band spectra of the target and the calibration stars}.}  
	\label{spectra_snr}
	\centering
	\begin{center}
		\begin{tabular}{ccc}
        \hline
		\hline 
			
Object Number & $J$-band SNR & $H$-band SNR \\   
\hline

Star 1        &  15.4  & 13.1   \\
Star 2        &  30.6  & 29.4  \\
Star 3        &  13.0  & 10.5   \\
Star 5        &  16.6  & 13.4  \\
%Star 6        &  19.6  & 12.5   \\
2M0050--3322 &  52.9  & 27.7   \\

			\hline			
		\end{tabular}
	\end{center}
\end{table}

\section{Light curve correction}\label{light_curves}

For the spectro-photometric data, we produced a $J$-band (12200 and 13200~{$\AA$}), a  $H$-band (15400 and 15800~$\AA$), and a $CH_{4}-H_{2}O$ (16200-16650~$\AA$) light curve for each object in the field. As these data were obtained from the ground, there might be other additional sources of non-astrophysical contamination affecting the shape of the light curve extracted for each object, such as varying atmospheric transparency, change in the water vapor content of the atmosphere, the seeing, variations in the outside temperature during the observation, wind speed and direction variations, airmass variations, etc. Thus, the science target light curve needs to be corrected for these potential sources of contamination. We corrected all the light curves of the target using the same wavelength range of the calibration stars spectra. Using this method, we expect to correct most of the spurious variability signal introduced by telluric contamination.

To perform the light curve correction of the photometric and spectro-photometric datasets, we followed a similar approach to \cite{Radigan2014}. Each light curve was corrected by dividing it by a calibration light curve produced by median combining the relative-flux light curves of all the other objects in the field, beside the science target. First, we normalized the light curves of all objects to the median flux of each of them. For each reference star, a calibration light curve was created by median combining the light curves of all the other objects, beside the science target. Then, the raw light curve of each calibration star was divided by its corresponding calibration light curve to obtain a corrected light curve. Finally, we measured the standard deviation, $\sigma$, of the corrected light curve of each calibration star. 

To perform an optimal correction of the 2M0050--3322 light curve we chose the {least intrinsically} variable calibration stars. We followed the best-selection criteria for the calibration stars used by \cite{Radigan2014}, for which they subtracted from the corrected light curve of each calibration star, a shifted version of itself, and divided it by $\sqrt{2}$ ($\sigma_{s}=[f_{cal}-f_{cal\_shifted}]/\sqrt{2}$). \cite{Radigan2014} then  identified  poor-quality calibration stars as those where $\sigma_{s}>1.5\times \sigma_{target}$, but in our case d. These criteria left only 5 stars as good calibrators for the $J$-band imaging data, and only stars 1 and 2 as valid correcting stars in the {$J$-band spectroscopic data}. For the $H$-band spectro-photometric light curve only star 6 was rejected. {For the  $CH_{4}-H_{2}O$ light curve no stars were rejected following \cite{Radigan2014} method, but stars 3, 5 and 6 did not improve their standard deviation after correcting them, so we decided to leave the, out as calibration stars (see Table \ref{stats_calstars_methane_water})}. In Tables~\ref{stats_calstarsJim}, \ref{stats_calstarsJ},  \ref{stats_calstarsH}, {\ref{stats_calstars_methane_water}}, and \ref{stats_target}, we show the standard deviation of all the calibration stars and 2M0050--3322 before and after correcting each light curve. {In the case of the spectrophotometric $J$-band light curves the $\sigma$ improves only marginally, probably due to the few points of the light curve.}

\begin{table}
	\caption{Standard deviation of the  $J$-imaging photometric light curve for the "good" calibration stars.}  
	\label{stats_calstarsJim}
	\centering
	\begin{center}
		\begin{tabular}{ccc}
        \hline
		\hline 
			
(x centroid, y centroid)  & $\sigma$ non-corrected LC & $\sigma$ corrected LC\\   
\hline

(1158.94, 202.79) &  0.0180 & 0.0055   \\
(212.59, 836.39) &  0.0150 & 0.0047 \\
(1423.89, 1181.28) &  0.0136 & 0.0055   \\
(829.97, 1280.67) &  0.0146  & 0.0042  \\
(1151.96, 1931.35) &  0.088  & 0.0084   \\

			\hline			
		\end{tabular}
	\end{center}
\end{table}

%--------------------------

\begin{table}
	\caption{Standard deviation of the  $J$-band spectro-photometric light curve for the calibration stars.}  
	\label{stats_calstarsJ}
	\centering
	\begin{center}
		\begin{tabular}{ccc}
        \hline
		\hline 
			
Object Number & $\sigma$ non-corrected LC & $\sigma$ corrected LC\\   
\hline

Star 1 &  0.0021 & 0.0023   \\
Star 2 &  0.0019 & 0.0017 \\
%Star 3 &  0.0091 & 0.010   \\
%Star 5 &  0.0008  & 0.0007  \\
%Star 6 &  0.0028  & 0.0027   \\

			\hline			
		\end{tabular}
	\end{center}
	
\end{table}

%--------------------------

\begin{table}
	\caption{Standard deviation of the  $H$-band spectro-photometric light curve for the calibration stars.}  
	\label{stats_calstarsH}
	\centering
	\begin{center}
		\begin{tabular}{ccc}
        \hline
		\hline 
			
Object Number & $\sigma$ non-corrected LC & $\sigma$ corrected LC\\   
\hline

Star 1 &  0.0077 & 0.0075   \\
Star 2 &  0.0083 & 0.0018 \\
Star 3 &  0.0066 & 0.0056   \\
Star 5 &  0.0084  & 0.0022  \\
%Star 6 &  0.0059  & 0.0096   \\

			\hline			
		\end{tabular}
	\end{center}
	
\end{table}

\begin{table}
	\caption{Standard deviation of the  $\mathrm{CH_{4}-H_{2}O}$-band spectro-photometric light curve for the calibration stars.}  
	\label{stats_calstars_methane_water}
	\centering
	\begin{center}
		\begin{tabular}{ccc}
        \hline
		\hline 
			
Object Number & $\sigma$ non-corrected LC & $\sigma$ corrected LC\\   
\hline

Star 1 &  0.0102 & 0.0056   \\
Star 2 &  0.0058 & 0.0038 \\
%Star 3 &  0.0064 & 0.0066   \\
%Star 5 &  0.0067  & 0.0068  \\
%Star 6 &  0.0051  & 0.0067   \\

			\hline			
		\end{tabular}
	\end{center}
	
\end{table}

\begin{figure*}
    \centering
    \includegraphics[width=0.48\textwidth]{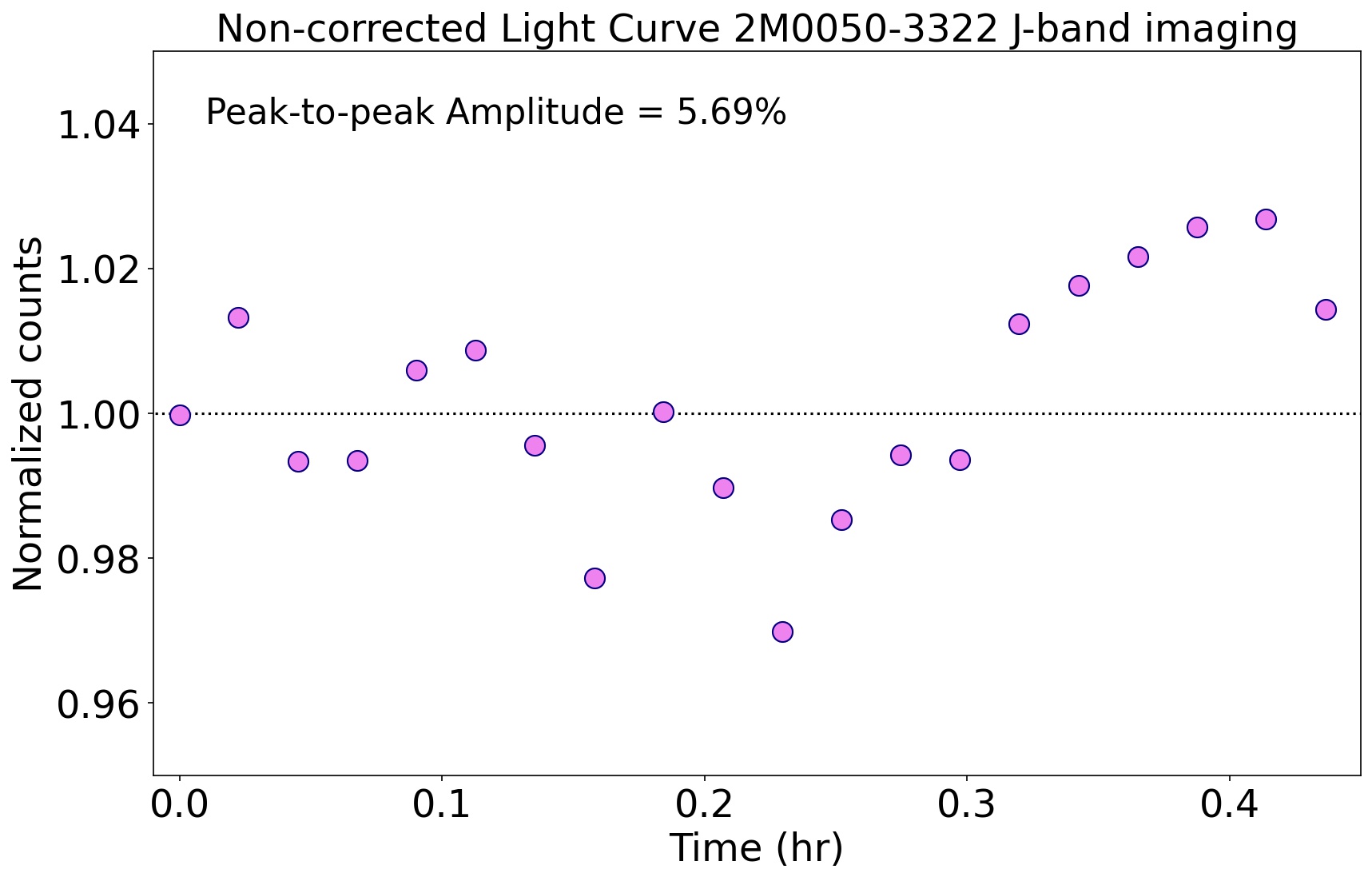}
    \includegraphics[width=0.48\textwidth]{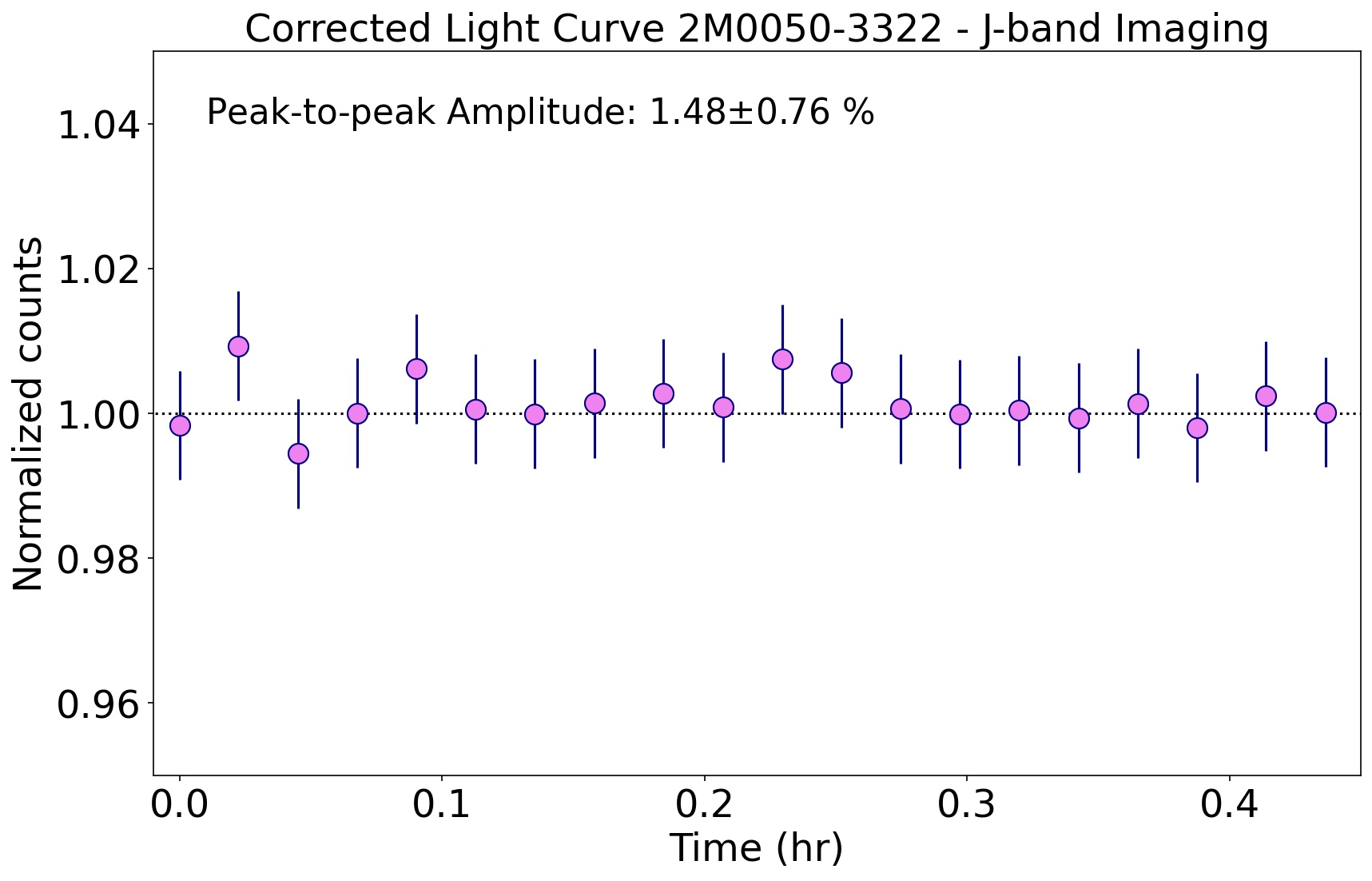}
    \caption{Normalized non-corrected (left) and corrected (right) $J$-band imaging light curves of  2M0050--3322. {We indicate  the peak-to-peak amplitudes of their fluctuations}.}
    \label{target_JimLCs}
\end{figure*}

\begin{figure*}
    \centering
    \includegraphics[width=0.48\textwidth]{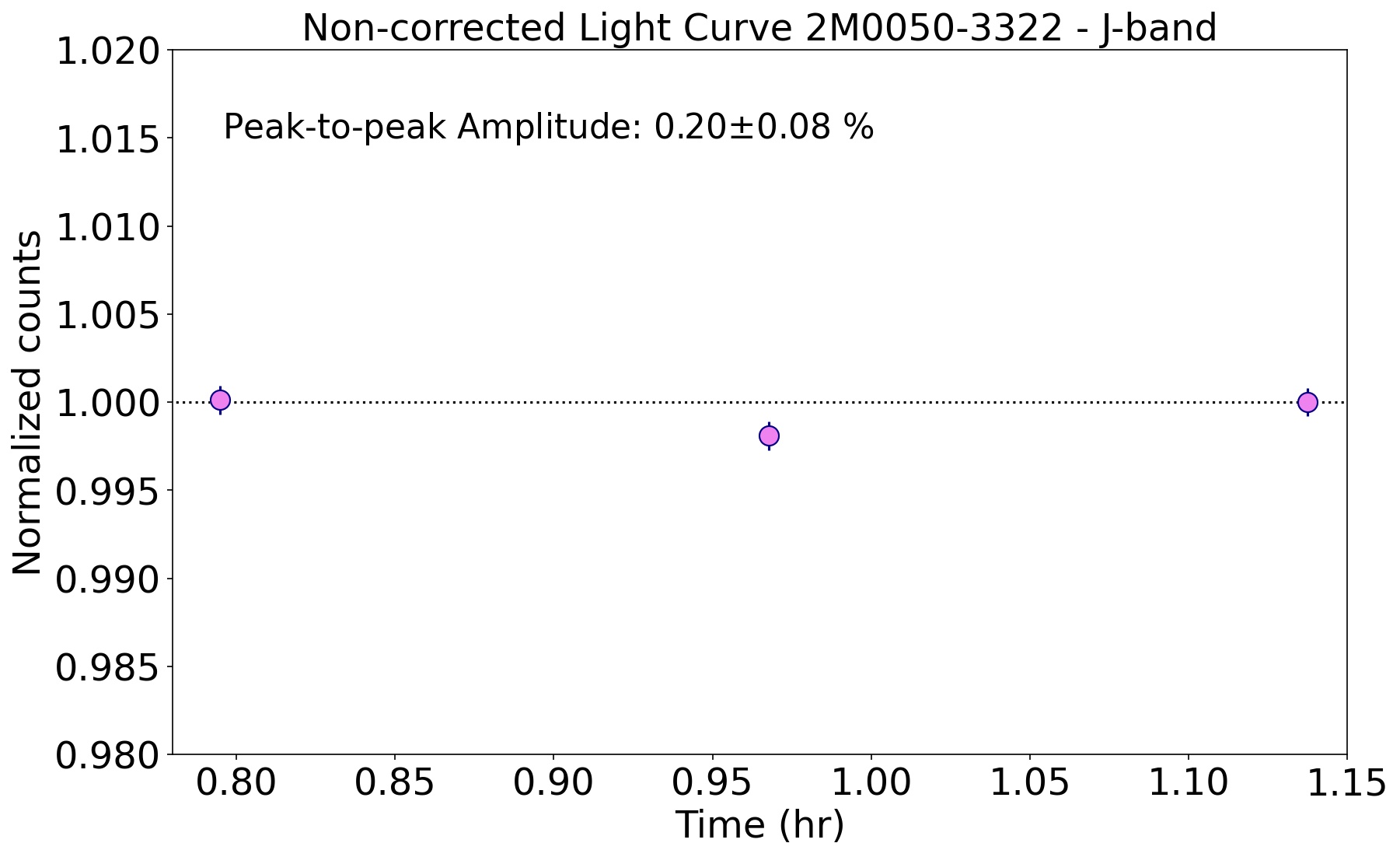}
    \includegraphics[width=0.48\textwidth]{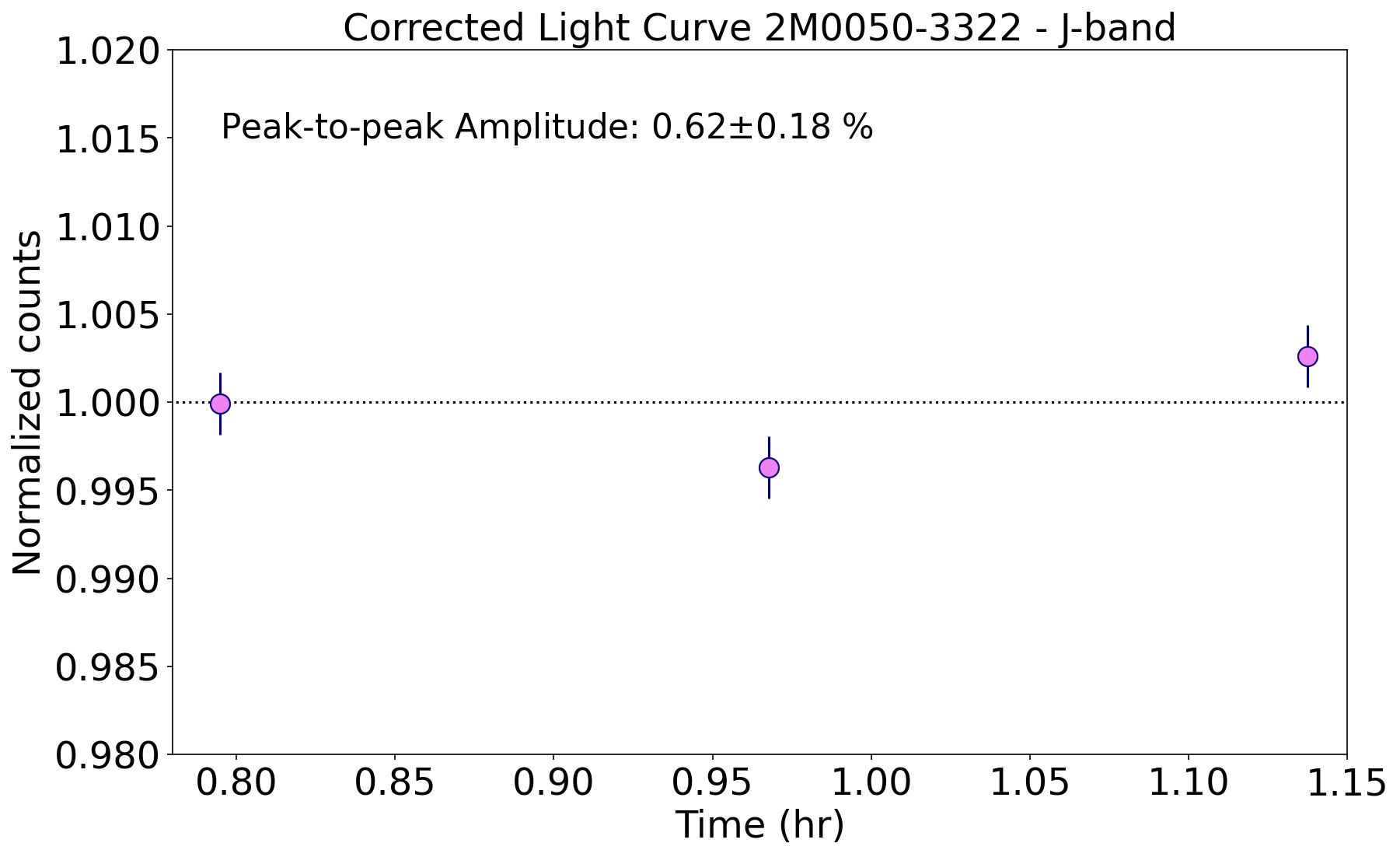}
    \caption{Normalized non-corrected (left) and corrected (right) $J$-band light curves of  2M0050--3322. {We indicate  the peak-to-peak amplitudes of their fluctuations}.}
    \label{target_JLCs}
\end{figure*}

\begin{figure*}
    \centering
    \includegraphics[width=0.48\textwidth]{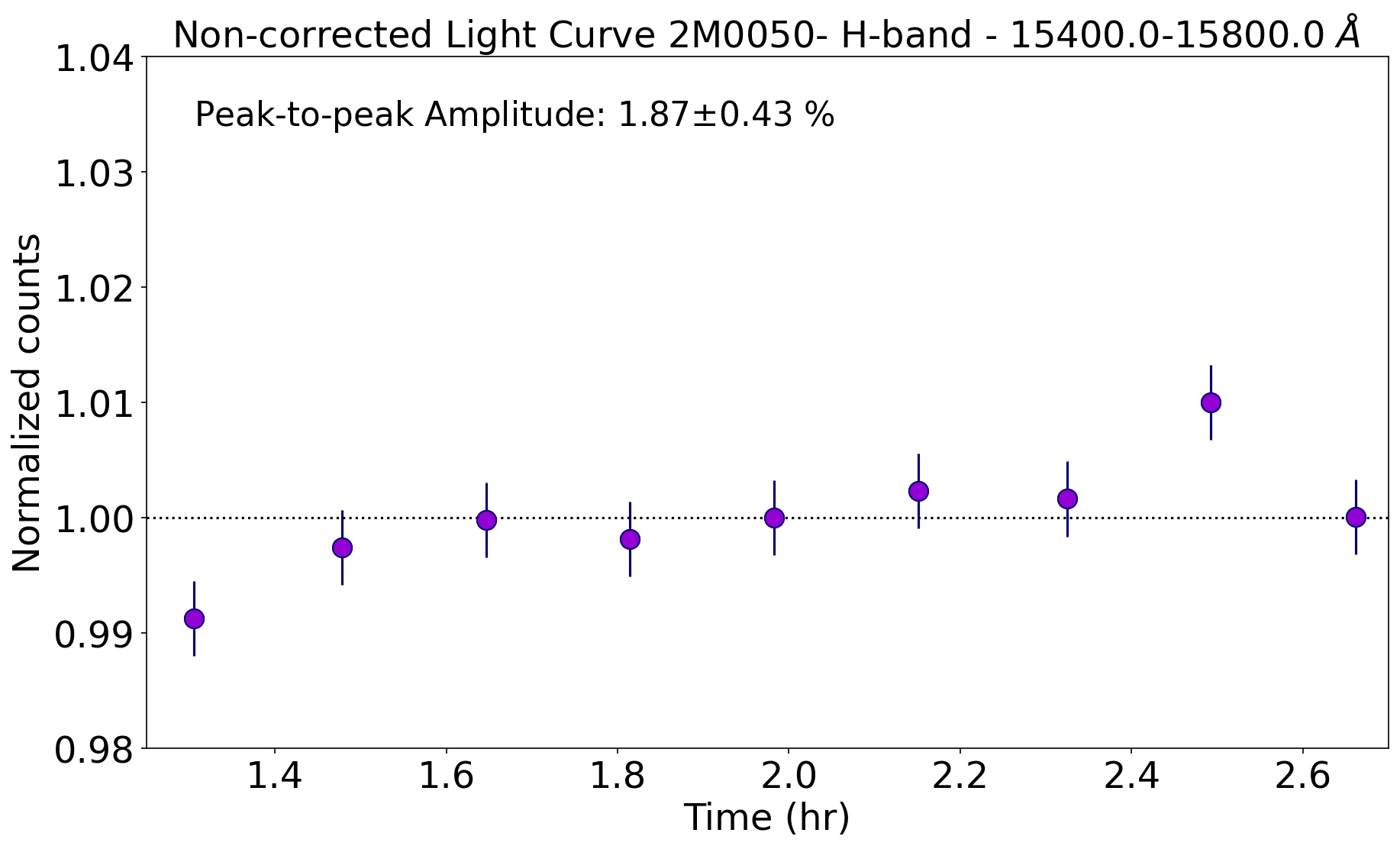}
    \includegraphics[width=0.48\textwidth]{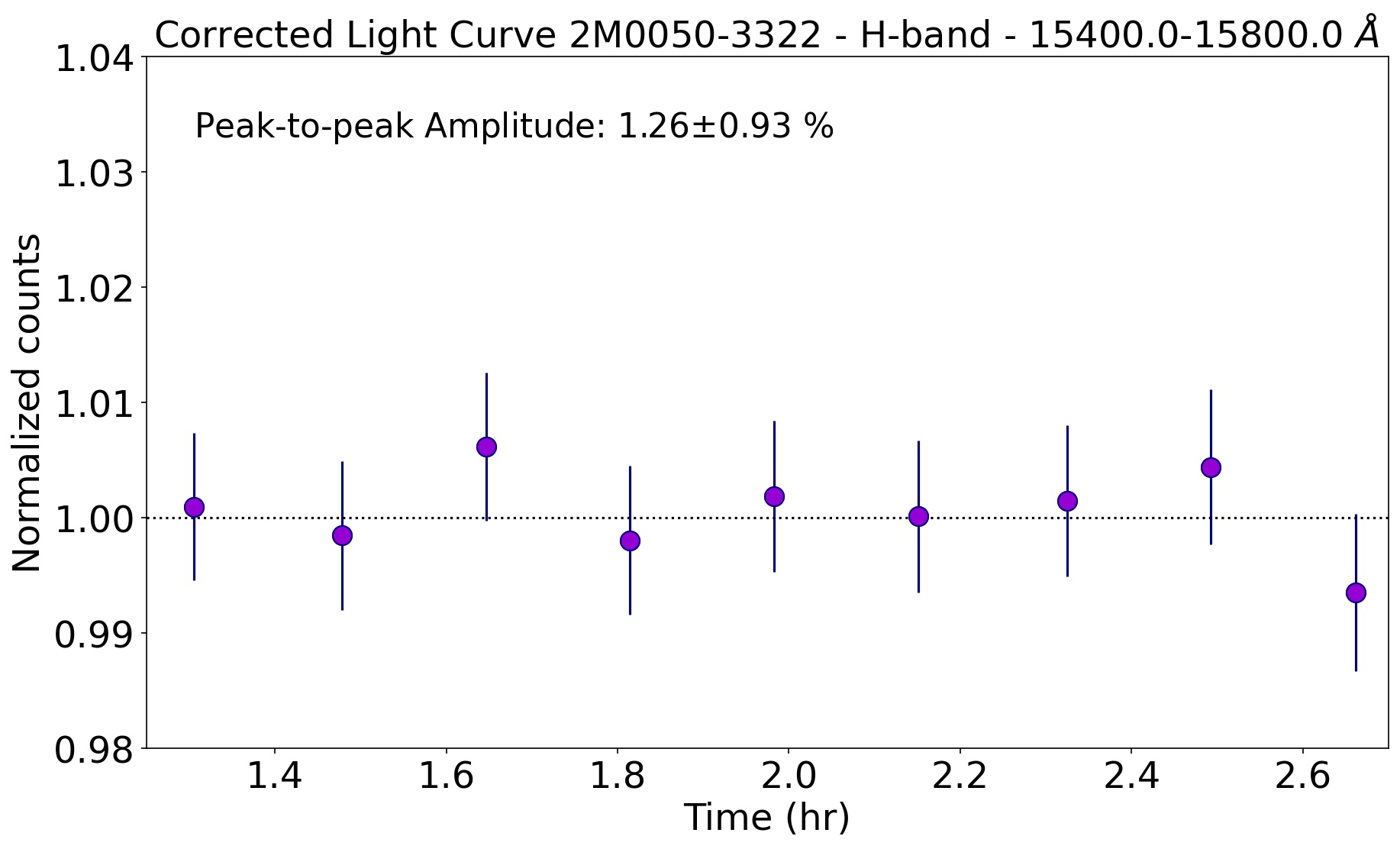}
    \caption{Normalized non-corrected (left) and corrected (right) $H$-band light curves of  2M0050--3322. {We indicate  the peak-to-peak amplitudes of their fluctuations}.}
    \label{target_HLCs}
\end{figure*}

\begin{figure*}
    \centering
    \includegraphics[width=0.48\textwidth]{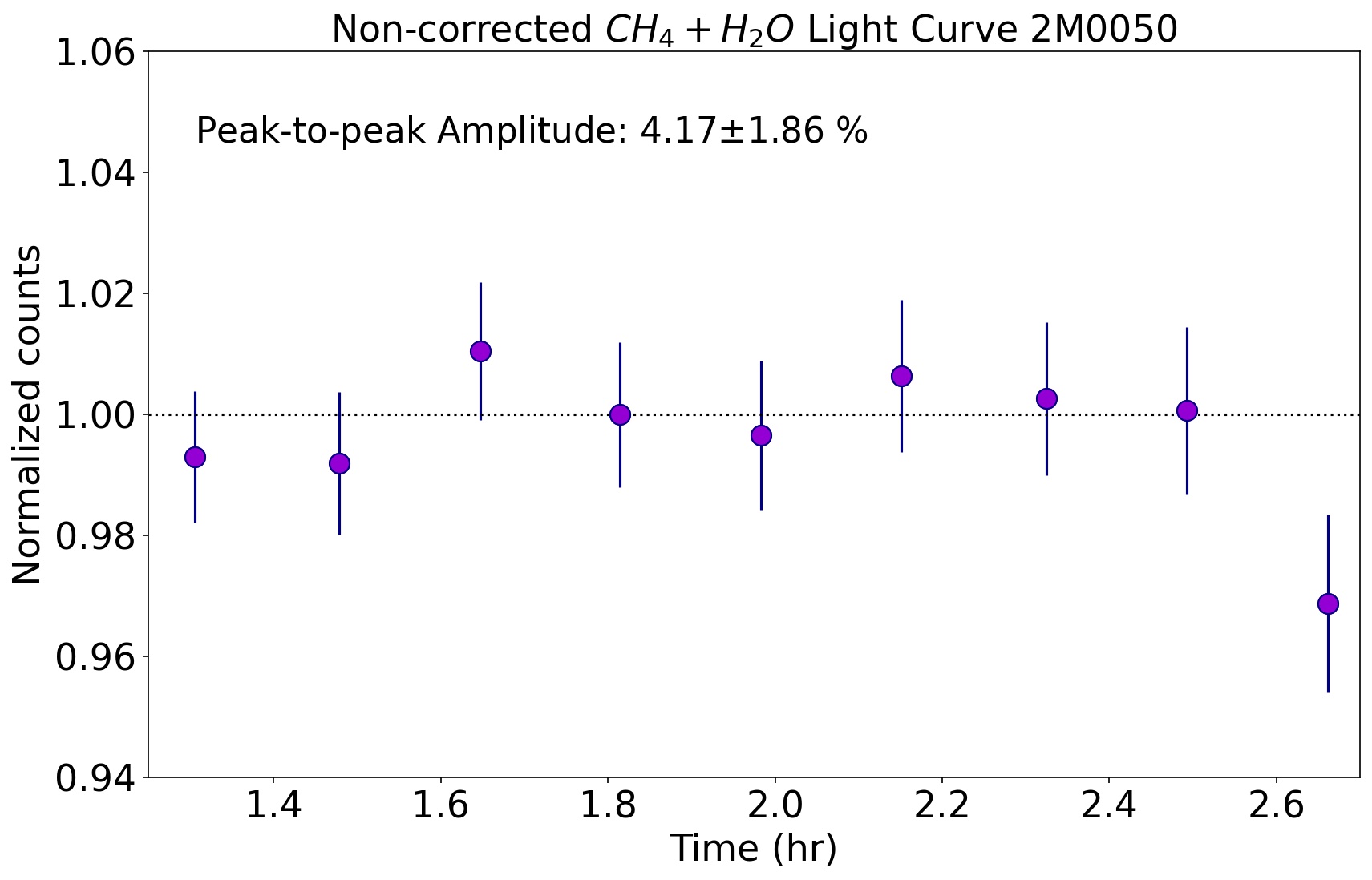}
    \includegraphics[width=0.48\textwidth]{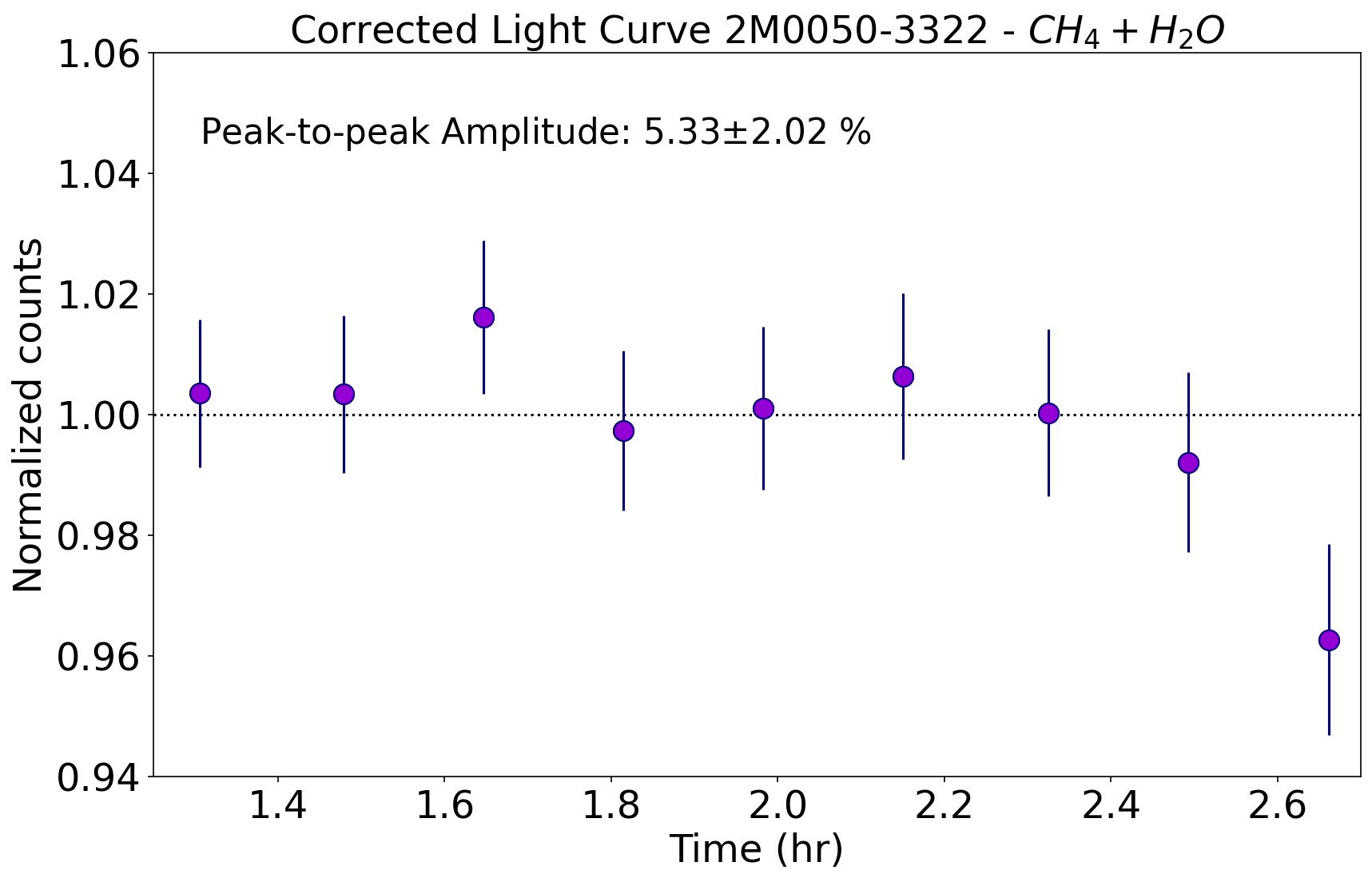}
    \caption{Normalized non-corrected (left) and corrected (right) $CH_{4}-H_{2}O$-band light curves of  2M0050--3322. {We indicate  the peak-to-peak amplitudes of their fluctuations}.}
    \label{target_CH4LCs}
\end{figure*}

\begin{figure*}
    \centering
    \includegraphics[width=0.98\textwidth]{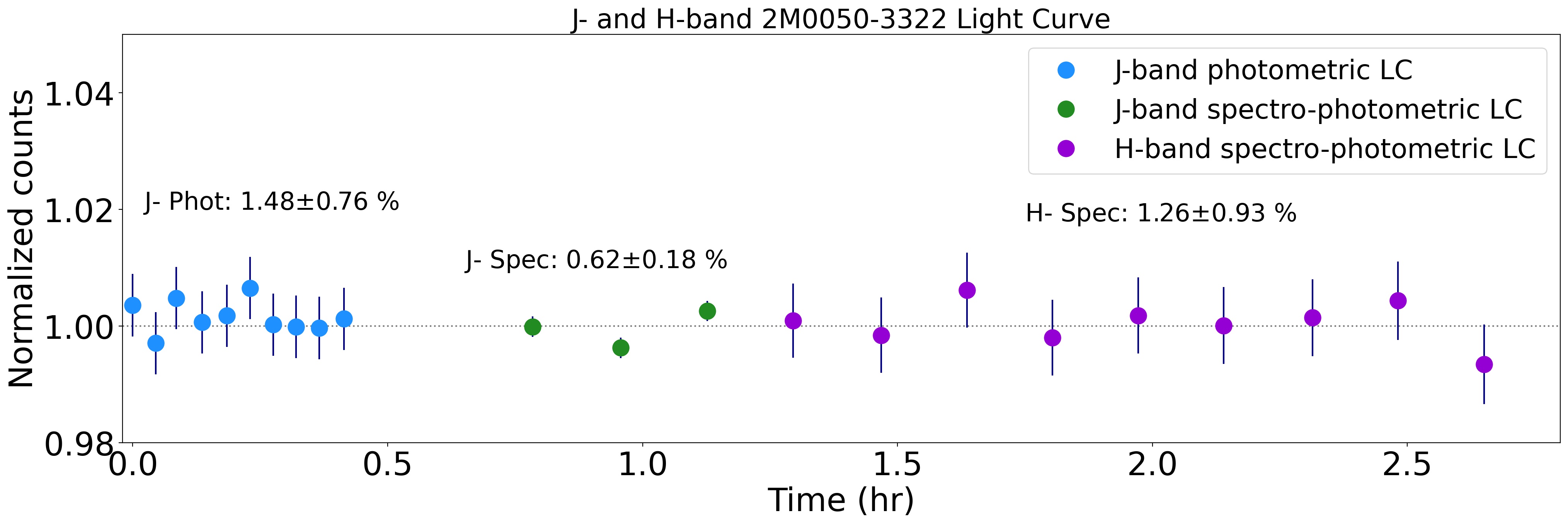}
    \includegraphics[width=0.98\textwidth]{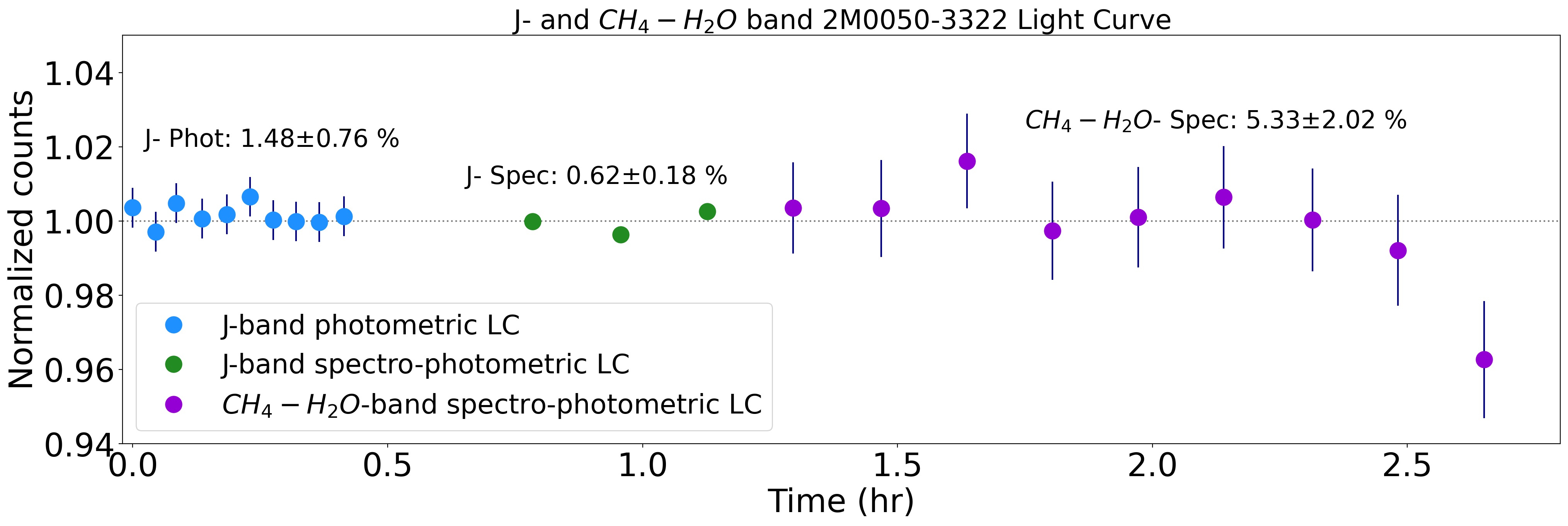}
    \caption{Corrected $J$-imaging, $J$-band, $H$-band and $\mathrm{CH_{4}-H_{2}O}$ light curves of  2M0050--3322, {and the peak-to-peak amplitudes of their fluctuations}.}
    \label{target_allLCs}
\end{figure*}

As in \cite{Manjavacas2021}, \textit{PypeIt}'s formal instrumental uncertainties underestimate the uncertainties of 2M0050--3322 light curve, since it does not necessarily account for spurious variability introduced by changes in the Earth's atmosphere during the observation. Thus, as we did in \cite{Manjavacas2021}, we added the mean of the $\sigma_{s}$ calculated for the target and the selected calibration stars as the uncertainty for each point in the light curve of the target ({0.0057}).
The non-corrected light curves of 2M0050--3322 are shown in Fig. \ref{target_JimLCs}, \ref{target_JLCs}, \ref{target_HLCs}, \ref{target_CH4LCs} left, and the corrected light curves in Fig. \ref{target_JimLCs}, \ref{target_JLCs}, \ref{target_HLCs} and \ref{target_CH4LCs} right. {All corrected and normalized light curves are shown together in Fig. \ref{target_allLCs}. In this figure {every two} data points of the original $J$-band imaging light curve in Fig. \ref{target_JimLCs} were combined for clarity}.

\begin{table*}
	\caption{Statistics of the $J$-band imaging, $J$- and $H$-band spectro-photometric light curves of 2M0050--3322, and {the peak-to-peak amplitudes of their fluctuations.}}  
	\label{stats_target}
	\centering
	\begin{center}
		\begin{tabular}{ccccc}
        \hline
		\hline 
			
Light curve (LC) & $\sigma$ non-corrected LC & $\sigma$ corrected LC & {Peak-to-peak  non-corrected LC} & {Peak-to-peak  corrected LC} \\   
\hline

$J$--imaging  & 0.0153  & 0.0034 & $\sim$5.69 \% &  1.48$\pm$0.75 \% \\
$J$--spec  & 0.0002 & 0.0039  & 0.20$\pm$0.08 \% & 0.62$\pm$0.18 \% \\
$H$--spec   & 0.0027  & 0.0036 &  1.87$\pm$0.43 \% & 1.26$\pm$0.93 \% \\
$\mathrm{CH_{4}-H_{2}O}$--spec   & 0.0095  & 0.0024 &  4.17$\pm$1.86 \% & {5.33$\pm$2.02} \% \\

			\hline			
		\end{tabular}
	\end{center}

\end{table*}

\section{Correlation Analysis}\label{correlation_analysis_section}

In this section, we analyze the different  sources of external spurious contamination in the $J$-imaging, $J$-band spectrophotometric data, the $H$-band spectrophotometric, and {the $\mathrm{CH_{4}-H_{2}O}$} data that could potentially lead to {non-astrophysical}, spurious variability introduced in the 2M0050--3322 light curves. {We  measure the Kendall $\tau$ correlation between different parameters, since it is a more robust correlation estimator than other estimators \citep{Croux2010}. In addition, we provide the significance (p-value) of each correlation, where numbers close to 0.0 suggest that the correlation obtained is significant. We consider that there is no correlation for Kendall $\tau$ values close to 0.0, a weak correlation for absolute Kendall $\tau$ values between 0.1 and 0.4, a moderate correlation for values between 0.4 and 0.7, and a strong correlation for values above 0.7}.

\subsection{{Comparison} between Stars and Target Light Curves}\label{corr_LC}
 
To evaluate the effects of potential contamination on the target's light curve due to the Earth's atmosphere, {we investigate the correlation between the non-corrected and corrected light curves of the target, and the comparison stars in the $J$- and $H$-band.  }

\subsubsection{$J$-band data}

{We calculate the Kendall's $\tau$ coefficients to estimate the correlation between the  target's $J$-band imaging data and the "good" calibration stars' light curves}. We find weak to {moderate} correlation for some calibration stars, with correlations varying from {-0.66 (p-value = 1.15e-05)}, to 0.65 (p-value = 1.75e-05). When we correct the calibration stars' and the target's light curves, we remove also most of the correlations between them, obtaining Kendall's $\tau$ coefficients between 0.07 (p-value = 0.67), and -0.34 (p-value = 0.03). In Fig. \ref{corr_noncorr_stars_Jband} and \ref{corr_corr_stars_Jband} of the Appendix we show the correlations of the "good" calibration stars light curves before and after correction.

We do not calculate these correlations for the $J$-band spectrophotometric data, since we have only three data points, which would not provide meaningful correlations.

\subsubsection{$H$-band  data}

{The Kendall's $\tau$ coefficients  suggests a weak to a moderate correlation between the $H$-band target's and "good" calibration stars light curves, depending on the   star}. The Kendall $\tau$ correlation coefficients vary between 0.44 (p-value = 0.12) for stars 1, 3 and 5, and 0.61 (p-value = 0.02) for the {calibration} star 2. In Fig. \ref{corr_noncorr_stars_Hband} in the Appendix, we show the correlation plots between the target and each of the stars that we use for calibration.

{After correcting the light curves of 2M0050-3322 and the calibration stars} using the method explained in Section \ref{light_curves}, we run the Kendall $\tau$ non-parametric correlation test again, finding correlation coefficients that range between 0.28 (p-value = 0.37) for star 1, and 0.55 (p-value = 0.04) for the calibration star 2, suggesting  non to a moderate correlation  (see Fig. \ref{corr_corr_stars_Hband} in the Appendix). We show all stars and the target's non-corrected and corrected light curves in Fig. \ref{all_LCs_plot} of the Appendix.

\subsubsection{$\mathrm{CH_{4}-H_{2}O}$ region}\label{methane_water_band}

%corr_noncorr_stars_CH4_H2O_band

{The Kendall's $\tau$ coefficients  suggests a weak  correlation between the $\mathrm{CH_{4}-H_{2}O}$ target's and "good" calibration stars light curves, depending on the   star. The Kendall $\tau$ correlation coefficients vary between 0.00 (p-value = 1.0) for star 2, and 0.22 (p-value = 0.47) for the {calibration} star 1. In Fig. \ref{corr_noncorr_stars_CH4_H2O_band} in the Appendix, we show the correlation plots between the target and  the stars that we use for calibration.}

{After correcting the light curves of 2M0050-3322 and the calibration stars using the method explained in Section \ref{light_curves}, we run the Kendall $\tau$ non-parametric correlation test again, finding correlation coefficients of 0.11 (p-value = 0.76) for star 1, and -0.17 (p-value = 0.61) for star 2  (see Fig. \ref{corr_corr_stars_CH4_H2O_band} in the Appendix). We show the stars and the target's non-corrected and corrected light curves in Fig. \ref{corr_corr_stars_CH4_H2O_band} of the Appendix.}

\subsection{Correlation with Full Width Half Maximum and centroids of the Spectra}\label{corr_fwhm}

For the $H$-band spectrophotometric data, we obtained spectra following an ABBA pattern. Thus, the slit losses might vary slightly at A and B positions, and also with the tracking of the telescope, and the flexure compensation of the instrument potentially influencing the measured variability of the target. These effects might be relevant since we observed at high airmasses (1.79--2.37). We investigated a potential relationship between the {fluctuations} found for 2M0050--3322,  the Full Width Half Maximum (FWHM) of the 2D spectra, and centroids of those 2D spectra  during the $\sim$1.4~hr of $H$-band monitoring with MOSFIRE. We do not perform this analysis for the $J$-band spectrophotometric data since we have only three data points. {We do not perform this analysis at the $\mathrm{CH_{4}-H_{2}O}$ region, since the signal-to-noise is relatively low in this region}.

We measured the FWHM at {approximately} the maximum flux of coadded {combined} ABBA spectra  between pixel x = 500, and  pixel x = 550 {(spatial direction)},  and then we calculated the FWHM at this position for each ABBA coadd using a {1D gaussian fit}. We show the evolution of the FWHM in Fig. \ref{evol_FWHM_centroids}, left of the Appendix. Similarly, we calculated the centroids of the 2D spectra using the same gaussian fit (Fig. \ref{evol_FWHM_centroids}, right of the Appendix). 

We calculated the Kendall $\tau$ correlation {between} the mean FWHM for each 2D spectrum and the non-corrected and corrected $H$-band light curve. We obtained a weak to moderate correlation between the FWHM and  the non-corrected ($\tau$ = -0.22, p-value = 0.48) and corrected 2M0050--3322 light curves ($\tau$ = -0.44, p-value = 0.12). These correlations are shown in Fig. \ref{corr_FWHM} of the Appendix.

{Finally, we obtained a strong correlation between the non-corrected 2M0050--3322 light curve and the measured centroids ({centers {as} obtained from the {1D} gaussian fit}) of the 2D spectra ($\tau$ = 0.94, p-value = 4.9e-5). This correlation disappeared when we corrected the target's  light curve ($\tau$ = 0.06, p-value = 0.91), see Fig. \ref{corr_pointing} of the Appendix}.

\subsection{Correlation with Atmospherical Parameters}\label{meteo}

The evolution of atmospheric conditions during the observation might influence the amount of flux collected by MOSFIRE, affecting simultaneously the target and the calibration stars. Namely, the most relevant factors that might potentially affect our observations are: the humidity content, the external temperature, wind speed, seeing and the airmass. The evolution of these parameters are registered in the header, and/or in the Mauna Kea Weather Center webpage (\url{http://mkwc.ifa.hawaii.edu}).

\subsubsection{{$J$-band  light curves}}

We calculated the $\tau$ Kendall correlation coefficient between the non-corrected and the corrected $J$-band imaging light curve for 2M0050--3322, and each of the atmospheric parameters mentioned above. We {found no to weak correlation} between the non-corrected $J$-band target's light curve and the humidity content ($\tau$ = 0.04, p-value = 0.79),  the external temperature ($\tau$ = 0.15, p-value = 0.35), and wind speed ($\tau$ = -0.16, p-value = 0.39). We found a weak correlation with the seeing ($\tau$ = -0.37, p-value = 0.023), and with the airmass ($\tau$ = 0.34, p-value = 0.04). The plot showing all correlations can be found in Fig. \ref{J_band_noncorr} in the Appendix. We did not find any correlation between the corrected $J$-band imaging light curve and the atmospheric parameters, as can be seen in Fig. \ref{J_band_corr} in the Appendix.

\subsubsection{{$H$-band  light curves}}

We calculated the $\tau$ Kendall correlation coefficient between the non-corrected, and corrected $H$-band spectrophotometric light curves for 2M0050--3322, and each of the atmospheric parameters mentioned above. We found a strong correlation between the airmass and the non-corrected $H$-band light curve of 2M0050--3322 ($\tau$ = 0.72, p-value = 0.005), a moderate correlation with the humidity content  ($\tau$ = 0.48, p-value = 0.07), and a weak anticorrelation with the seeing ($\tau$ = -0.39, p-value = 0.18), the wind speed ($\tau$ = -0.29, p-value = 0.29), and the external temperature ($\tau$ = -0.22, p-value = 0.48). The correlation plots can be found in Fig. \ref{H_band_noncorr} of the Appendix.

After correcting the $H$-band light curve, we found  a moderate  correlation with the external temperature ($\tau$ = 0.44, p-value = 0.12), and a weak anticorrelation with the humidity content ($\tau$ = -0.20, p-value = 0.46), and the seeing ($\tau$ = -0.16, p-value = 0.61). A weak correlation with the wind speed ($\tau$ = 0.11, p-value = 0.67),  and a no correlation with the airmass ($\tau$ = -0.06, p-value = 0.92). The plots showing all correlations can be found in Fig. \ref{H_band_corr} in the Appendix.

\subsubsection{{$\mathrm{CH_{2}-H_{2}O}$  light curves}}

{We calculated the $\tau$ Kendall correlation coefficient between the non-corrected, and corrected $\mathrm{CH_{2}-H_{2}O}$  spectrophotometric light curves for 2M0050--3322, and each of the atmospheric parameters mentioned above. We found no correlation between the airmass and the non-corrected $\mathrm{CH_{2}-H_{2}O}$ light curve of 2M0050--3322 ($\tau$ = 0.0, p-value = 1.0), a no correlation with the humidity content  ($\tau$ = 0.08, p-value = 0.75), and the seeing ($\tau$ = 0.0, p-value = 1.0). We measure a weak correlation with the wind speed ($\tau$ = -0.11, p-value = 0.67), and the external temperature ($\tau$ = -0.39, p-value = 0.18). The correlation plots can be found in Fig. \ref{CH4-H2O_noncorr} of the Appendix.}

{After correcting the $\mathrm{CH_{2}-H_{2}O}$ light curve, we found  a moderate  correlation with the external temperature ($\tau$ = 0.61, p-value = 0.02), and the seeing ($\tau$ = 0.44, p-value = 0.12), and a moderate anticorrelation with airmass ($\tau$ = -0.44, p-value = 0.12). We find a weak anticorrelation with the humidity content ($\tau$ = -0.25, p-value = 0.35), and a weak correlation with the wind speed ($\tau$ = 0.11, p-value = 0.67). The plots showing all correlations can be found in Fig. \ref{CH4-H2O_corr} in the Appendix. }

\subsection{Differential Chromatic Refraction}

The differential chromatic refraction (DCR) produces an astrometric shift in the position of a source when observed at high airmasses with respect the same source observed at low airmasses, because of the dependence of the refractive index of the air with wavelength \citep{Filippenko1982}. The DCR effect is higher at higher airmasses, and at shorter wavelengths, being very relevant in the blue optical. The DCR also depends on the external temperature, atmospheric pressure and the spectral type (colors) of the astronomical sources \citep{Stone1996}.

 A shift in the position of the spectra in the MOSFIRE detector could cause spurious flux changes due to pixel-to-pixel variation effects. Since our observations were performed at high airmasses, between 1.79 and 2.37, we investigate the magnitude that this effect might have in our spectro-photometric measurements in the $H$-band. We used the equation in {\cite{Stone1996}} to estimate the magnitude of the DCR effect in a maximum of $\sim$10~mas, which is about 1/20 of a MOSFIRE pixel (MOSFIRE scale plate of  0.1798”/pix). Given that 2M0050--3322 does not have very distinct colors from the rest of the calibration stars in the field (see Table \ref{colors}), we expect {the effect to affect all sources similarly}.

\begin{table*}
    \caption{2MASS $J$, $H$, and $K$ magnitudes and $J-H$, $J-K$ and $H-K$ colors for 2M0050--3322 and calibration stars.}
    \label{colors}
    \begin{center}
    \begin{tabular}{ccccccccccccc}

       \hline
       \hline
                   & $J$      & e$J$    & $H$      & e$H$    & $K$      & e$K$    & $J-H$  & e$J-H$  & $J-K$   & e$J-K$  & $H-K$   & e$H-K$   \\
       \hline
2M0050---3322  & 15.928 & 0.070 & 15.828 & 0.191 & 15.241 & 0.185 & 0.100 & 0.203 & 0.687 & 0.198 & 0.587 & 0.266  \\
Star 1   & 15.642 & 0.062 & 15.274 & 0.113 & 14.892 & 0.150 & 0.368 & 0.129 & 0.750 & 0.162 & 0.382 & 0.188  \\
Star 2   & 14.926 & 0.047 & 14.475 & 0.057 & 14.310 & 0.088 & 0.451 & 0.074 & 0.616 & 0.100 & 0.165 & 0.105  \\
Star 3    & 16.561 & 0.160 & 15.905 & 0.207 & 15.891 & 0.336 & 0.656 & 0.262 & 0.670 & 0.372 & 0.014 & 0.395  \\
Star 5    & 16.146 & 0.098 & 15.524 & 0.152 & 15.178 &       & 0.622 & 0.181 & 0.968 & 0.098 & 0.346 & 0.152  \\
%Star 6   & 15.642 & 0.062 & 15.274 & 0.113 & 14.892 & 0.150 & 0.368 & 0.129 & 0.750 & 0.162 & 0.382 & 0.188 \\
\hline
    \end{tabular}
    \end{center}
\end{table*}

\section{Results} \label{results}

\subsection{Photometric  {fluctuations}}\label{phot_variability}

{The total time of $J$- and $H$-band monitoring is $\sim$2.6~hr, with a gap of $\sim$25 min between the $J$-band imaging and spectroscopic data. We observed $\sim$0.45~hr in $J$-imaging mode, $\sim$0.35~hr in $J$-band spectrophotometric mode, and 1.4~hr in $H$-band spectrophotometric mode}, covering almost two rotational periods of 2M0050--3322 (P = 1.55$\pm$0.02~hr, \citealt{Metchev2015}). As mentioned above, the three corrected light curves are shown in Fig. \ref{target_allLCs}. {We measured a peak-to-peak {(maximum to minimum)} amplitude on their fluctuations of 1.48$\pm$0.75\% for the $J$-band imaging light curve, 0.62$\pm$0.18\% $J$-band spectrophotometric light curve, and a 1.26$\pm$0.93\% for the $H$-band spectrophotometric light curve}. The $CH_{4}-H_{2}O$ light curve shows  relatively higher  amplitude of {5.33$\pm$2.02\%} (see Fig. \ref{target_CH4LCs}, right).

\subsubsection{BIC Test for Significant Variability} \label{BIC_test}

To test the significance of the observed fluctuations in the light curve of 2M0050--3322, we used the Bayesian Information Criterion (BIC) as in \cite{Manjavacas2021}, \cite{Vos2020} and \cite{Naud2017}. The BIC is defined as:

\begin{equation}
    \mathrm{BIC}=-2~\mathrm{ln}~\mathcal{L}_\mathrm{max} + k ~\mathrm{ln}~ N
\end{equation}
where $\mathcal{L}_\mathrm{max}$ is the maximum likelihood achievable by the model. {The likelihood function, $\mathcal{L}$, is given by} 
\begin{equation}
  \mathcal{L} =  \prod_{i=1}^{n} \frac{1}{\sqrt{2 \pi s^2}}  e ^ {\frac{-(y_i - f(\theta))^2}{2s^2}}	  
\end{equation}
{where $s$ is the flux uncertainty, $y_i$ is the measured flux, $f(\theta)$ is the model flux. $\mathcal{L}_\mathrm{max}$ is obtained by maximizing this function. } $k$ is the number of parameters in the model. {The flat model has one parameter -- the constant relative flux value. The sinusoidal model has four parameters -- the constant flux value as well as the amplitude, period and phase. The BIC thus penalizes the sinusoidal model for having additional parameters compared with the flat model.} $N$ is the number of data points used in the fit. {$\Delta\mathrm{BIC}>0$ indicates that the sinusoidal model is favored and $\Delta\mathrm{BIC}<0$  indicates that the non-variable, flat model is favored. A $|\Delta\mathrm{BIC}|$ value between 0 and 6 indicates that one model is positively favored over the other, a value between 6 and 10 indicates that one model is strongly favored over the other and values above 10 indicate that one model is very strongly favored over the other \citep{Schwarz1978}.}

\begin{figure}
    \centering
    \includegraphics[width=0.50\textwidth]{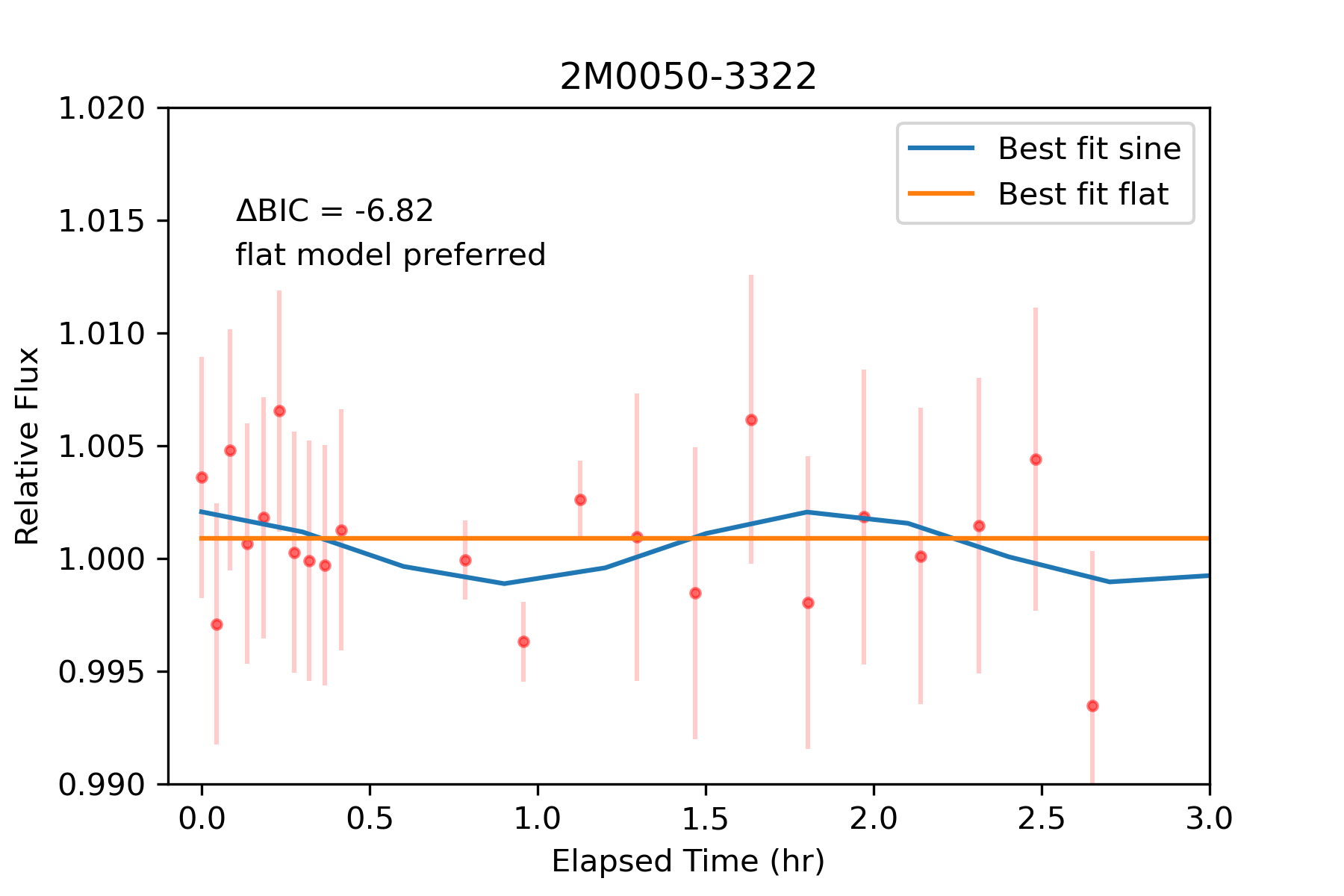}
   \caption{{2M0050--3322 light curve is shown by the orange points, with the best-fit non-variable (flat) model shown in orange, and variable (sinusoidal) model shown in blue. The BIC test shows that the flat model is strongly favored by the light curve.}}
    \label{BIC_plot}
\end{figure}

We calculate $\Delta\mathrm{BIC} = \mathrm{BIC}_{flat} - \mathrm{BIC}_{\mathrm{sin}}$ to assess whether a variable sinusoidal or non-variable flat model is favored by the data.   The sinusoidal and flat model are shown in Fig.~\ref{BIC_plot}. We obtained a $\Delta\mathrm{BIC}$ value of -6.82, that implies that the flat model is  strongly preferred over a sinusoidal  model. A similar result was obtained if we run the BIC test on the individual $J$, $H$-band and $CH_{4}-H_{2}O$ light curves. {The fact that the fluctuation of the $CH_{4}-H_{2}O$ light curve are higher than for the $J$- and $H$-band light curves, is due to the last point of the $CH_{4}-H_{2}O$ light curve having a lower flux than the rest, which explains why the BIC test still finds no significant variability.}

To understand which conditions the 2M0050--3322 light curve should meet so that the BIC analysis favors the sinusoidal model versus the flat model, we artificially modified {the} 2M0050-3322 light curve {by} changing  the uncertainties of each point (brighter target). {To obtain a statistically significant result, we use a Monte Carlo simulation to produce synthetic light curves similar to 2M0050--3322 light curve. We generated 100 synthetic light curves using the original light curve redefining each data point using a Gaussian random number generator. The mean of the Gaussian is the measured flux of the original light curve, and the standard deviation is the photometric uncertainty of each point in the light curve, but divided by a factor. These factors are 1 (similar light curve to the original one), 2, 3 and 4. We generate 100 light curves with each factor used to reduce the uncertainties of the light curve. Then we carry the BIC analysis again, and find which conditions  the light curve {must meet} to be considered significantly variable.  For the light curves generated using the uncertainties of the original 2M0050 light curve, we found that ~48\% of the times the light curve was flagged as variable, and ~52\% as non-variable. For the light curves generated using the uncertainties divided by 2, we found that ~2\% of the light curves are significantly variable, and ~98\% non-variable. For the light curves generated with the uncertainties of 2M0050 divided by 3, and by 4, we found that \textit{none} of the light curves was found significantly variable.}

%{Secondly, the  amplitudes we measured in each individual light curve is limited by the time of the observation for each of them, which does not even cover one 2M0050--3322 rotation, thus, the total peak-to-peak  amplitude could be higher. To mitigate the lack of a longer monitoring, we artificially generate a longer 2M0050--3322 light curve using a Monte Carlo simulation. We generate 99 synthetic light curves using the original light curve redefining each data point using a Gaussian random number generator. The mean of the Gaussian is the measured flux of the original light curve, and the standard deviation is the photometric uncertainty of each point. Then we end up with a $\sim$300~hr simulated light curve for 2M0050--3322. Finally, we performed the BIC analysis on this light curve, concluding that the BIC also strongly favors the flat model over the sinusoidal model ($\Delta$BIC = -27,012).}

\subsubsection{MOSFIRE sensitivity} \label{MOSFIRE_sensitivity}

\begin{figure}
    \centering
    \includegraphics[width=0.51\textwidth]{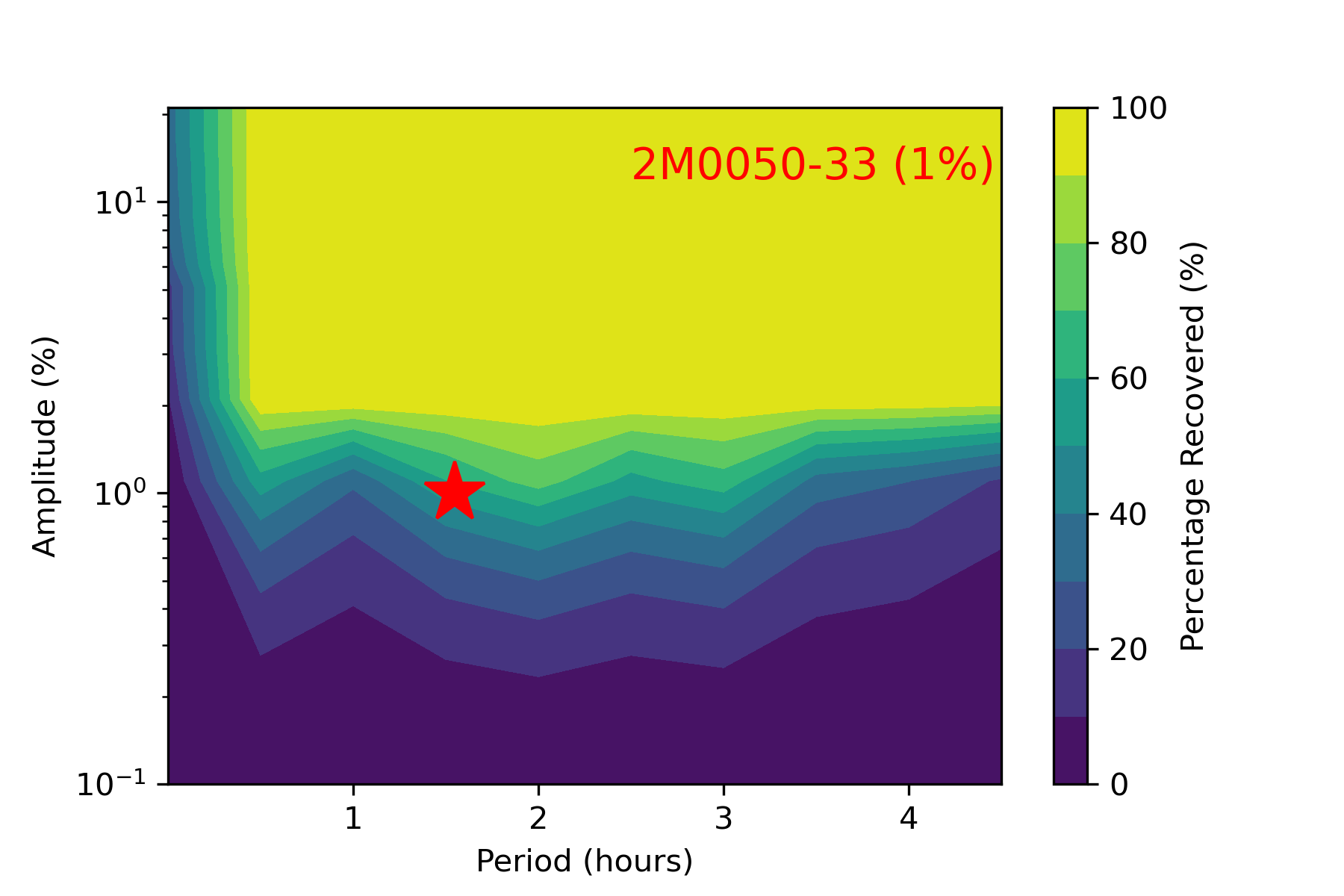}
    \caption{{Sensitivity plot for Keck\,1/MOSFIRE. MOSFIRE would be able to measure the predicted light curve for a T7.0 brown dwarf similar to 2M0050--3322  50-60\%  of the times with the observation strategy followed in this work. }}
    \label{sensitivity_plot}
\end{figure}

We estimated the sensitivity of MOSFIRE to detect {a light curve with a significant variability amplitude, and with a similar amplitude to the fluctuations of the 2M0050--3322 light curve ($\sim$1\%)}. This test  allow us to determine the upper limit of the variability amplitude detectable with MOSFIRE. To create the sensitivity plot, we followed the same approach as for \cite{Vos2020, Vos2022}, injecting sinusoidal curves into artificial light curves created using a Gaussian-distributed noise.
{Simulated amplitudes were in the range $0.01-22\%$ in steps of 1\% and period were in the range $0.001-5~$hr in steps of $0.5~$hr. We also simulated 1000 random phases for each set of (A,P) paramters, resulting in a total of 220,000 simualted light curves. }
Then, using a Lomb-Scargle periodogram, we estimated the probability of detection as the percentage of light curves with a given variability amplitude and period that produce a periodogram power above  a given threshold {(1\% false-alarm probability)}. We show the sensitivity plot in Fig. \ref{sensitivity_plot}.  We placed a red star in Figure \ref{sensitivity_plot} for {the peak-to-peak  amplitudes of the fluctuations} measured for the $J$- and $H$-band light curves  with a rotational period of 1.55$\pm$0.02~hr \citep{Metchev2015}.

The sensitivity plot concludes that MOSFIRE would be able to detect a 1.0\% variability amplitude light curve  between 50--60\% of the times. Thus, given the sensitivity of MOSFIRE with the observation strategy followed in this work, we might not be able to detect significant variability in 40--50\% of the cases. {MOSFIRE would be able to detect between 90--100\% of the cases a light curve with a minimum variability amplitude of 2\% for an object with a similar magnitude to 2M0050--3322.}

%We estimate the rotational period of 2M0050--3322 using the $J$- and $H$-band light curve (Fig. \ref{light_curves}) using  a Bayesian Generalized Lomb-Scargle (BGLS) periodogram \citep{Mortier2015}, that takes into account the uncertainties of the flux, and provides most robust results for light curves with gaps, like the one presented here. The  BGLS periodogram a period of $\sim$1.94~hr (see Fig. \ref{LS_periodogram} in the Appendix). {Although this estimated period is relatively close to the 1.55$\pm$0.02~hr period calculated by \cite{Metchev2015}, we need to consider that our data are relatively scarce, since we do not cover two rotational periods of the target, and that the maximum peaks of both periodograms are not much more prominent that other peaks shown in both periodograms (see Fig. \ref{LS_periodogram} in the Appendix). In addition, the BGLS periodogram also suggests the possibility of another period longer than 5~hr. Longer monitoring is needed to confirm either rotational period}.

%{We do not have enough time coverage to observe a full rotational period of the target (3.5$\pm$0.2~hr, \citealt{Metchev2015}), but still we  searched for other periods on the $J$-band light curve using a Lomb-Scargle periodogram \citep{Lomb1976, Scargle1982, Horne_Baliunas1986}, and a Bayesian Generalized Lomb–Scargle (BGLS) Periodogram \citep{Mortier2015} which did not find any periodicity in the $J$-band light curve}.

\subsection{Spectral  {fluctuations}} \label{spec_variability}

%We explored the amplitude of the variability as a function of the wavelength by comparing the maximum and the minimum flux spectra among the 13 spectra obtained. In Figure \ref{ratio}, left, we show the brightest and faintest  spectrum, indicating the molecular and atomic absorption features for 2M2208+2921. 

\begin{figure*}
\vspace{0.7cm}
    \centering
    \includegraphics[width=0.45\textwidth]{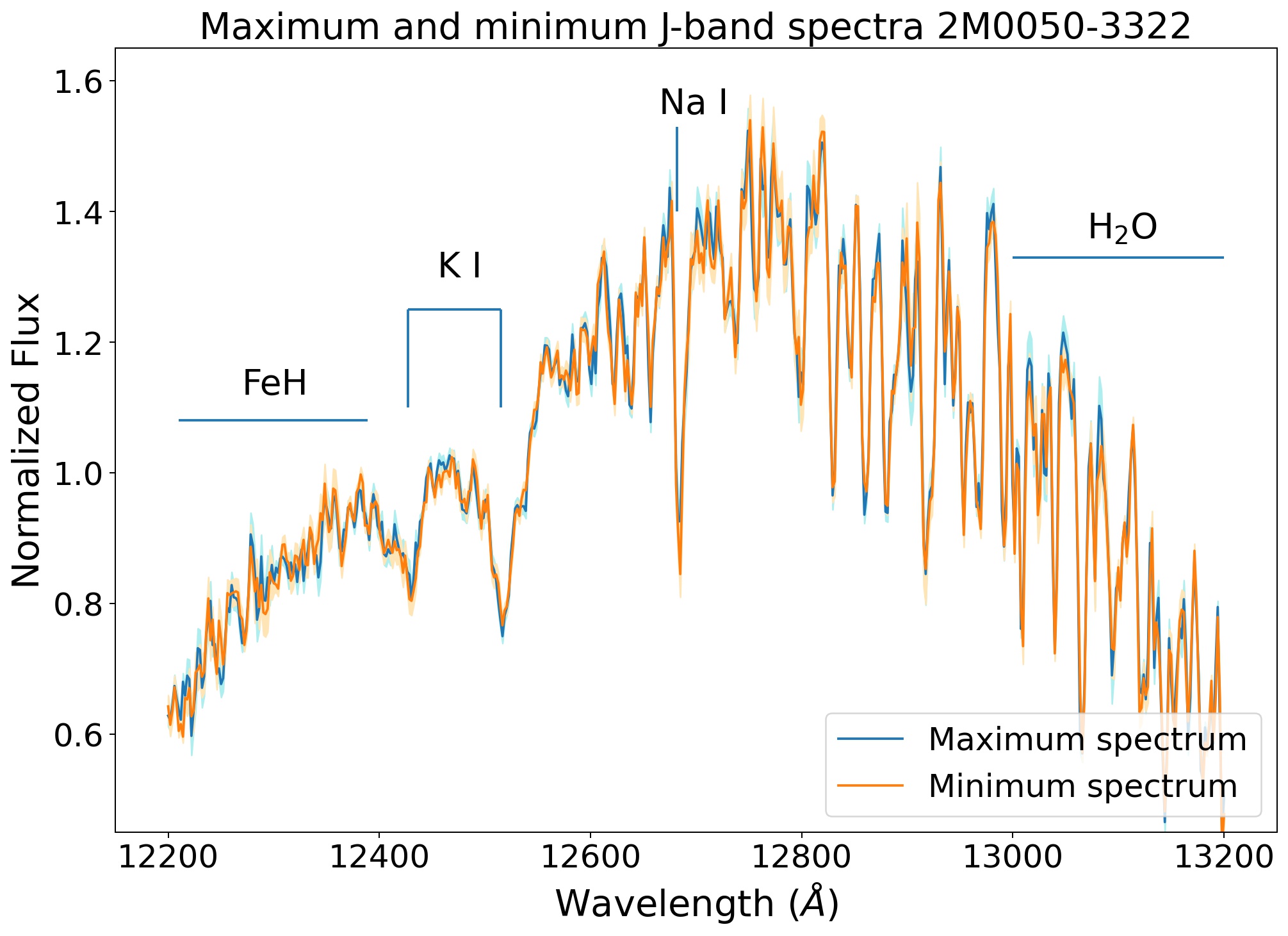}
    \includegraphics[width=0.45\textwidth]{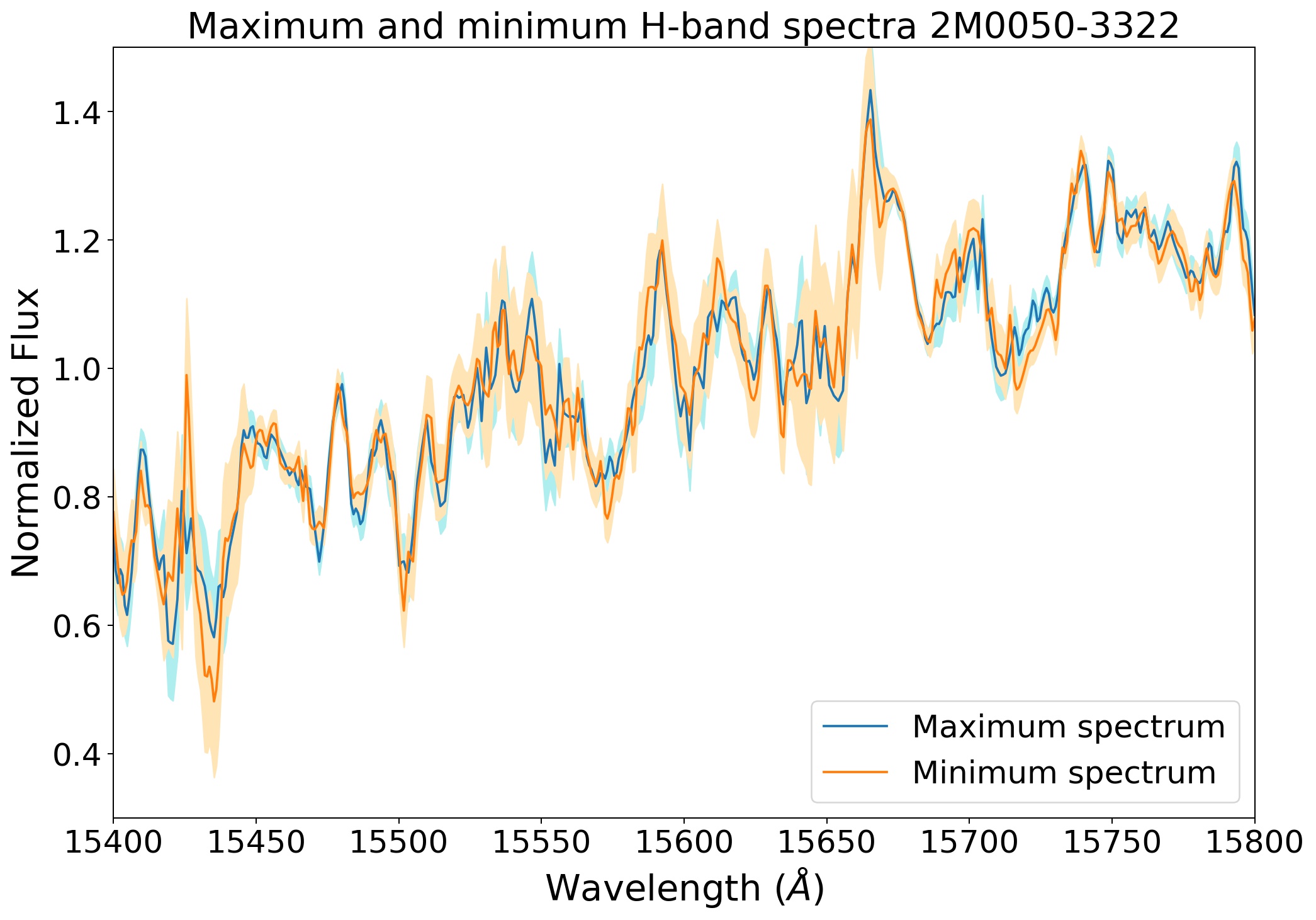}
    \caption{Left: Maximum flux $J$-band spectrum for 2M0050-332 (orange), and minimum  $J$-band spectrum for 2M0050-332 (blue). Right: Maximum flux $H$-band spectrum for 2M0050-332 (orange), and minimum  $H$-band spectrum for 2M0050--3322 (blue).}
    \label{max_min}
\end{figure*}

\begin{figure*}
\vspace{0.7cm}
    \centering
    \includegraphics[width=0.45\textwidth]{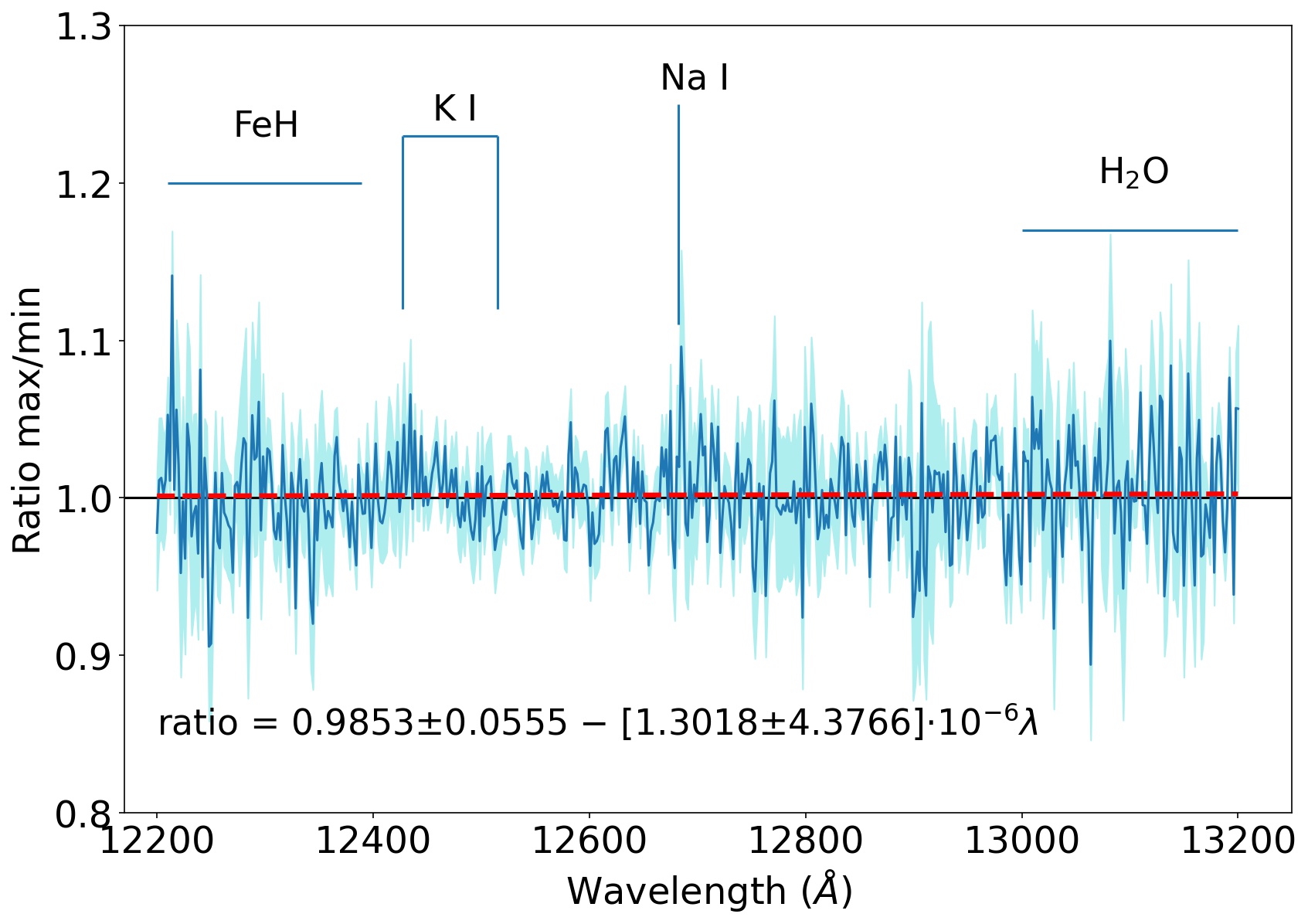}
    \includegraphics[width=0.45\textwidth]{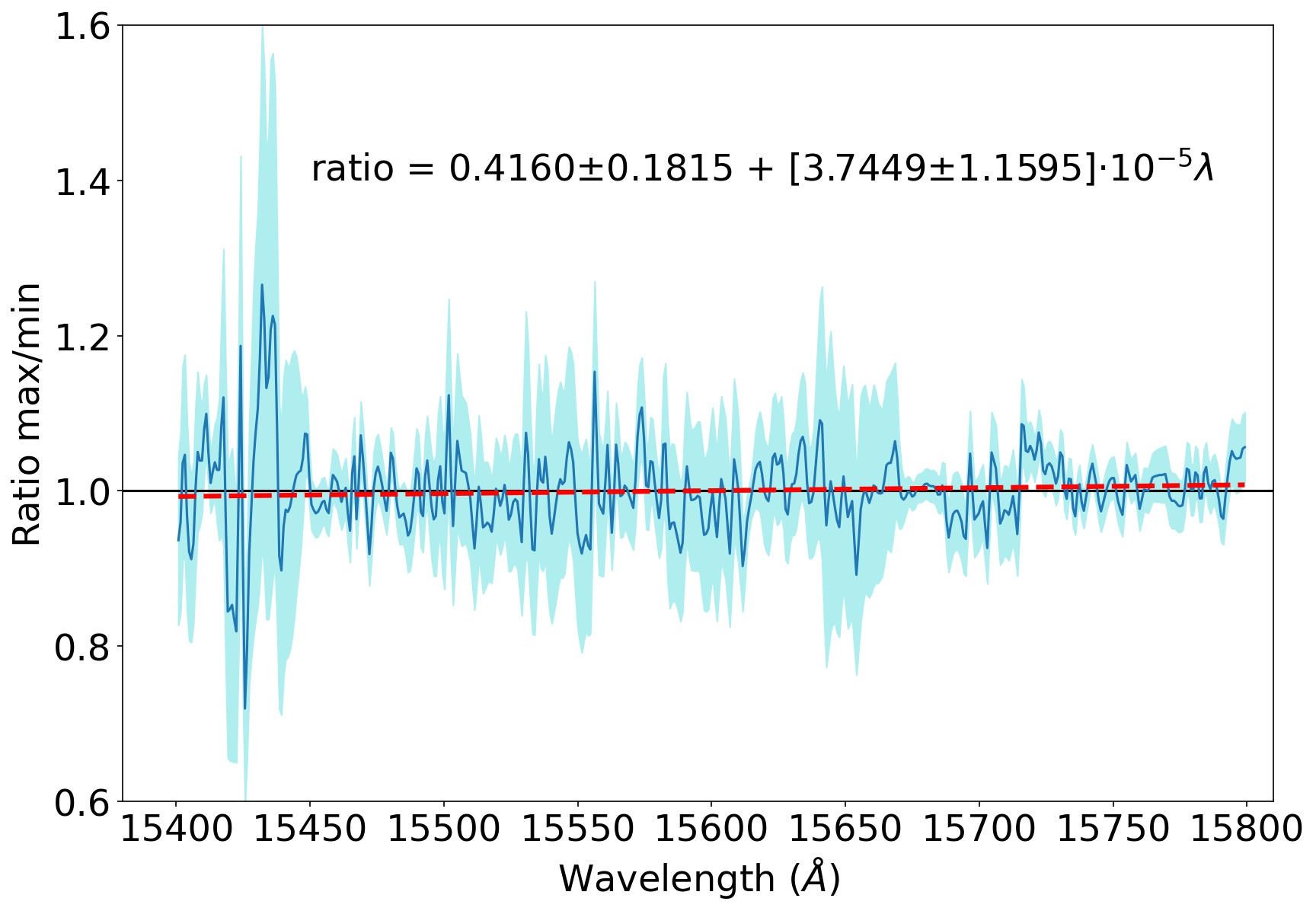}
    \caption{Left: Ratio of the maximum divided by the minimum spectra of 2M0050--3322 in the $J$-band. Right: Ratio of the maximum divided by the minimum spectra of 2M0050--3322 in the $H$-band. }
    \label{ratio}
\end{figure*}

{Given that we did not detect significant variability for 2M0050--3322 in Section \ref{phot_variability}, the spectral peak-to-peak fluctuations measured, and its wavelength dependence presented in this section are only  tentative}. We measured the {peak-to-peak fluctuations dependence as a function of wavelength} by {dividing} the maximum and minimum flux spectra in the $J$ and $H$-band spectrophotometric data. In Fig. \ref{max_min} we show the maximum and minimum flux spectra for both filters. In Fig. \ref{ratio} we show the ratio of those spectra in the $J$ and $H$-band, i.e. the {relative peak-to-peak fluctuations} across the wavelength range of those bands with their uncertainties, indicating the molecular and atomic absorption features of 2M0050--3322. We fit a line to the ratio of the maximum and minimum flux spectra using the \textit{numpy.polyfit} \textit{Python} library to the $J$ and $H$-band ratio independently. We obtained a slope consistent with zero for the $J$-band ratio ($ratio = 0.9853\pm0.0555 -- [1.3018\pm4.3766]\times 10^{-6}\lambda$, see Fig. \ref{ratio}, left), and a tentative positive slope (3-$\sigma$) to the $H$-band ratio ($ratio = 0.4160\pm0.1815+[3.7449\pm1.1595]\times 10^{-5}\lambda$, see Fig. \ref{ratio}, right), suggesting that the peak-to-peak fluctuations might increase slightly from 15400 to 15800~$\AA$. {Within the 15400 to 15800~$\AA$ wavelength range the values of the ratio are consistent within the uncertainties. The value of the ratio at 15400~$\AA$ is 0.9927$\pm$0.2547 and 1.0077$\pm$0.2579 at 15800~$\AA$. Thus, we cannot measure a significant ratio increase within this wavelength range, although the slope of the $H$-band ratio is significantly positive}.

\section{{Discussion}}\label{discussion}

{The 2M0055--3322 $J$-band imaging light curve shows fluctuations with a peak-to-peak amplitude  of 1.48$\pm$0.75\%. The $J$-band spectroscopic light curve shows fluctuations with a  peak-to-peak amplitude of 0.62$\pm$0.18\% {(with just three data points)}. 
The  $H$-band spectrophotometric light curve has fluctuations with peak-to-peak amplitude of 1.26$\pm$0.93\%, compatible with the amplitude derived for the $J$-band imaging and spectrophotometric light curves (Fig. \ref{target_allLCs})}. Finally, the fluctuation of the $\mathrm{CH_{4}-H_{2}O}$ light curve is 5.33$\pm$2.02\%. \cite{Radigan2014} did not detect significant variability in their imaging $J$-band light curve for 2M0050--3322, but provided an upper limit of 1.1\% peak-to-peak amplitude, with the WIRC camera in the Du Pont telescope in Las Campanas observatory, which is a 2.5~m telescope. \cite{Metchev2015} measured a variability amplitude of $<$0.59\% for the [3.6] Spitzer channel light curve, and an amplitude of 1.07$\pm$0.11\% in the [4.5] Spitzer channel light curve. {The last value is consistent with the amplitudes measured for the $J$-band and $H$-band light curves}, even though these bands trace deeper layers of the atmosphere of 2M0050--3322 than the Spitzer mid-infrared channels \citep{Yang2016}, {and the observations were separated in time by several years}.

{The BIC analysis presented in Section \ref{BIC_test} favors the flat light curve model versus a sinusoidal model, indicating that the peak-to-peak fluctuations measured in the 2M0050--3322 light curve might suggest in the best case  tentative low-level variability.}

%In addition, as explained in Section \ref{spec_variability}, there is no wavelength dependence for the variability found in the $J$-band, since the slope derived for the ratio between the maximum and minimum spectra is compatible with zero. In contrast, in the $H$-band, there is a weak wavelength dependence of the variability, that increases slightly with wavelength. 

Few mid and late-T dwarfs have been monitored for variability at all. For completeness we include a list of all mid and late-T dwarfs with upper limits or measurements of variability (see Table \ref{table_variables}).{ Furthermore, only two other late-T dwarfs have been monitored for spectro-photometric variability: 2MASS~J22282889--431026 (T6.5,  \citealt{Buenzli2012}), and Ross~458c (T7.5, \citealt{Manjavacas2019b}), both with HST/WFC3 and its G141 grism.
Comparing the light curves of these two late-T brown dwarfs, their variability amplitudes, and wavelength variability dependence we find several differences and commonalities between them. 2MASS~J22282889--431026 (2M2228-4310) and Ross~458c have sinusoidal light curves {(from  \citealt{Buenzli2012} and \citealt{Manjavacas2019b}, respectively)}.  Their  $J$-band variability amplitude is smaller than  the $H$-band (1.85$\pm$0.07\% vs 2.74$\pm$0.11\% for 2M2228-4310, and 2.62$\pm$0.02\% vs 3.16$\pm$1.26\% for Ross~458c). The rotational period of 2M2228-4310 is shorter  (1.41$\pm$0.01~hr, \citealt{Buenzli2012}) than for Ross~458c (6.75$\pm$1.58~hr, \citealt{Manjavacas2019b}). Ross~458c shows a dependence of its variability amplitude with wavelength, but only in the $J$-band. 2M2228-4310 does not show any wavelength dependence {in the $J$ or $H$-bands}. Both 2M2228-4310 and Ross~458c show phase shifts between the $J$- and $H$-band light curves, and with the $\mathrm{CH_{4}-H_{2}O}$ light curve. With the MOSFIRE $J$- $H$- and $\mathrm{CH_{4}-H_{2}O}$ light curves presented here, we were not able to confirm or discard if the shape of 2M0050--3322 light curve is also sinusoidal. We could not confirm the rotational period of 1.55$\pm$0.02~hr measured by \cite{Metchev2015}, and we could not confirm {nor rule out} phase shifts between the $J$- and $H$-band light curves 2M0050--3322. The peak-to-peak amplitudes of the fluctuations measured in the $J$- and $H$-band, and $\mathrm{CH_{4}-H_{2}O}$ 2M0050--3322 light curves are smaller than the amplitudes measured for 2M2228-431 and Ross~458c. A higher sample of mid to late-T dwarfs with photometric or spectro-photometric variability information would help to confirm similarities between the light curves of mid to late-T dwarfs, and further investigate and explain their differences.}

\begin{table*}\label{table_variables}
\centering
\caption{List of {T-dwarfs with spectral type $>$T5.0} with upper levels or measurements of photometric or spectro-photometric variability up-to-date.}
\small
\begin{tabular}{lccccccccc}
       \hline
       \hline
Name                      & RA          & DE           & SpT   & A ($J$-band)         & A ($H$-band)          & A({[}3.6{]})    & A({[}4.5{]})    & Period                         & Ref. \\
                          &             &              & (NIR) &  (\%)                    & (\%)                & (\%)            & (\%)            & (hr)                           &   \\
       \hline
{J034807--602227}  & {03 48 07.72} & {-60 22 27.00} & T7.0    & … & …                   & … & 1.5$\pm$1.4    & 1.080$\pm$1.075                & {(9)} \\
J005019--332240   & 00 50 19.94 & -33 22 40.27 & T7.0    & $< 0.7 \pm 0.4 $ & …                   & \textless{}0.59 & $1.07\pm0.11$     & $1.55 \pm 0.02$                & (2)(4)      \\
J024313--245329    & 02 43 13.71 & -24 53 29.83 & T6.0    &     $0.040 \pm 0.011$               &     & …               & …               & …                              & (4)(6)      \\
J051609--044549 & 05 16 09.45 & -04 45 49.93 & T5.5  & …                  & …                   & \textless{}0.83 & \textless{}0.81 & …                              & (2)         \\
J081730-−615520    & 08 17 29.99 & -61 55 15.65 & T6.0    & $\sim$0.6              & $1.93\pm0.005$ & …               & …               & $2.8$                          & (4)(7)    \\
J104753+21242 & 10 47 53.85 & +21 24 23.46 & T6.5  &  …  & …  &  …  & 0.5 & $1.741\pm0.007$                & (8)      \\
Ross 458c                 & 13 00 42.08 & +12 21 15.05 & T8.5  & $2.62\pm0.02 $ & …                   & \textless{}1.37 & \textless{}0.72 & $6.75 \pm 1.58$                & (2)(3)      \\
J141623+134836  & 14 16 23.94 & +13 48 36.32 & T7.5  & …                  & …                   & \textless{}0.91 & \textless{}0.59 & …                              & (2)         \\
J154627--332511   & 15 46 27.18 & -33 25 11.18 & T5.5  & $0.014\pm0.008$   &     & …               & …               & …                              & (5)(4)      \\
J182835--484904   & 18 28 35.72 & -48 49 04.63 & T5.5  &   $0.9\pm0.1$         &        & …               & …               & $5.0\pm0.6$ & (5)(6)      \\
J222828--431026   & 22 28 28.89 & -43 10 26.27 & T6.5  & $1.85\pm0.07 $ & $2.74\pm0.11$     & $4.26\pm0.2$  & $1.51 \pm 0.15$ & $1.41 \pm 0.01$                & (1)(2)      \\
\hline
\end{tabular}
\footnotesize{Variability references: (1) \cite{Buenzli2012}, (2) \cite{Metchev2015}, (3) \cite{Manjavacas2019b}, (4) \cite{Radigan2014} (5)  \cite{Clarke2002}, (6) \cite{Buenzli2014}, (7) \cite{Artigau2009}, (8) \cite{allers2020measurement}, (9) \cite{tannock2021weather}}.
\end{table*}

\section{Interpretation}\label{interpretation}

\subsection{Brown Dwarf Storms}\label{storms}

The most likely cause of photometric and spectro-photometric variability in brown dwarfs is due to the existence of heterogeneous clouds in their atmospheres  (e.g. \citealt{Apai2013}) that evolve rapidly within few hours. {These clouds patterns are not {static}, and can show beating {patterns}. Thus, the variability amplitude of brown dwarfs might vary with time \citep{Apai2017}.}

\begin{figure}
    \centering
    \includegraphics[width=0.5\textwidth]{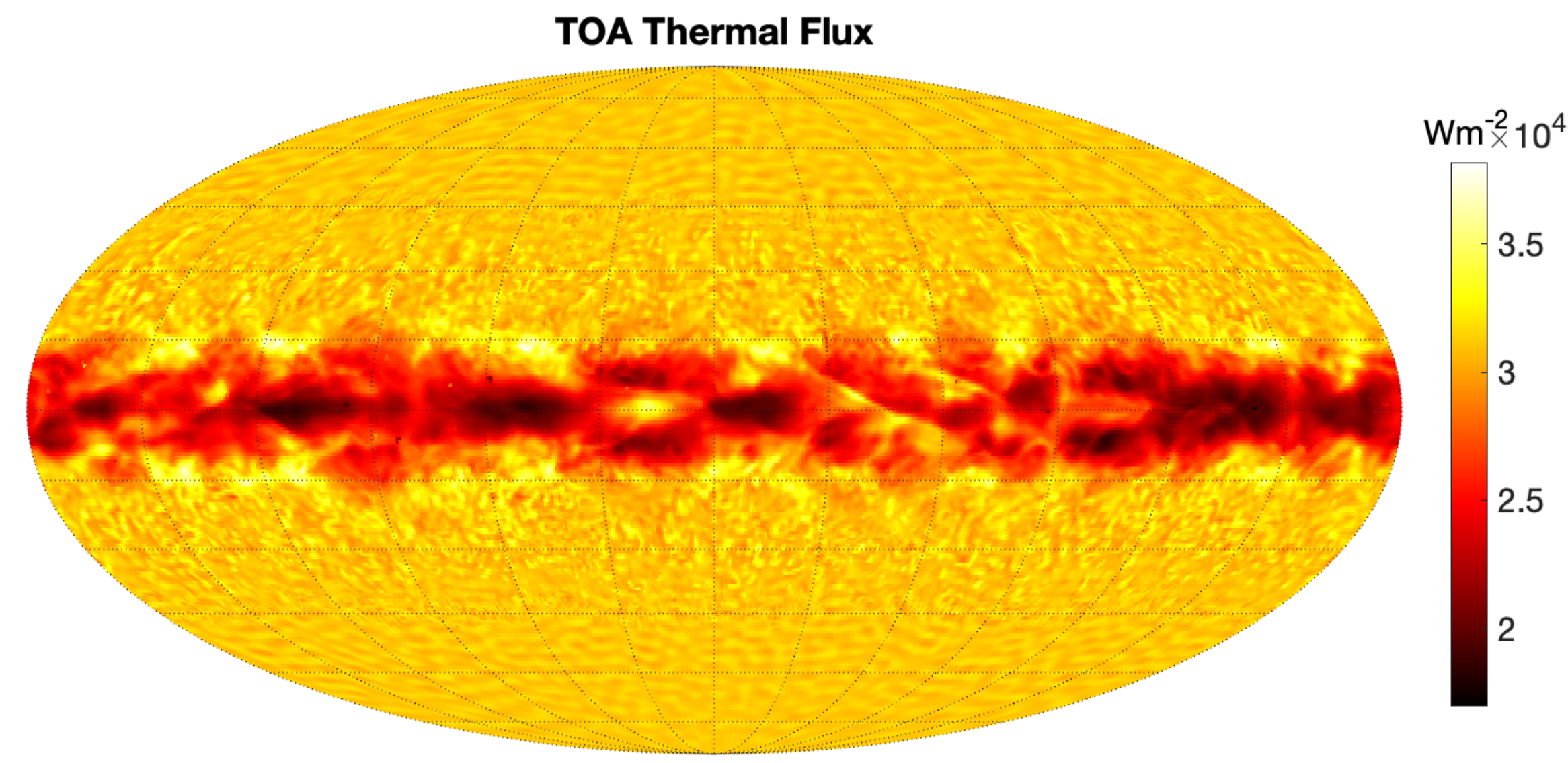}
    \caption{Edge-on map for 2M0050--3322 predicted by General Circulation Models \citep{Tan2021global}}
    \label{GCM}
\end{figure}

\begin{figure}
    \centering
    \includegraphics[width=0.5\textwidth]{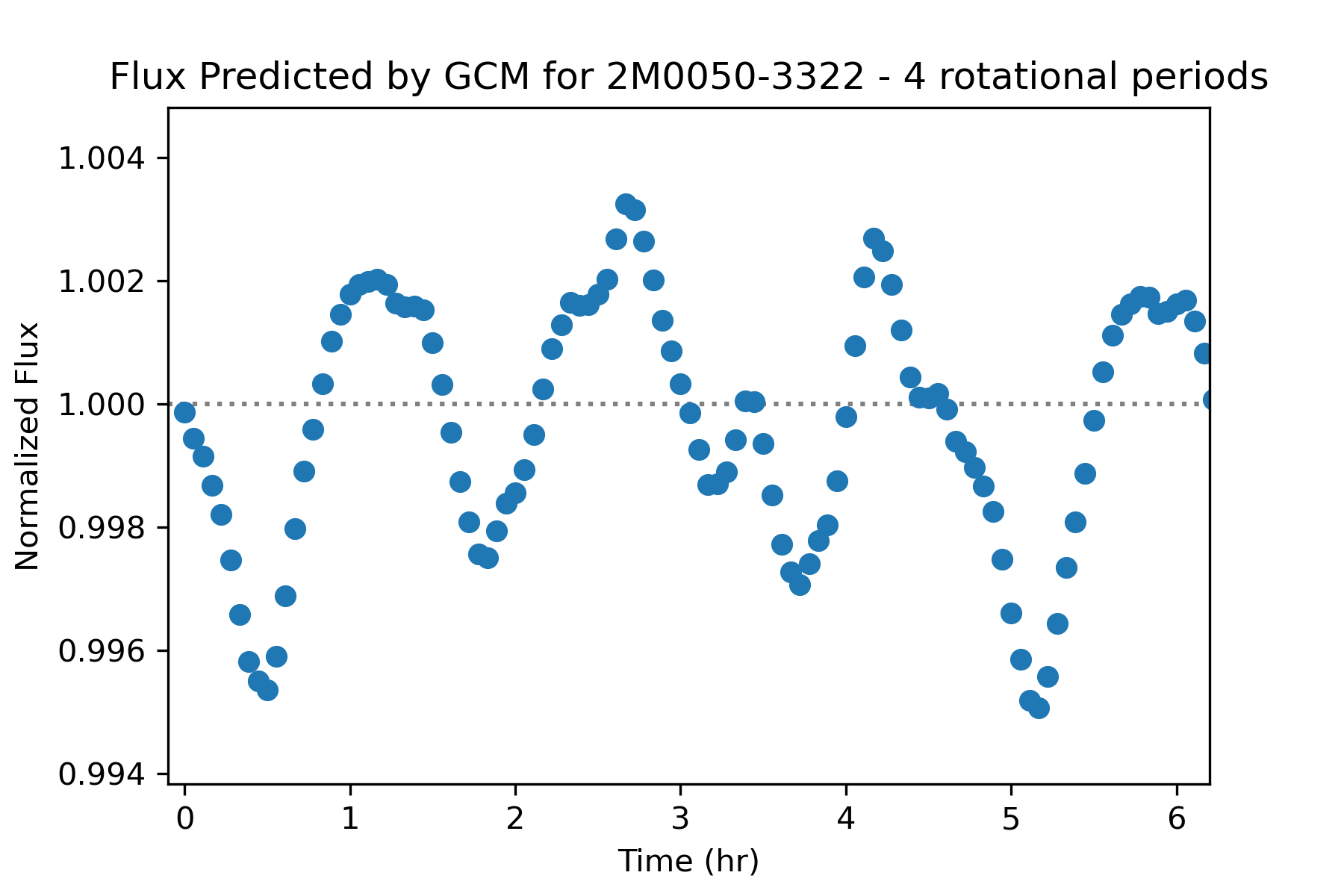}
    \caption{Normalized light curve for an edge-on 2M0050--3322 as predicted by General Circulation Models \citep{Tan2021global}. We show four consecutive periods of the light curve.}
    \label{LC_GCM}
\end{figure}

{We use  General Circulation Models (GCM) developed by \cite{Tan2021global} to predict the top-of-atmosphere thermal flux for 2M0050-3322}. For a T7.0 brown dwarf with log~g=5.0, like 2M0050--3322, GCM  can provide  instantaneous {model} maps for the top-of-the atmosphere thermal flux, and predict the most likely light curve for such an object. Model details are described in Appendix \ref{ch.gcm}. For 2M0050--3322, the vigorous  dynamical features and the associated heterogeneous clouds, driven by the radiative feedback of  MnS and $\mathrm{Na_{2}S}$ clouds, are expected to be near the equator (see Fig. \ref{GCM}). These equatorial features are large-scale equatorially trapped waves that propagate either eastward or westward relative to mean flow.   The GCM predicts a somewhat sinusoidal and regular light curve over a few rotational timescale, with a nearly 1\% peak to peak flux variation when the object is viewed equator-on {(see Fig. \ref{LC_GCM})}. Meanwhile, due to the multiple longitudinal-dependent wave patterns, the light curve also exhibits certain complexities out of a pure sinusoid (see the light curve Figure {\ref{LC_GCM}}). {\cite{Vos2017} found  a correlation between {close to edge on view}, and high variability amplitudes in the $J$-band and in the \textit{Spitzer} channels because of an increased path length through the atmosphere at {low} inclination angles. Thus, the GCM provide the variability amplitudes at different inclination angles}. For inclinations of 0$^{\circ}$ {(edge on)}, 30$^{\circ}$, 60$^{\circ}$, and 90$^{\circ}$, the GCM predict variability amplitudes of 1.00\%, 0.85\%, 0.52\% and 0.22\%, respectively, in agreement with \cite{Vos2017}. 
Statistics of the dynamical activities in the model, including the characteristic sizes and numbers of storms and cloud thickness, are  stable over the  long-term model integration. We do not find evidence of sudden decrease or increase of storm activities that could result in dramatic light curve evolution, which is in {agreement to light-curve observations for this object in \cite{Metchev2015} and with the peak-to-peak amplitudes of the fluctuations of the light curves presented in this work. }
 
The peak-to-peak variability amplitude predicted {by} the GCM \citep{Tan2021global} {agrees} with the peak-to-peak amplitude {of the fluctuations} measured for the  $J$ and $H$-band   light curves of 2M0050--3322, {even though the variability detected for our target is not significant.}

%\textcolor{red}{Xianyu: feel free to add any more explanations in this paragraph for the GCM predictions.}

Similarly, radiative-transfer models {with solar-metallicity} \citep{Saumon_Marley2008} provide the vertical cloud structure expected for the atmosphere of an object with similar effective temperature and surface gravity as 2M0050--3322 {($\mathrm{T_{eff}}$ = 800~K, log~g = 5.0, from \citealt{Filippazzo2015}}). {Further details on the radiative-transfer models, and how the contribution functions were obtained can be found in \cite{Manjavacas2021}, their Section 8.}

\begin{figure}
    \centering
    \includegraphics[width=0.45\textwidth]{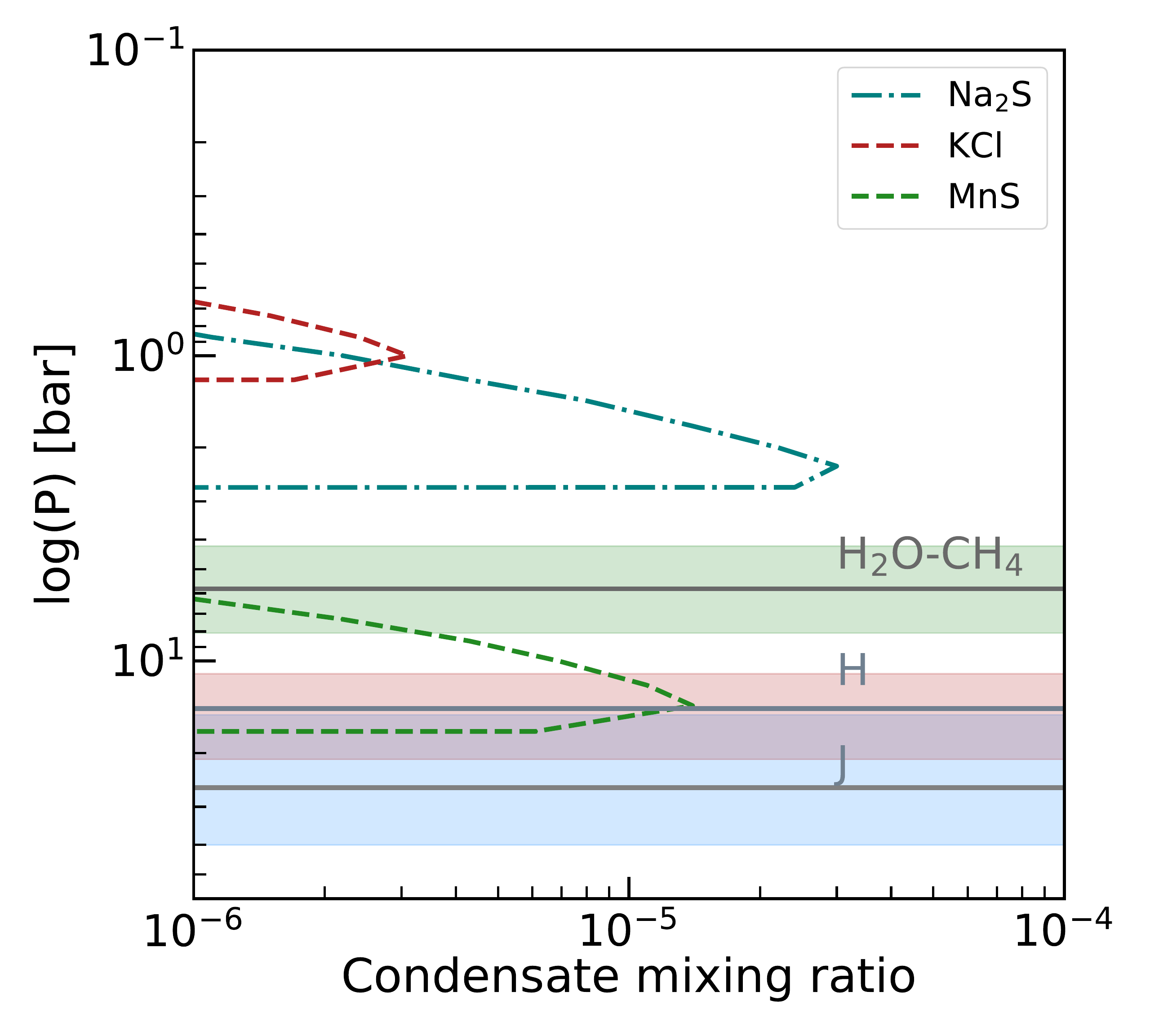}
    \caption{{Condensate mixing ratio (mole fraction) of MnS, KCl and $\mathrm{Na_{2}S}$ clouds for a T7.0 brown dwarf with log~g~=~5.0. The mole fraction for each cloud types provide us with the vertical cloud structure of 2M0050--3322.}}
    \label{vertical_structure}
\end{figure}

\begin{figure}
    \centering
    \includegraphics[width=0.45\textwidth]{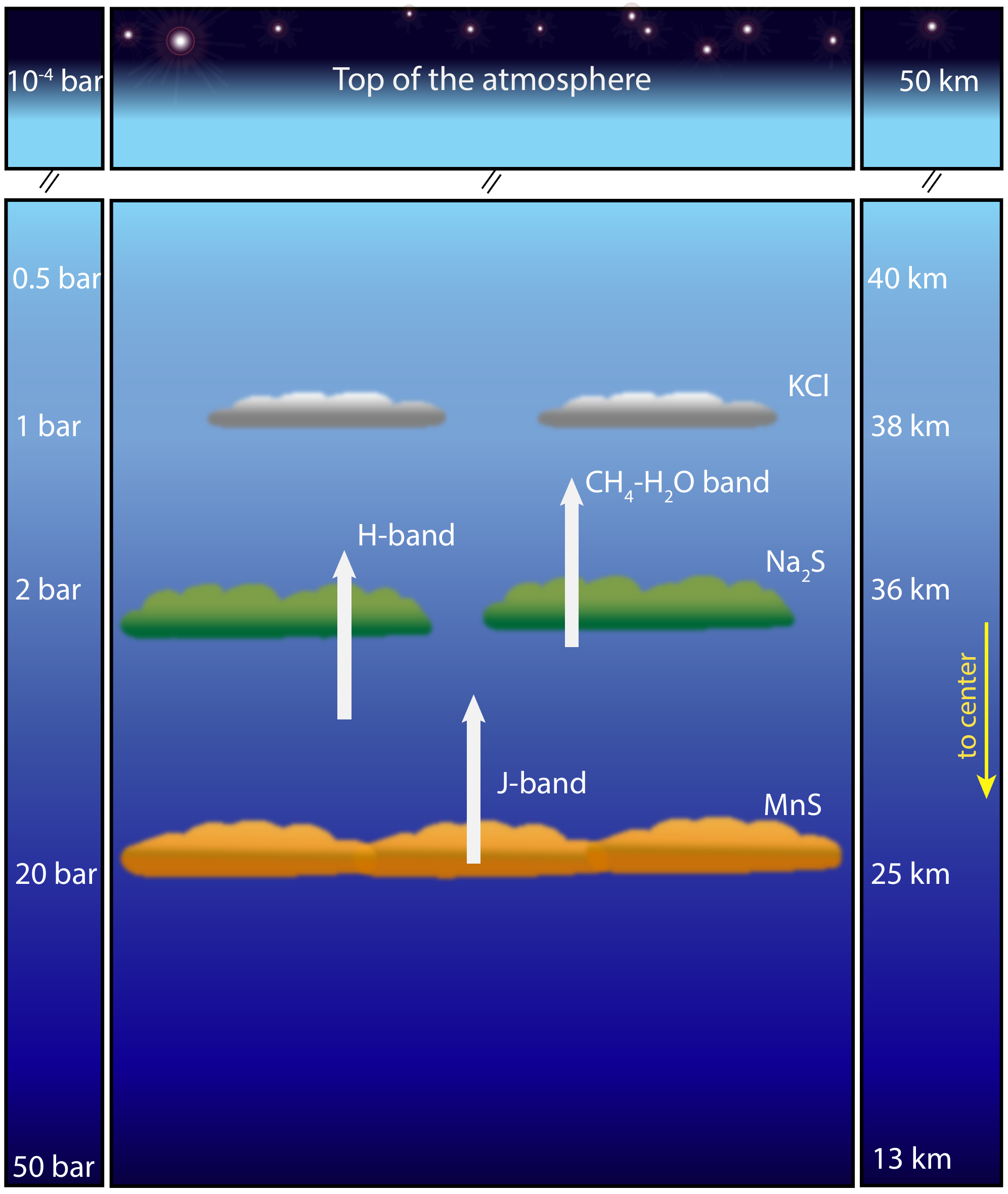}
    \caption{{Visual representation of the different cloud layers in the atmosphere of 2M0050--3322 showing the pressure levels at which each of them condensates, and the top of the atmosphere for reference. We show the traced pressure levels by the different near-infrared wavelengths in our MOSFIRE $J$- and $H$-band spectra.}}
    \label{traced_presures}
\end{figure}

In Fig. \ref{vertical_structure}, we show the predicted condensate mixing ratios by radiative-transfer models ({mole fraction}) of MnS, KCl and $\mathrm{Na_{2}S}$ clouds for a T7.0 brown dwarf with log~g~=~5.0. As seen in Fig. \ref{vertical_structure}, at the bottom of the atmosphere, at about $\sim$20~bar, we find the bottom of the MnS cloud. At $\sim$3~bar, we find the bottom of the $\mathrm{Na_{2}S}$ cloud, at $\sim$1.5~bar we find the bottom of the KCl cloud. In Fig. \ref{traced_presures} {we show a visual representation of the vertical structure of 2M0050--3322}, and  the pressure levels probed by different near-infrared wavelength ranges. The $J$-band traces the deepest levels, between 20--40~bars, the $H$-band traces the middle levels of the atmosphere, between 10--20~bar, and the $\mathrm{CH_{4}-H_{2}O}$ band {traces} the medium-upper layers of the atmosphere, around 4--8~bar. Thus, the $J$-band would trace the deepest cloud layers of the atmosphere and above, while the $H$-band would trace the medium layers of the atmosphere, and the $\mathrm{CH_{4}-H_{2}O}$ just the upper layer of the atmosphere. {Longer monitoring in time is needed to confirm higher fluctuations in the $\mathrm{CH_{4}-H_{2}O}$ light curve.}
%Since two of the three clouds (ZnS KCl and $\mathrm{Na_{2}S}$) that we have considered in the radiative-transfer models have their bottom of the atmosphere around the same pressure levels that the $\mathrm{CH_{4}-H_{2}O}$ band traces, between 4 and 9~bar (see Fig. \ref{traced_presures}), it might explain the highest variability amplitude measured in this pressure level.

%\begin{figure}
%    \centering
%    \includegraphics[width=0.45\textwidth]{2m0050_cf.pdf}
%    \caption{Traced pressure levels by different near-infrared wavelengths.}
%    \label{traced_presures}
%\end{figure}

\subsection{Auroral Emission}\label{auroras}

{Until now, the most common mechanism to explain the existence of {aurorae} in the Giant planets of the Solar System were charged particles  from the {solar} wind that interact with the atmosphere's particles exiting them, and producing an emission \citep{Clarke2013}. Nevertheless, Saturn has been recently found to have {aurorae} also connected to storms or vortexes in its atmosphere (in the polar region) or weather-driven {aurorae} \citep{chowdhury2022}. \cite{chowdhury2022} used spectral observations of auroral $H_{3}^{+}$ emission lines in near-infrared high resolution data, a dominant molecular ion species in Saturn's ionosphere, to trace the electrical dynamo region of the ionosphere. Using the $H_{3}^{+}$ emission lines \cite{chowdhury2022} detected twin-vortex flows in the upper atmosphere of Saturn, consistent with theories that predict the presence of such a polar feature. This is the first time that a connection between {aurorae} and weather patterns have been observed in a Solar System giant planet, with the subsequent implications for cold brown dwarfs, where weather is the most likely explanation for potential aurorae. }

{A handful of brown dwarfs have been  found to have radio emissions potentially connected to the existence of {aurorae}}. The first one was LSRJ1835+3259, a M8.5 brown dwarf at the lithium-burning boundary, for which radio and optical auroral emission were detected \citep{Hallinan2015}. Since then, radio emissions  have been found in other brown dwarfs, some of them rapid rotating late T-dwarfs like 2M0050--3322. {This is the case of WISEPC J112254.73+255021.5 (T6.0), that shows radio emissions with a period of 0.28~hr \citep{Route_Wolszczan2016}, and 2MASS J10475385+2124234 (T6.5), with a periodic emission of 1.741$\pm$0.007~hr \citep{Allers2020} suggesting the existence of {aurorae} in both objects}. 2M0050--3322 shares {a similar} spectral type (T7.0) and  rapid rotational period of 1.55$\pm$0.02~hr as measured by \cite{Metchev2015}. {No radio emission has been searched for 2M0050--3322, but aurorae could potentially have an influence in the fluctuations measured in the $J$-, $H$-bands, and the $\mathrm{CH_{4}-H_{2}O}$ band, as recently found for Saturn \citep{chowdhury2022}}.

\section{Conclusions}\label{conclusions}

\begin{enumerate}

    \item We have used MOSFIRE at the Keck\,I telescope to monitor  2MASS~J00501994--3322402, a field (0.5--10~Myr) T7.0 brown dwarf over $\sim$2.6~hr in the $J$- and $H$-bands.
    
    \item We {measured peak-to-peak fluctuations} of 1.48$\pm$0.75\% in the $J$-band imaging light curve,  0.62$\pm$0.18\% in the $J$-band spectrophotometric light curve, a 1.26$\pm$0.93\% in the $H$-band, and a {5.33$\pm$2.02\%} in the $\mathrm{CH_{4}-H_{2}O}$ band. The $J$- and $H$-band {peak-to-peak fluctuation amplitudes} agree with those measured by \cite{Metchev2015} {in the mid-infrared}.
    
    \item {The BIC analysis concluded that the 2M0050--3322 $J$-, $H$- and $\mathrm{CH_{4}-H_{2}O}$ band light curves  does not show significant variability.}
    
    \item We produced a sensitivity plot for our Keck/MOSFIRE observations, concluding that with the current observational strategy we would miss a $\sim$1\% variability amplitude light curve in 40-50\% of the cases. {To detect a significant light curve with MOSFIRE, the variability amplitude of an object similar to 2M0050--3322 should be at least 2\%}.
    
    %\item Using a Lomb-Scargle periodogram and a Bayesian Generalized Lomb-Scargle periodogram, we estimated an approximate rotational period of 1.8-1.9~hr, which is close to the 1.55$\pm$0.02~hr measured by \cite{Metchev2015}.
    
    \item {We tentatively measured the wavelength dependence of the fluctuations in the $J$- and $H$-band light curves. We found no wavelength dependence on the $J$-band {fluctuation}. In contrast, we measured a {tentative} slight wavelength dependence in the $H$-band}.
    
    \item The {peak-to-peak  amplitudes of the fluctuations} measured for 2M0050--3322 in the $J$- and $H$-bands agree with the predictions of General Circulation Models for a T7.0 brown dwarf with log~g~=~5.0 {\citep{Tan2021global}}.
    
    \item {We show the predicted vertical structure of 2M0050--3322 by radiative-transfer models. At the bottom of the atmosphere ($\sim$20~bar) we find we bottom of the MnS cloud. At $\sim$3~bar, the bottom of the $\mathrm{Na_{2}S}$ cloud, and at $\sim$1.5~bar we find the bottom of the KCl cloud. Given that two of the three clouds used in the radiative-transfer models have the bottom of their cloud between above 4~bar, and that the $\mathrm{CH_{4}-H_{2}O}$ band traces approximately these pressures levels, it might explain why the {fluctuations} are  higher for the $\mathrm{CH_{4}-H_{2}O}$ light curve. Longer monitoring in time is needed to confirm higher fluctuations in the $\mathrm{CH_{4}-H_{2}O}$ light curve.}
    
    \item {Aurorae probably {linked} to weather patterns have been discovered for other late-T, fast rotating brown dwarfs similar to 2M0050--3322. Thus the existence of aurorae might contribute to the low-level variability measured for 2M0050--3322}.

    %{Our observations  highlight the importance of high resolution spectroscopy to understand the atmospheric variability and 3D structures of brown dwarfs and giant exoplanets ground-based, with multi-object spectrographs like Keck\,I/MOSFIRE, or EMIR at the Gran Telescopio de Canarias (GTC) telescope, but also from space-based telescopes like HST/WFC3. The just lanched James Webb Space Telescope (JWST)  is expected to produce ground-braking discoveries in the field of brown dwarfs and exoplanets. NIRSpec (Near Infrarred Spectrograph) and NIRISS (Near Infrared Imager and Slitless Spectrograph) on-board JWST will provide high signal-to-noise and resolution, and broad-wavelength spectroscopic observations, that will enable the detection of variability in multiple pressure layers, allowing us to probe the vertical structure of brown dwarf and imaged exoplanet atmospheres with an unprecedented accuracy. }

\end{enumerate}

\acknowledgments

We thank our anonymous referee for the thorough review of our manuscript and helpful comments that helped to improve our paper.

The authors wish to recognize and acknowledge the very significant cultural role and reverence that the summit of Mauna Kea has always had within the indigenous Hawaiian community.  We are most fortunate to have the opportunity to conduct observations from this mountain.

We would like to acknowledge the \textit{PypeIt} Development team for developing a pipeline that was able to reduce our challenging MOSFIRE data with extremely wide slits, in particular to Dr. Joe Hennawi for his efficient support.

We acknowledge the MOSFIRE/Keck I Instrument Scientist, Dr. Josh Walawender, for his advises and recommendations on the preparation of the observations, and the reduction of the data. Thanks to his idea of taking "skylines spectra" of our mask with narrower slits we could calibrate in wavelength the spectra presented in this paper.

\facilities{MOSFIRE (W. M. Keck Observatory)}


\software{astropy \citep{2013A&A...558A..33A}}
\software{Pypeit \citep{Prochaska2019,Prochaska2020}}



\newpage

\appendix

\section{Correlation between parameters}

\begin{figure}
    \centering
    \includegraphics[width=0.48\textwidth]{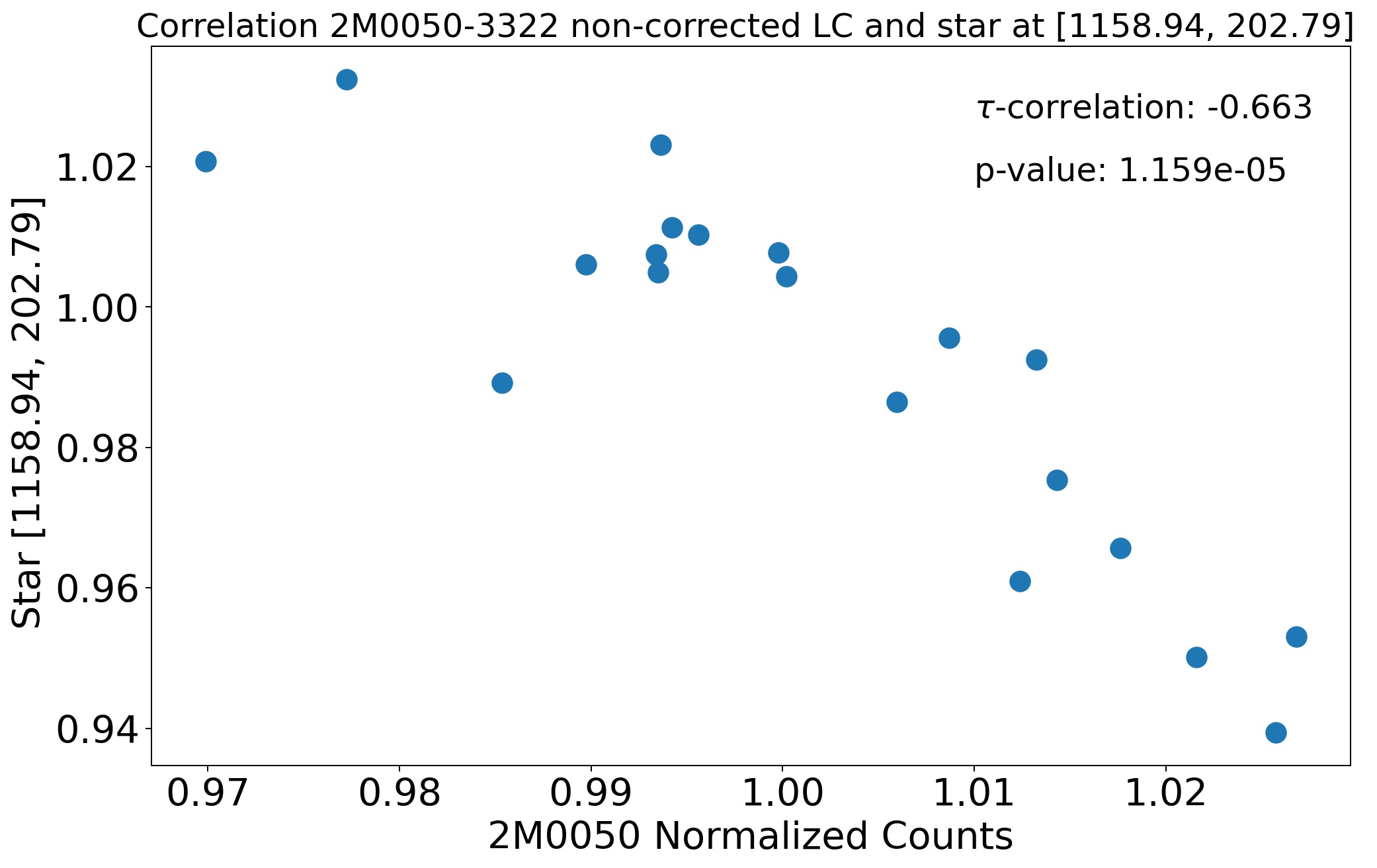}
    \includegraphics[width=0.48\textwidth]{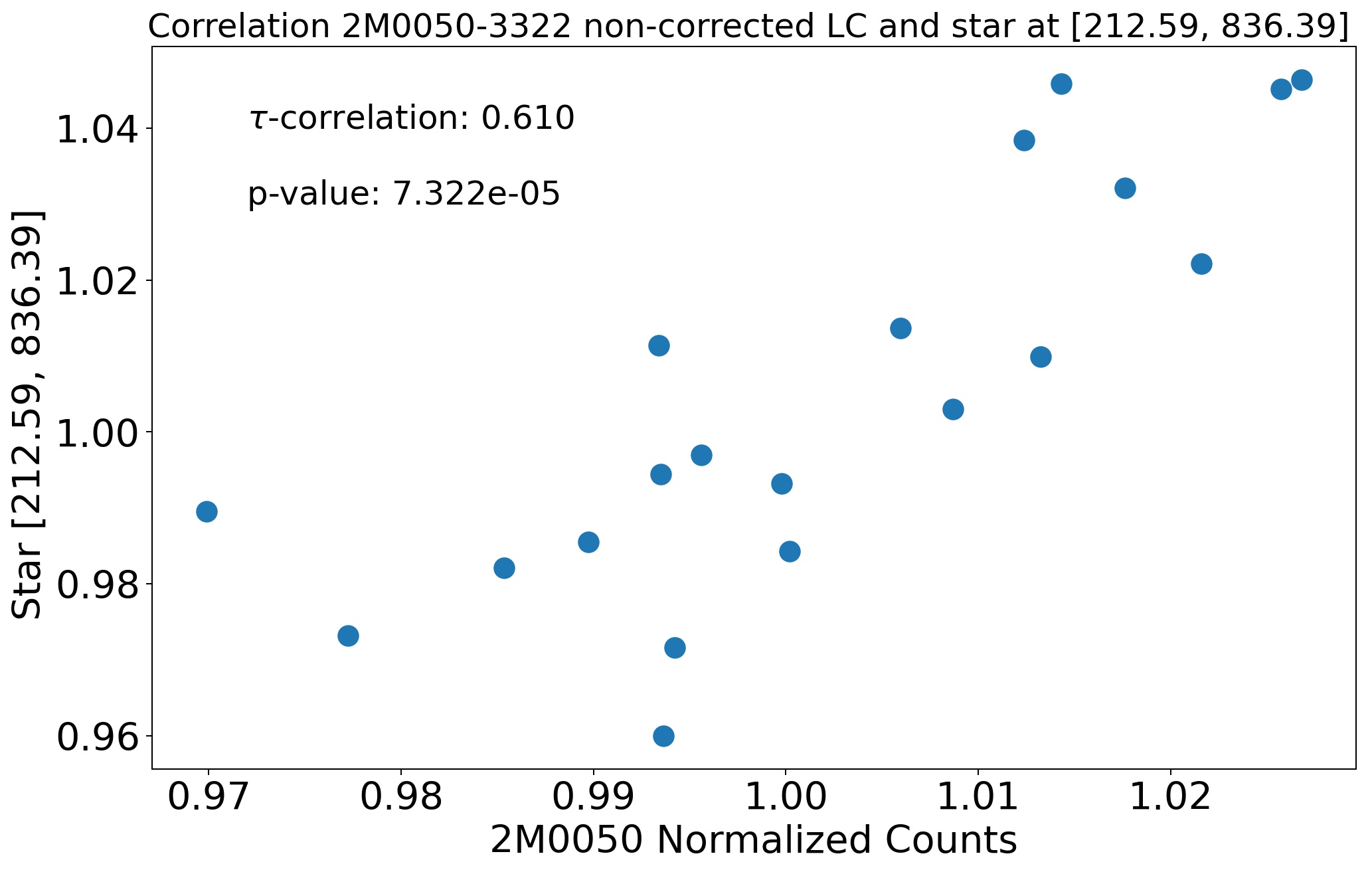}
    \includegraphics[width=0.48\textwidth]{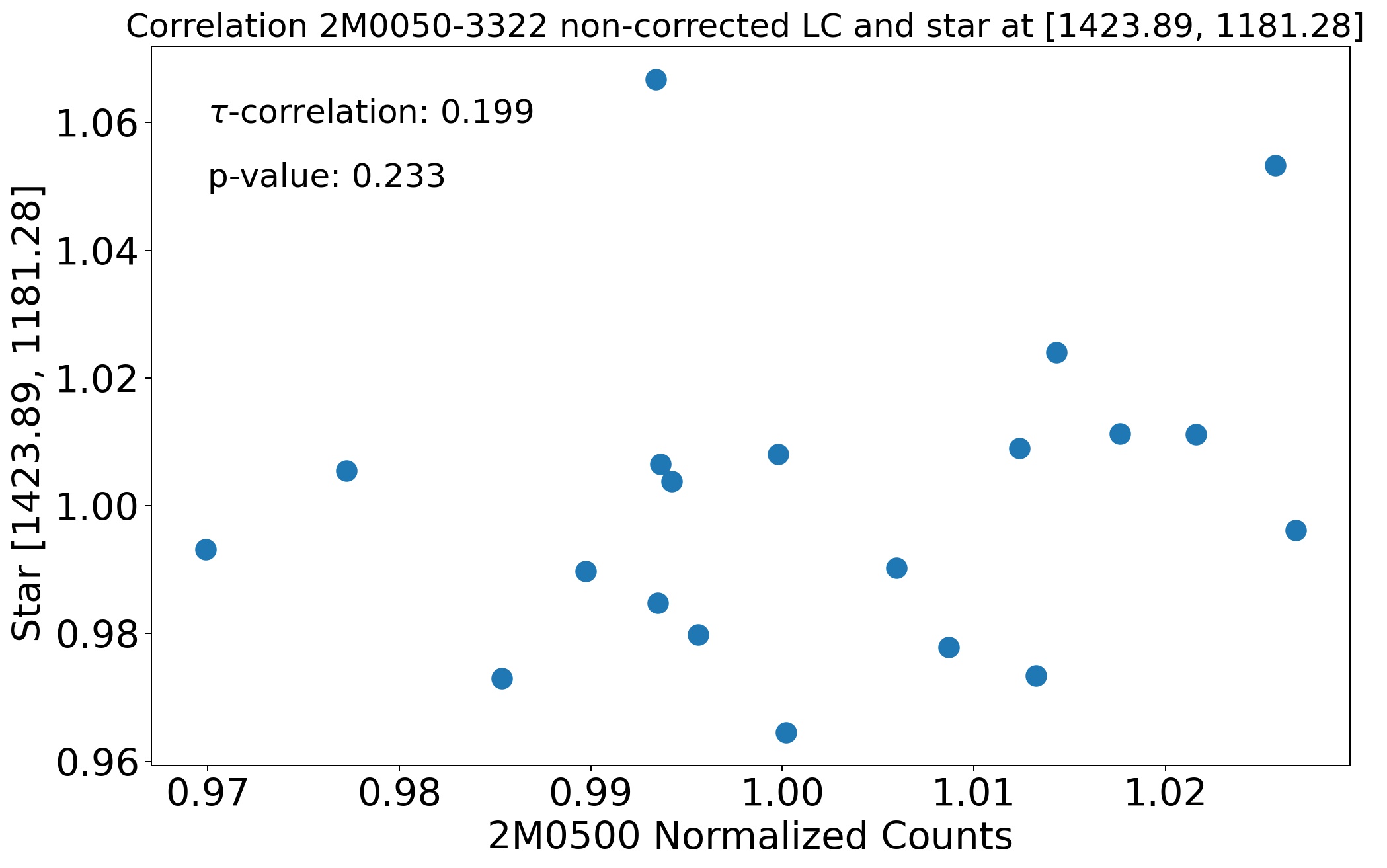}
    \includegraphics[width=0.48\textwidth]{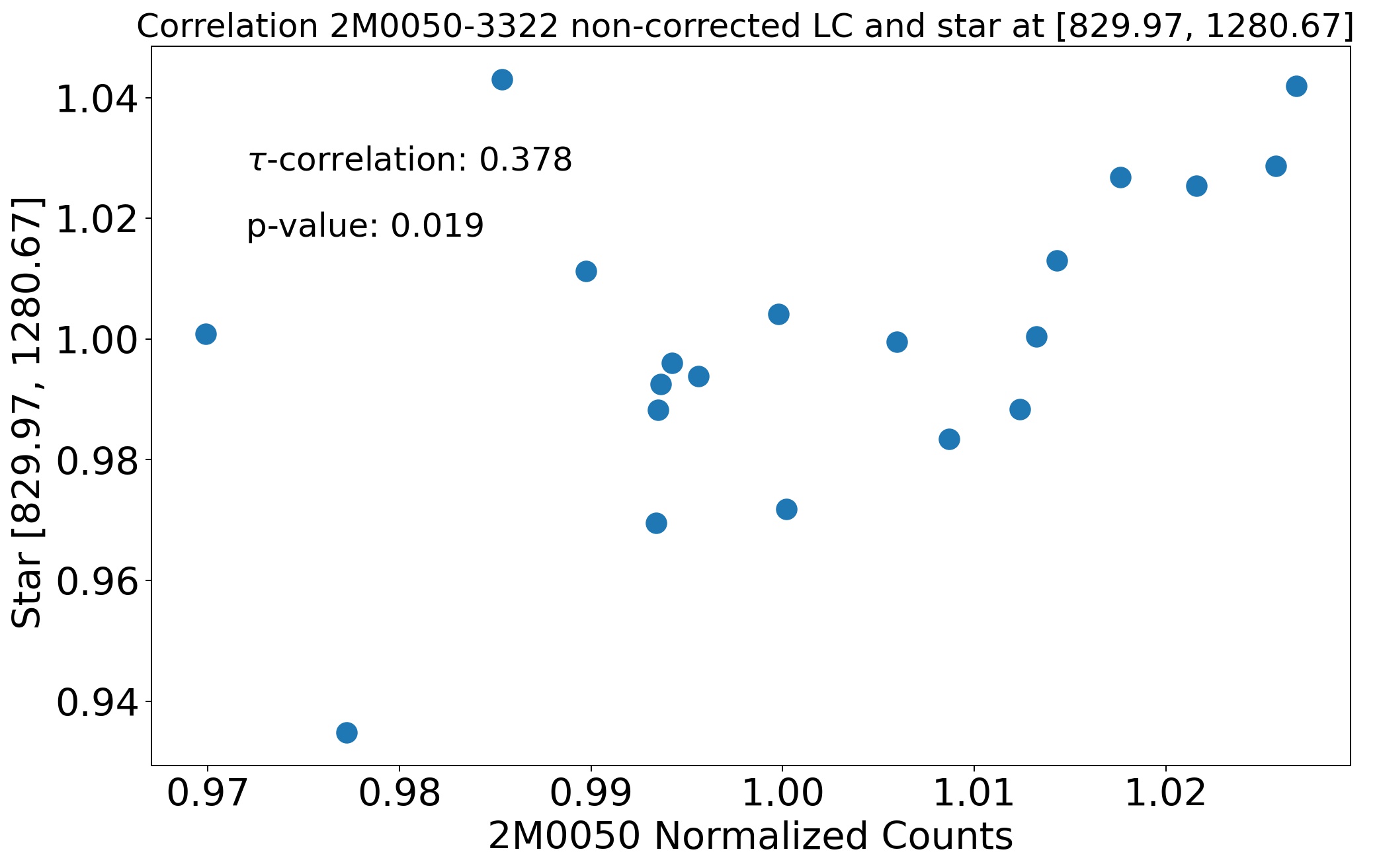}
    \includegraphics[width=0.48\textwidth]{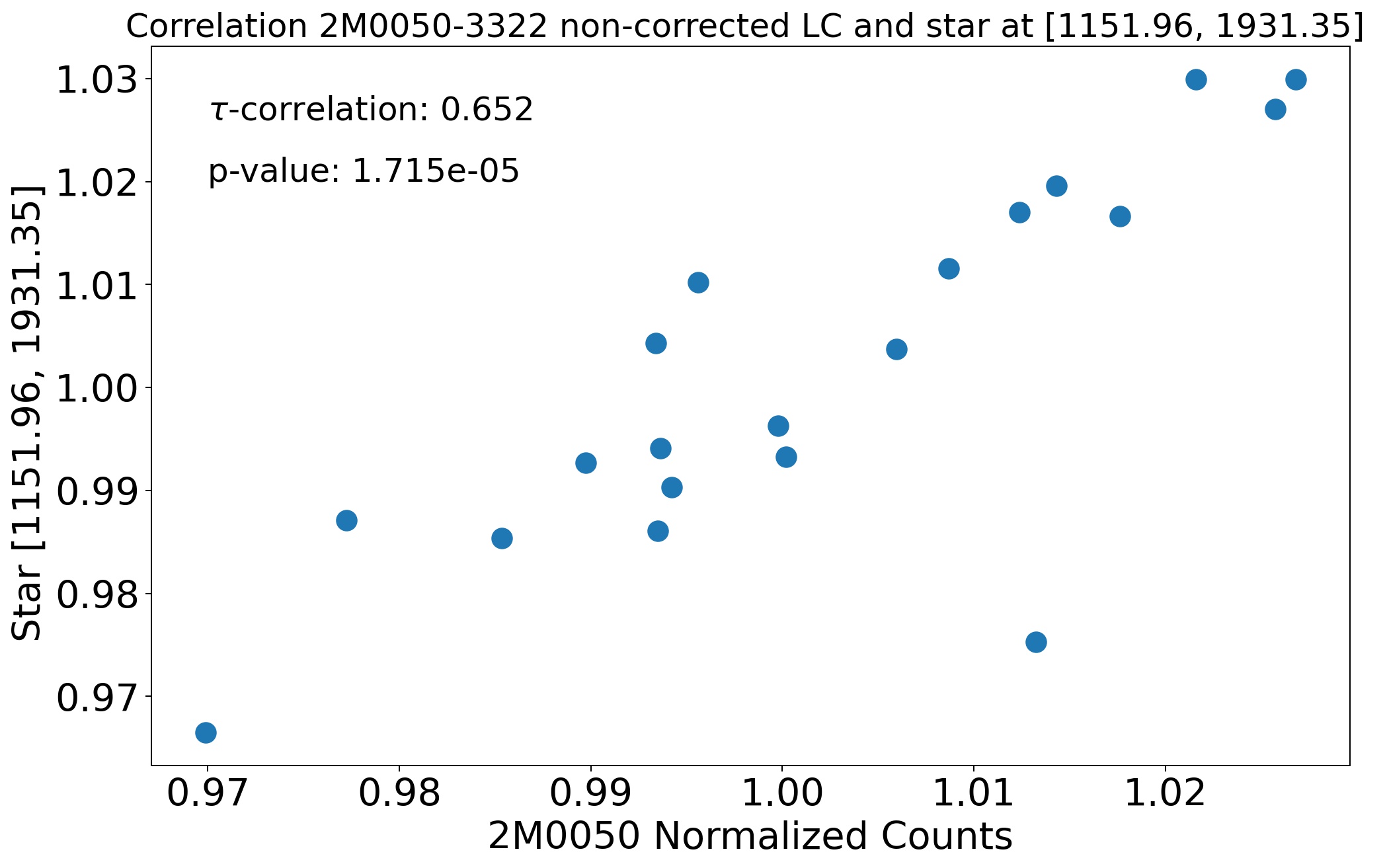}
    \caption{Correlation between the imaging $J$-band target's non-corrected light curve, and the non-corrected calibration stars light curves.}
    \label{corr_noncorr_stars_Jband}
\end{figure}

\begin{figure}
    \centering
     \includegraphics[width=0.48\textwidth]{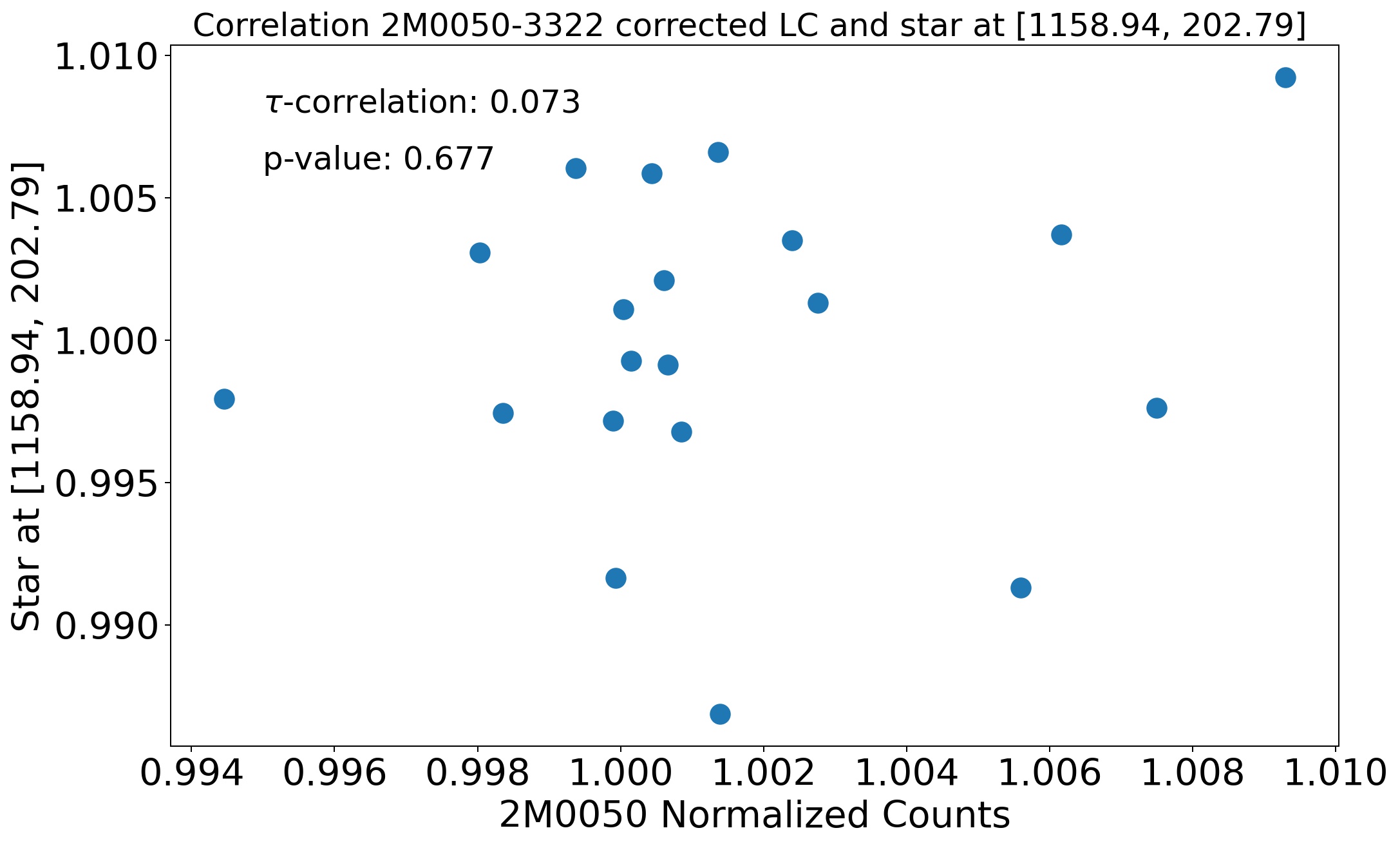}
    \includegraphics[width=0.48\textwidth]{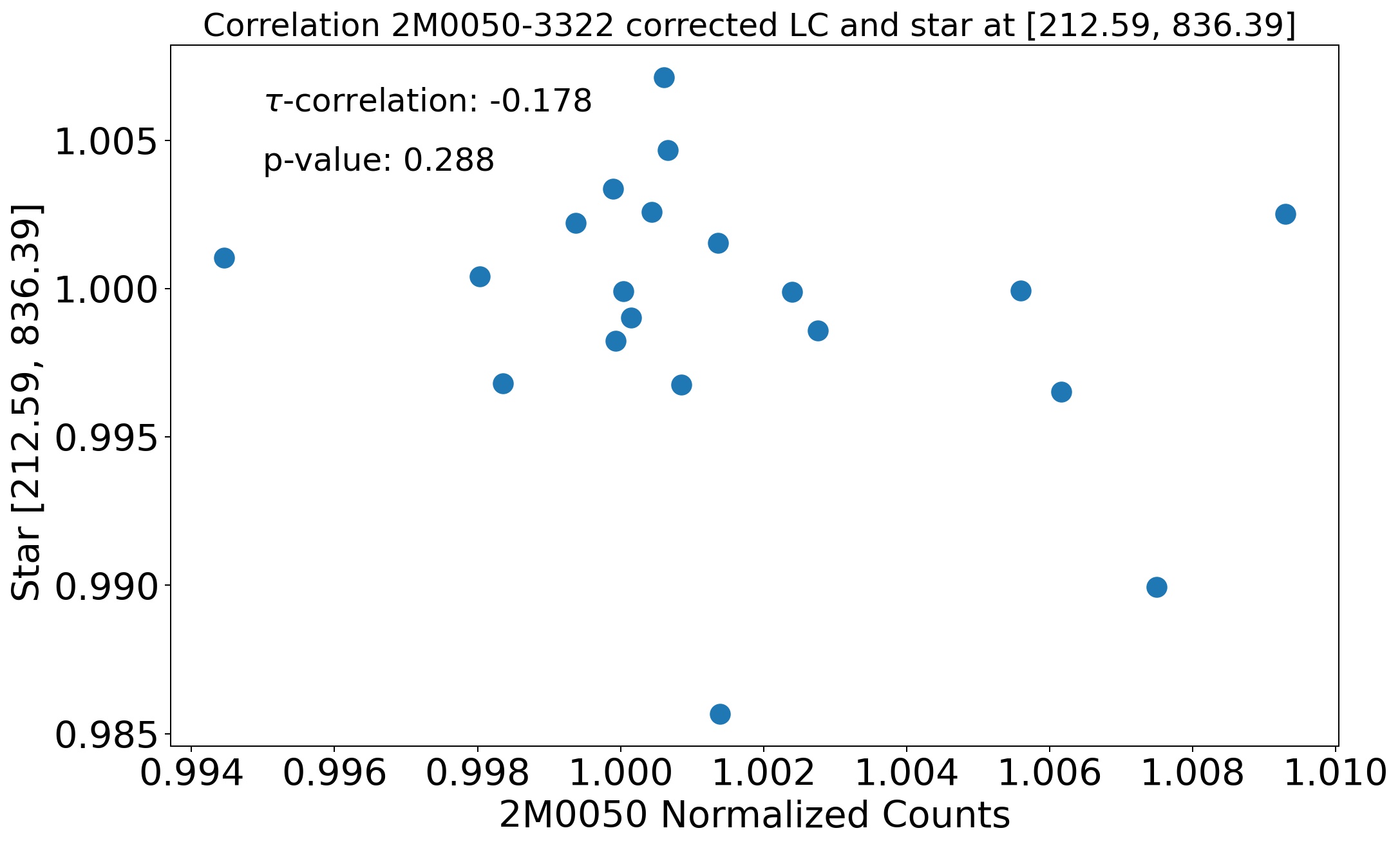}
    \includegraphics[width=0.48\textwidth]{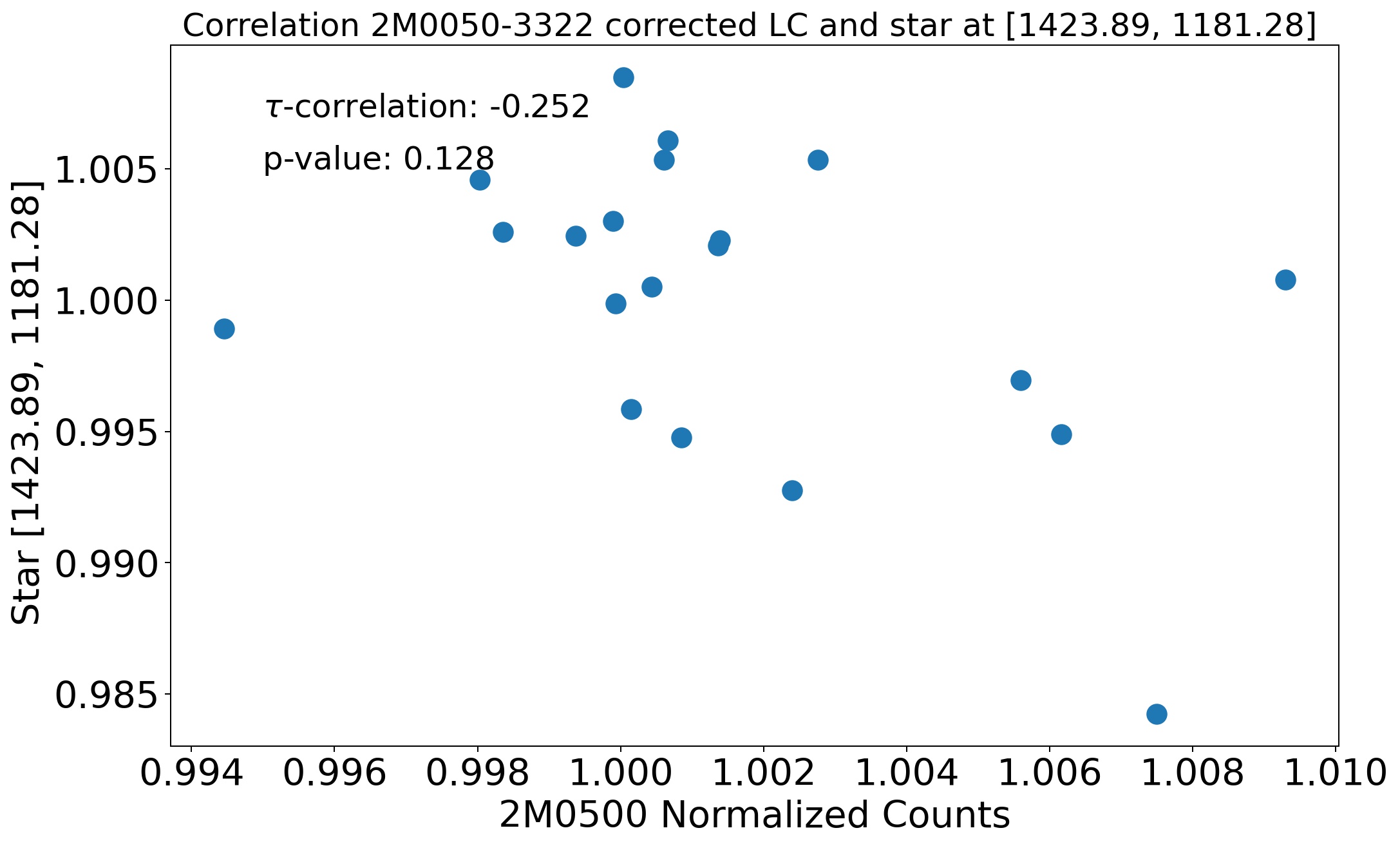}
    \includegraphics[width=0.48\textwidth]{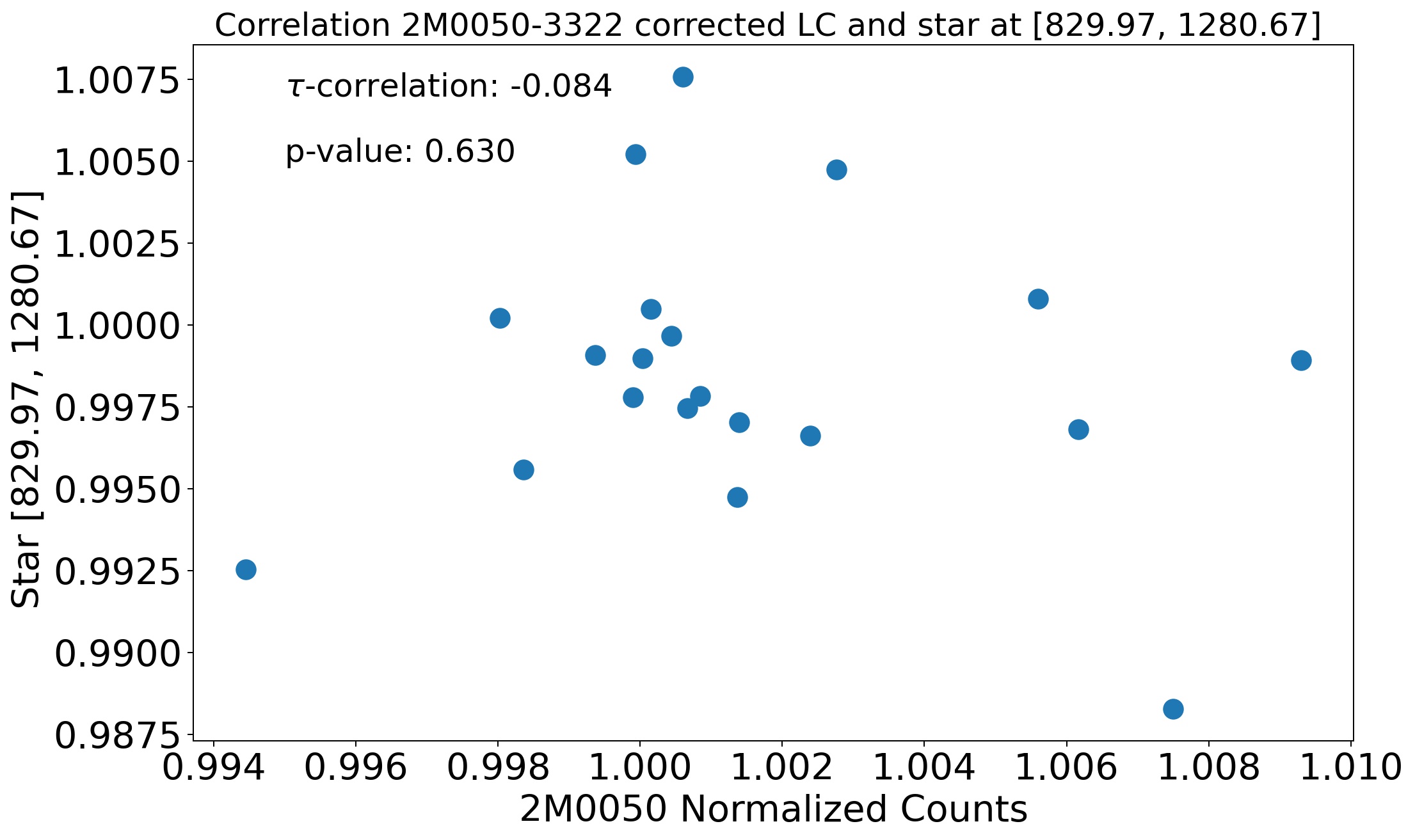}
    \includegraphics[width=0.48\textwidth]{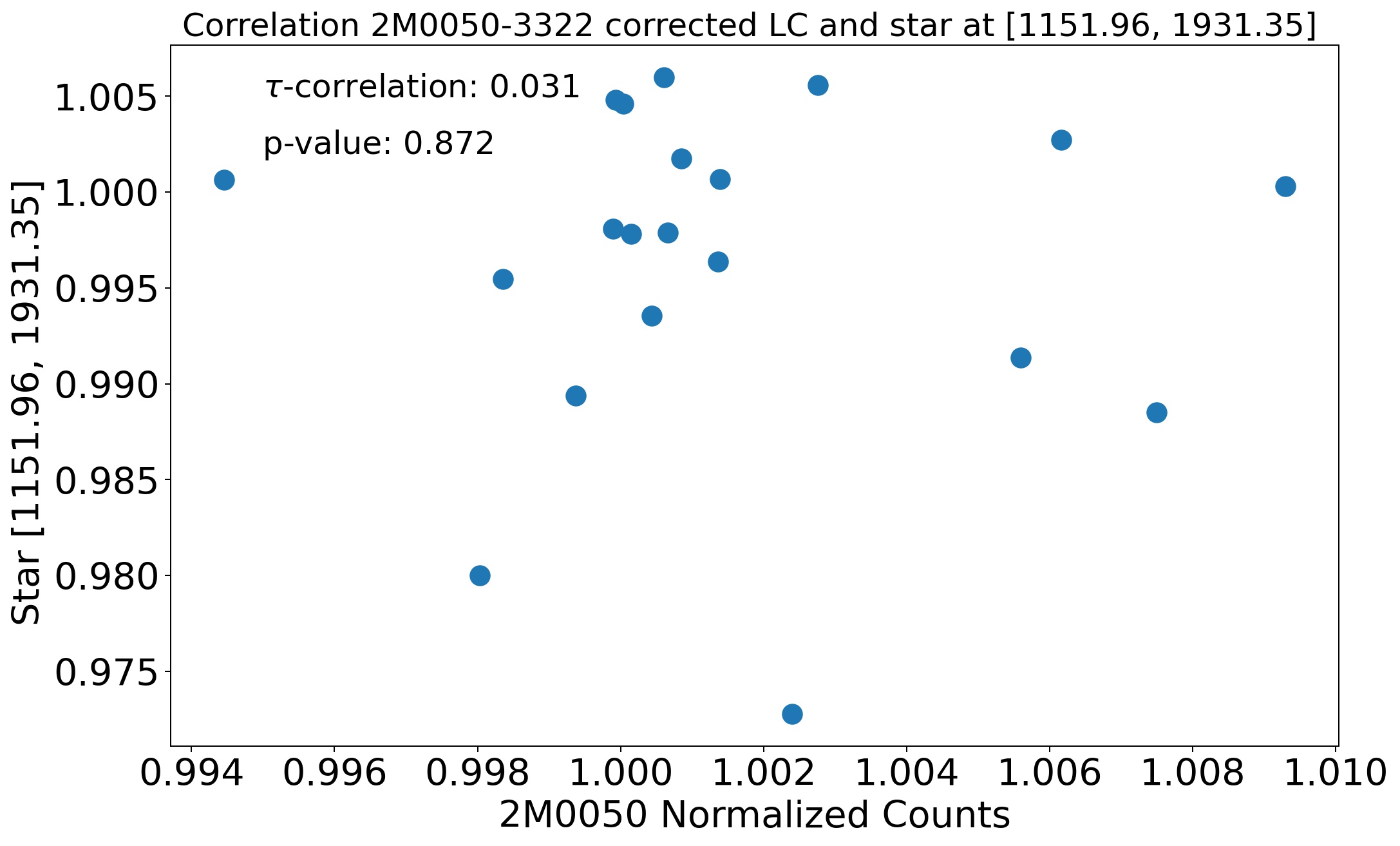}
     
    \caption{Correlation between the imaging $J$-band target's corrected light curve, and the corrected calibration stars light curves.}
    \label{corr_corr_stars_Jband}
\end{figure}

\begin{figure}
    \centering
    \includegraphics[width=0.48\textwidth]{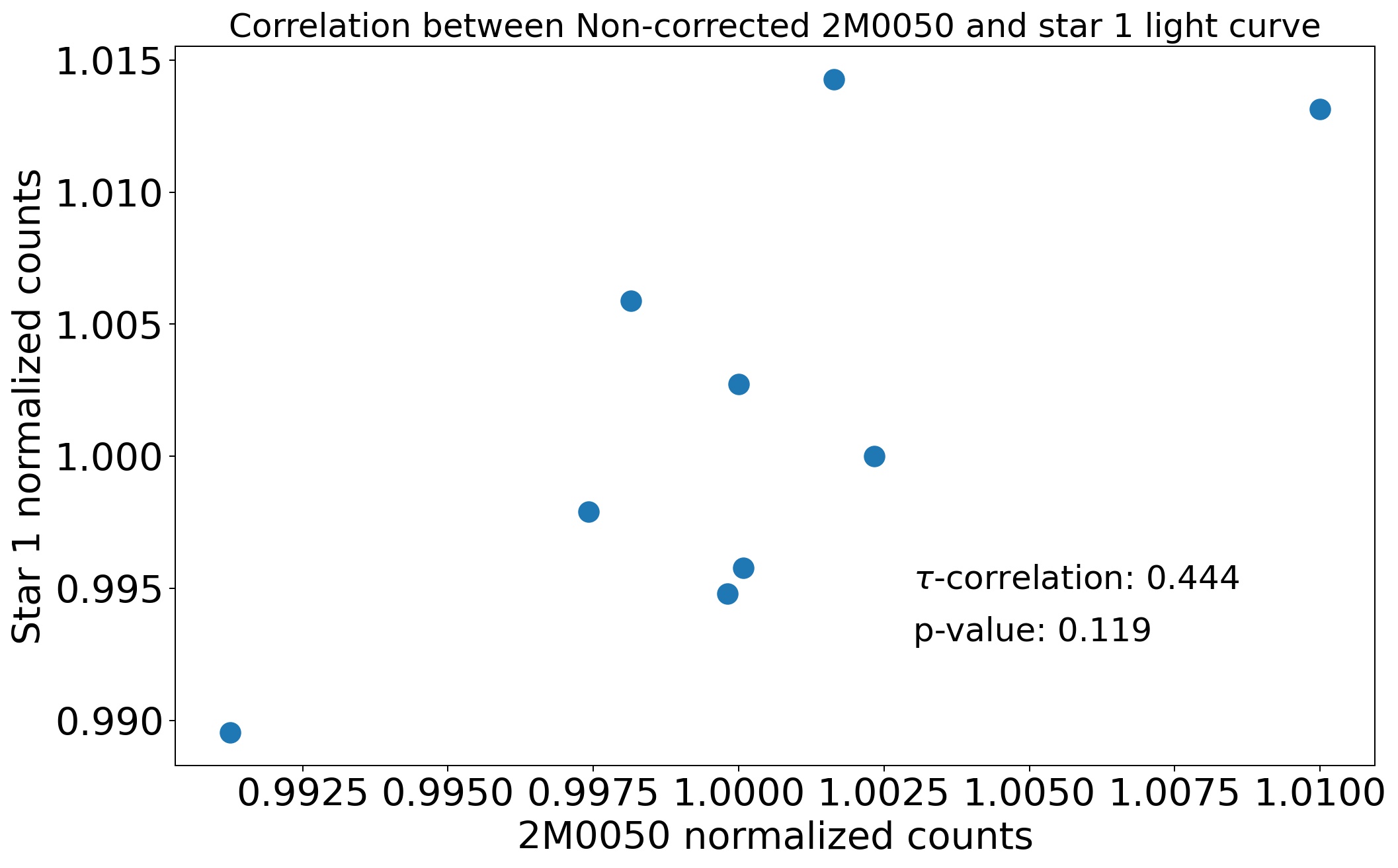}
    \includegraphics[width=0.48\textwidth]{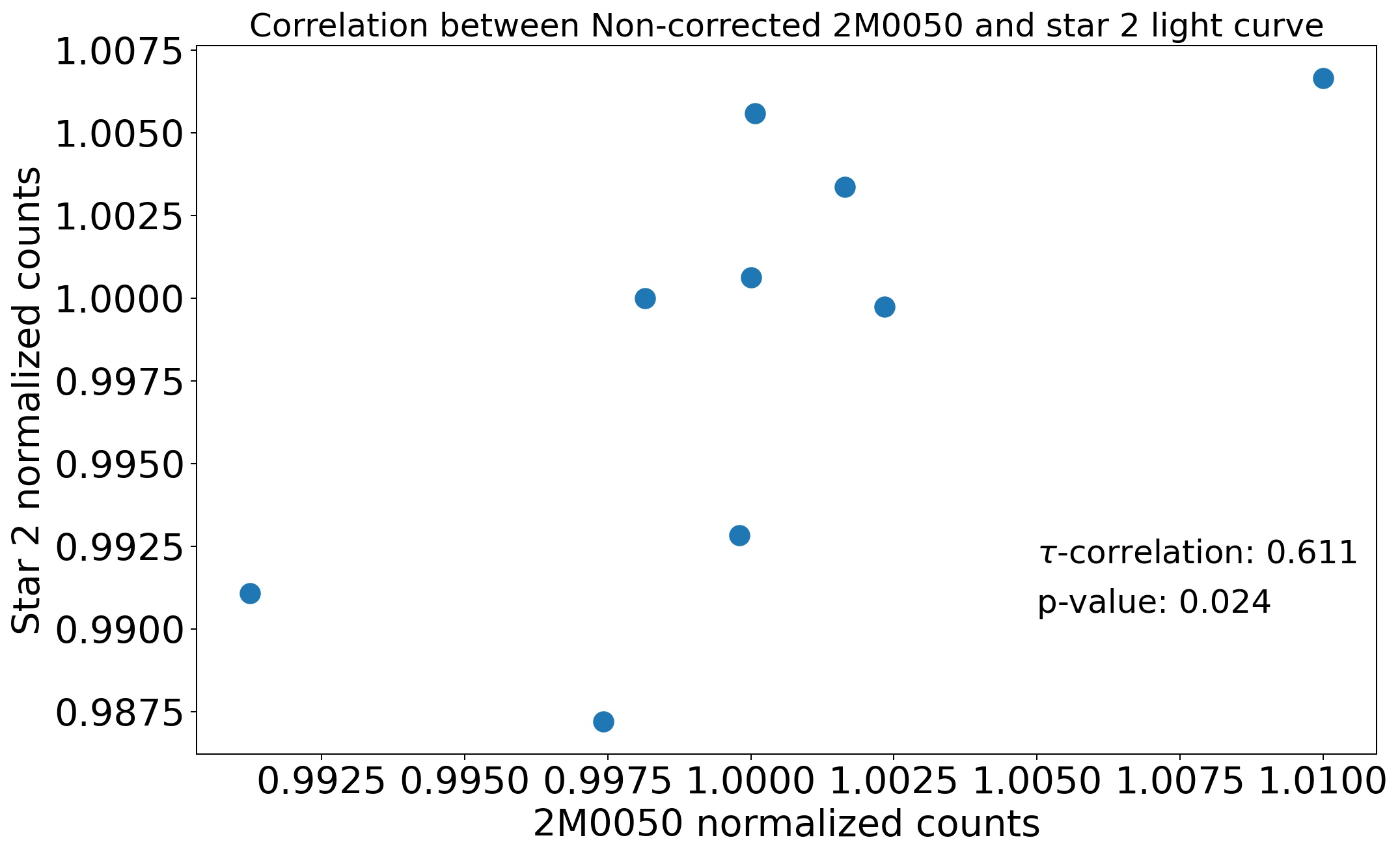}
    \includegraphics[width=0.48\textwidth]{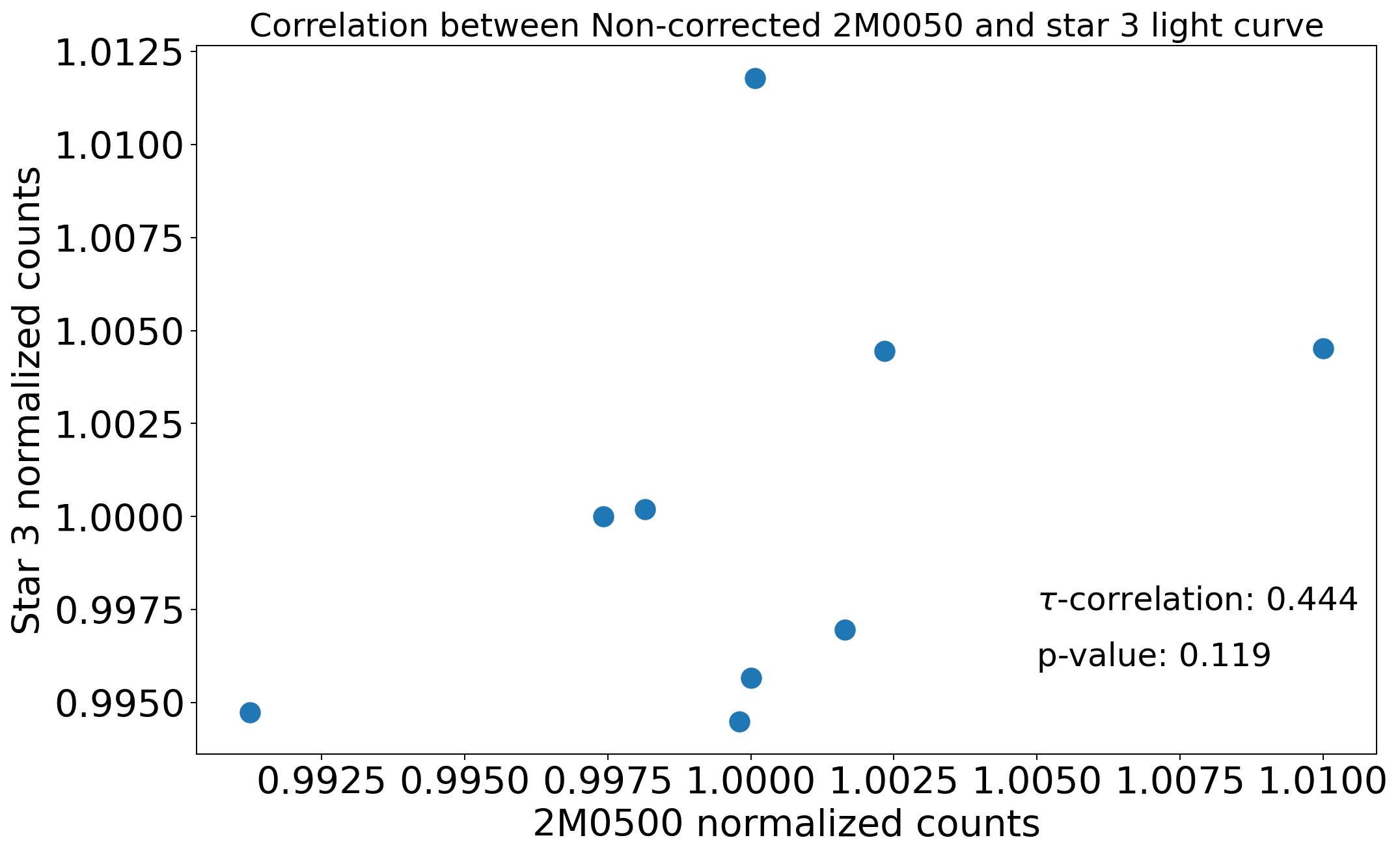}
    \includegraphics[width=0.48\textwidth]{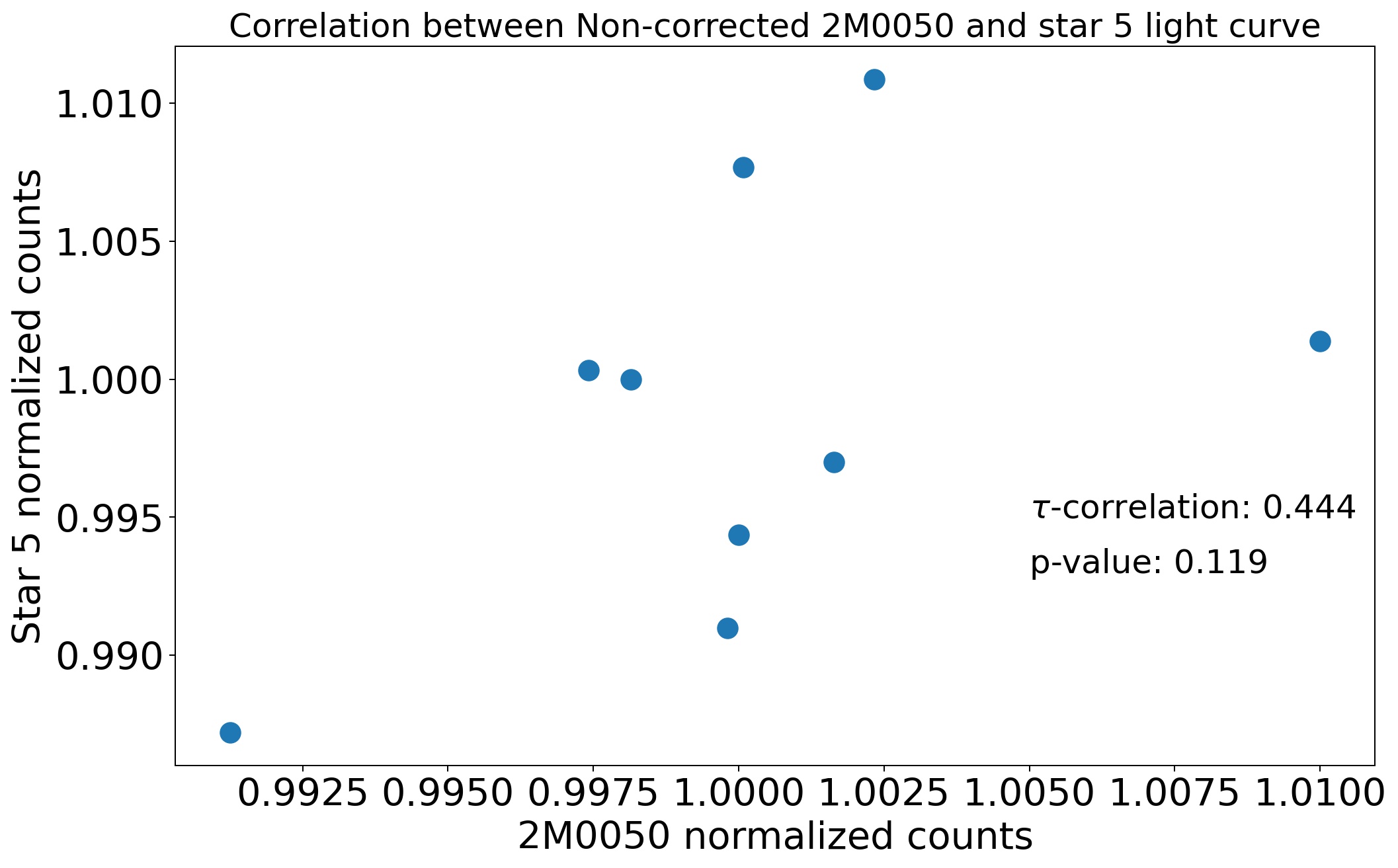}
    \caption{Correlation between the $H$-band target's non-corrected light curve, and the non-corrected calibration stars light curves.}
    \label{corr_noncorr_stars_Hband}
\end{figure}

\begin{figure}
    \centering
    \includegraphics[width=0.48\textwidth]{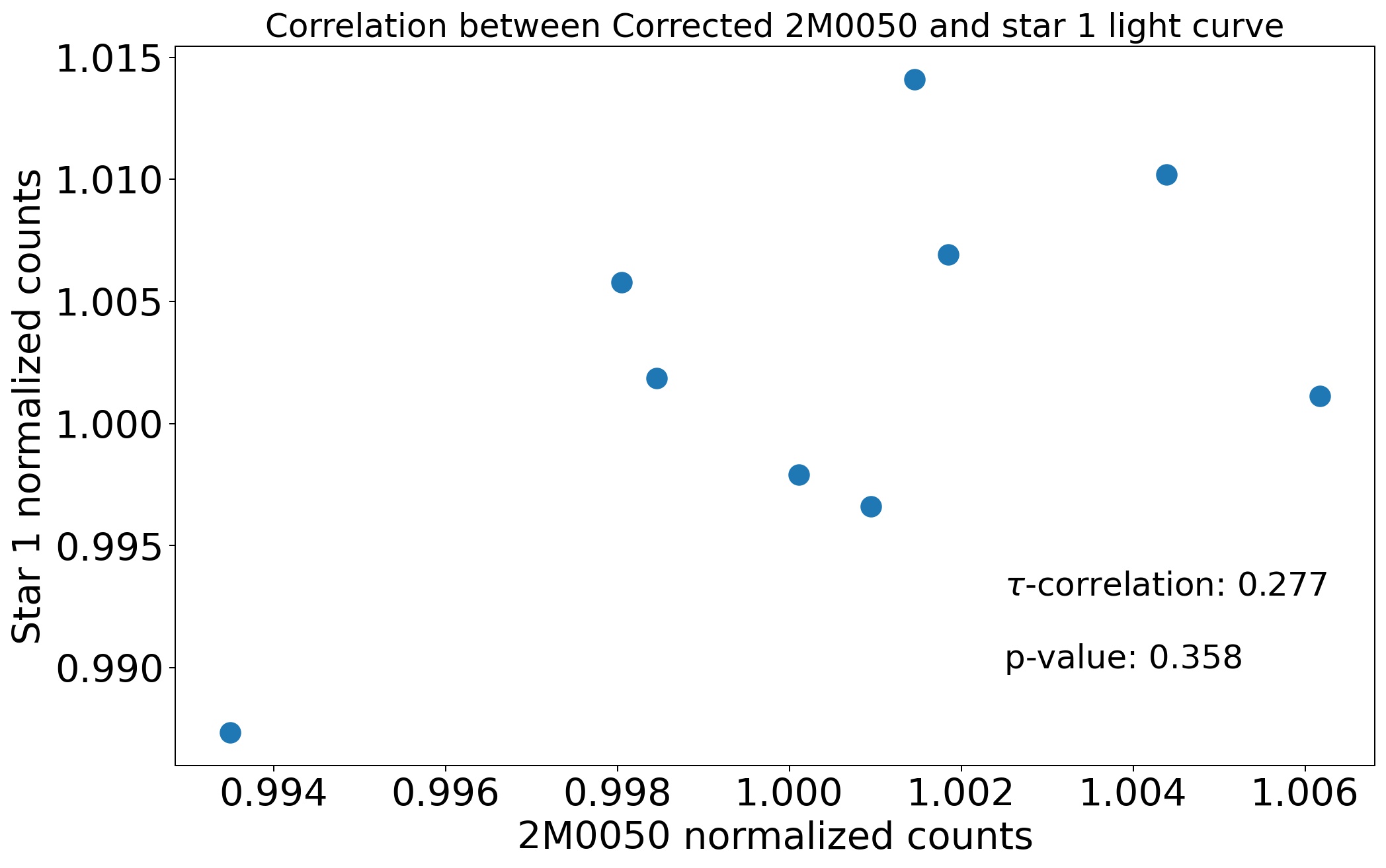}
    \includegraphics[width=0.48\textwidth]{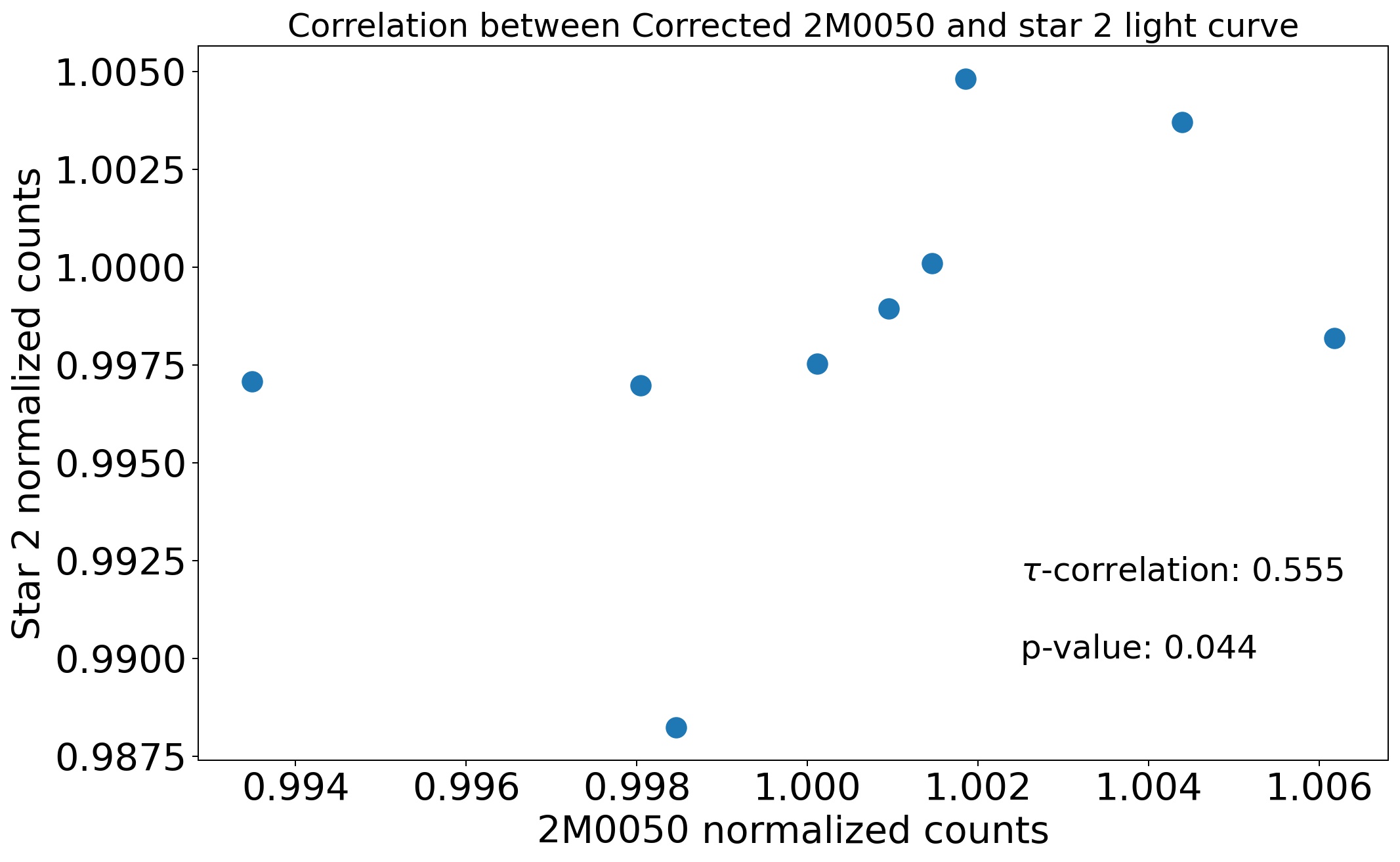}
    \includegraphics[width=0.48\textwidth]{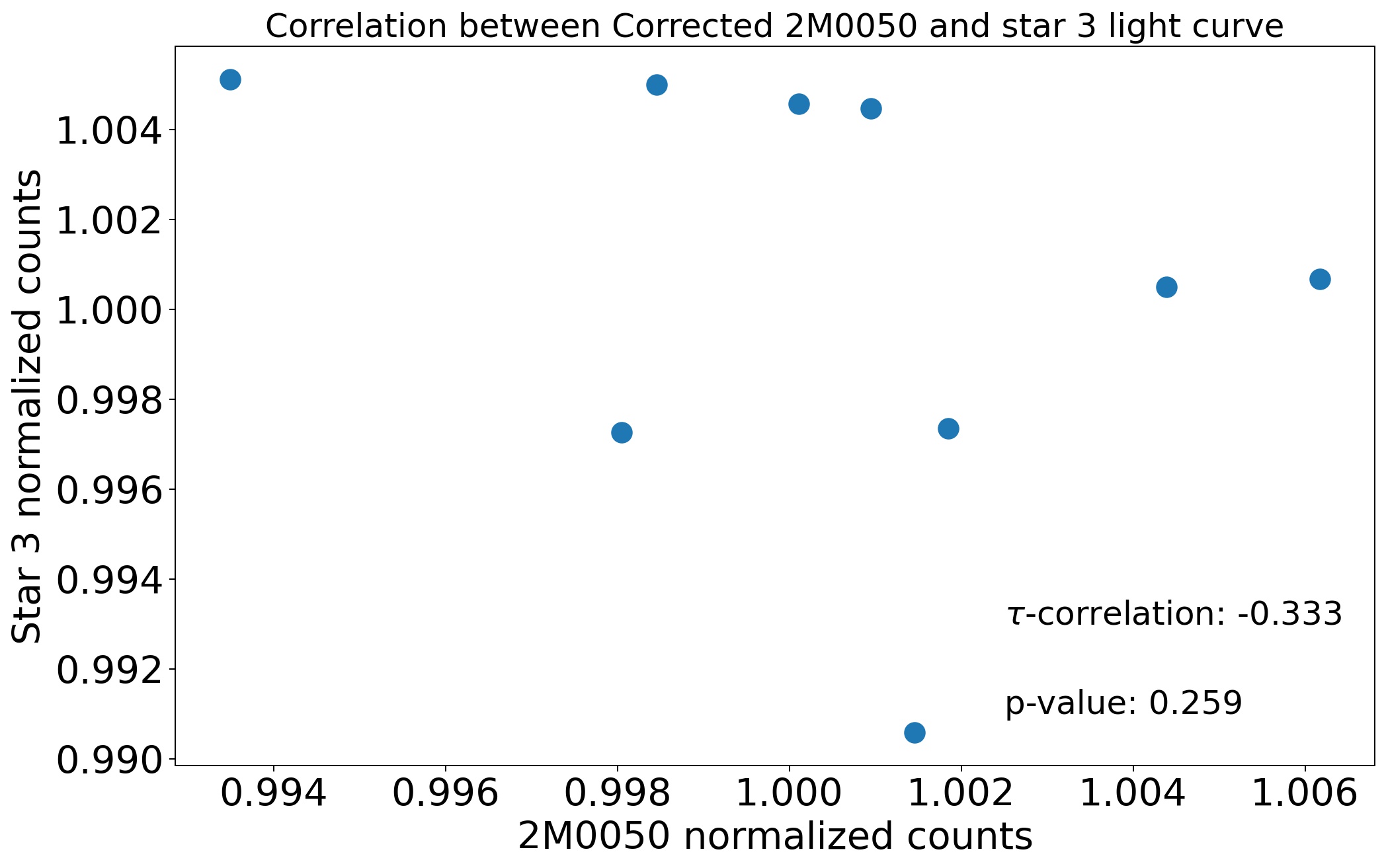}
    \includegraphics[width=0.48\textwidth]{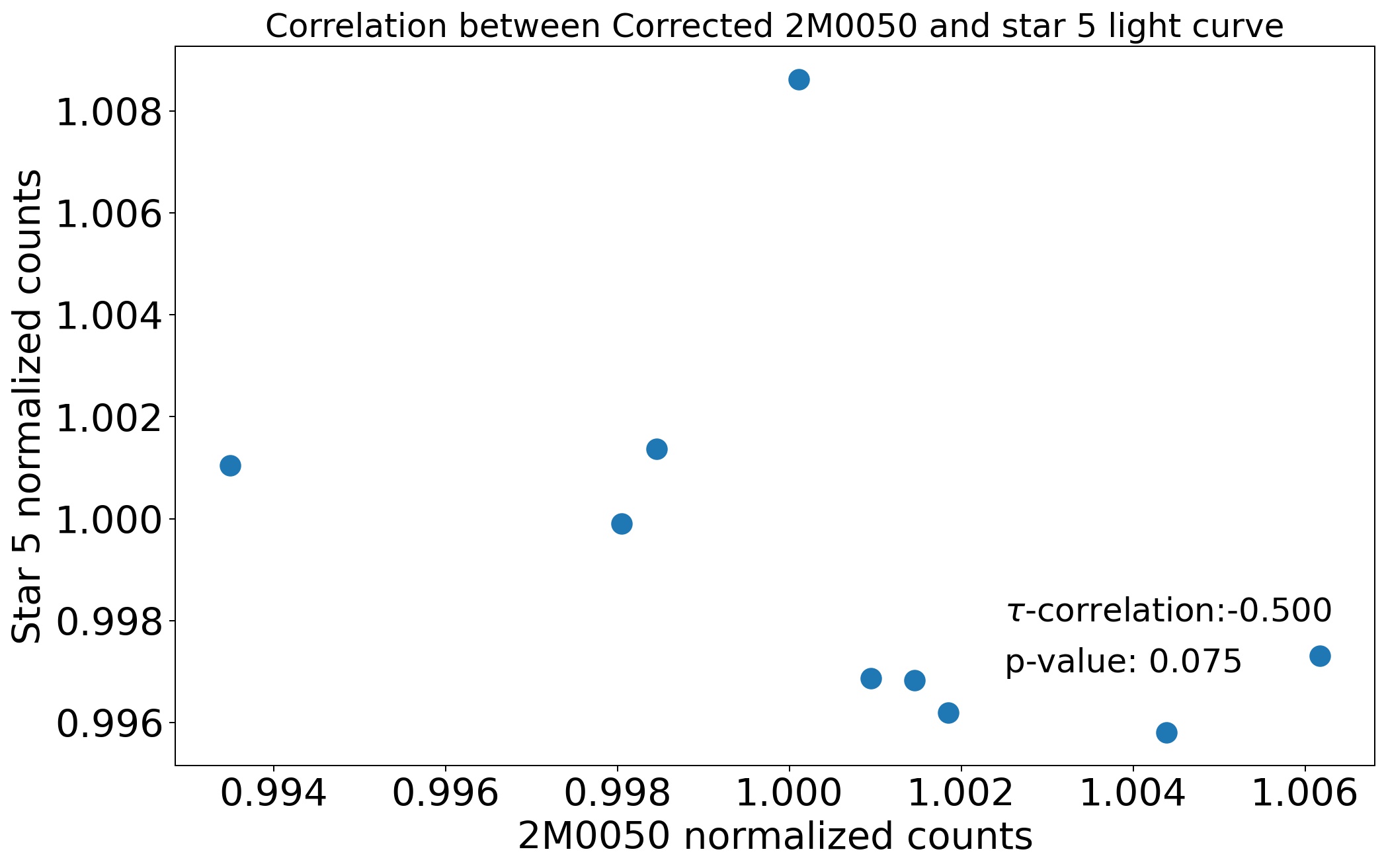}
    \caption{Correlation between the $H$-band target's corrected light curve, and the corrected calibration stars light curves.}
    \label{corr_corr_stars_Hband}
\end{figure}

\begin{figure}
    \centering
     \includegraphics[width=0.49\textwidth]{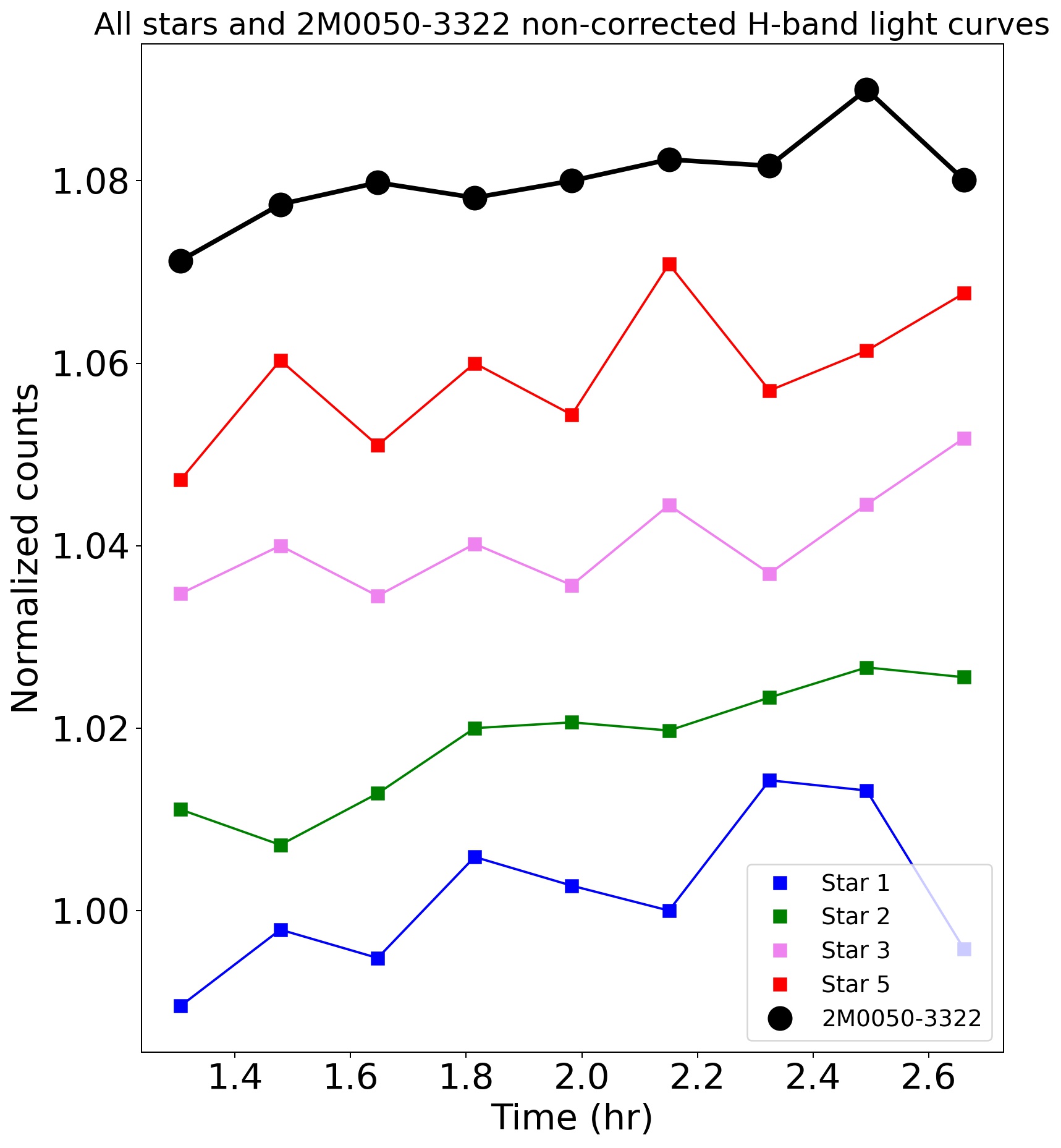}
     \includegraphics[width=0.48\textwidth]{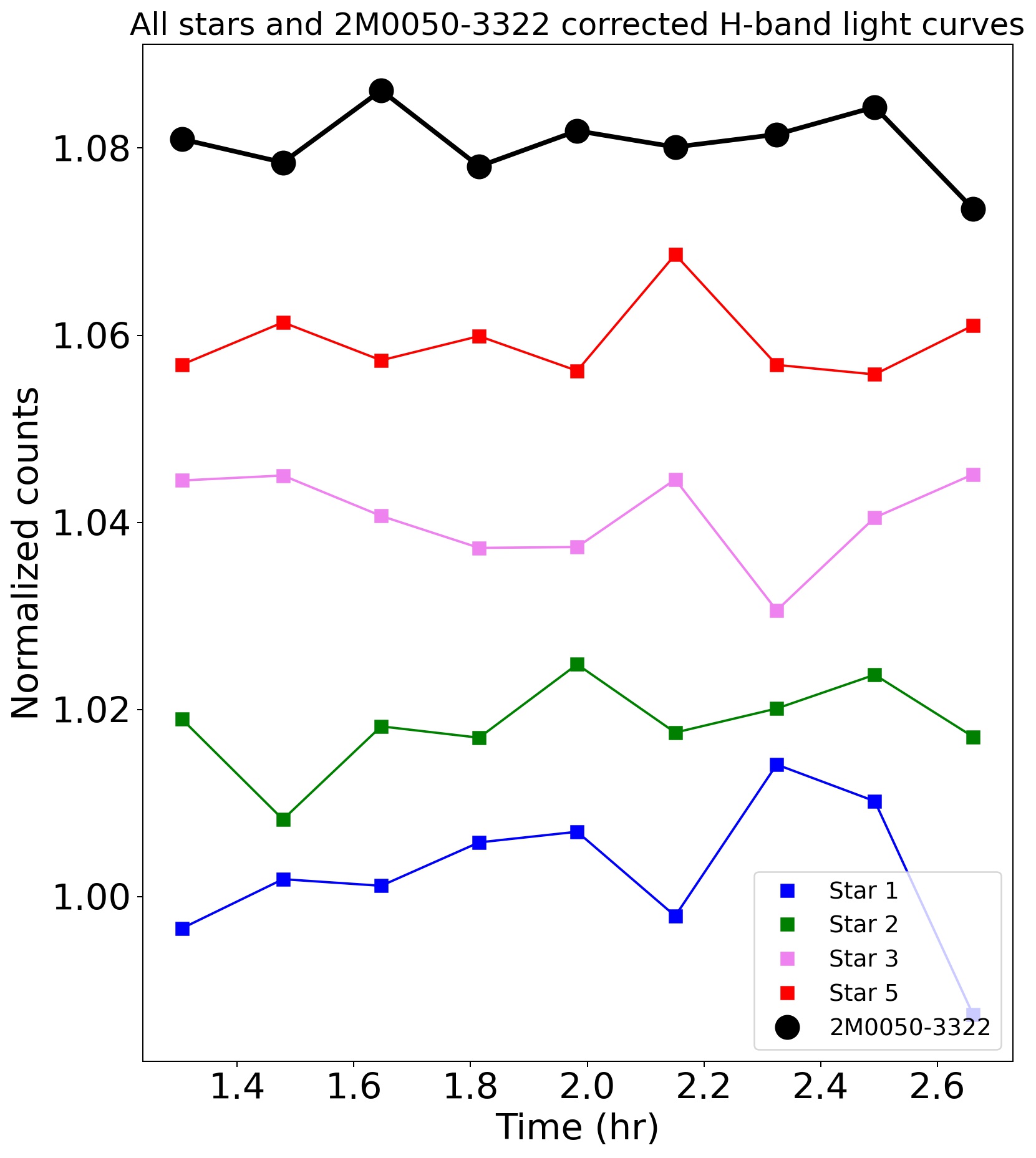}
    \caption{Non-corrected (left)  and corrected (right) $H$-band light curves of all stars and the target (black line).}
    \label{all_LCs_plot}
\end{figure}

\begin{figure}
    \centering
    \includegraphics[width=0.48\textwidth]{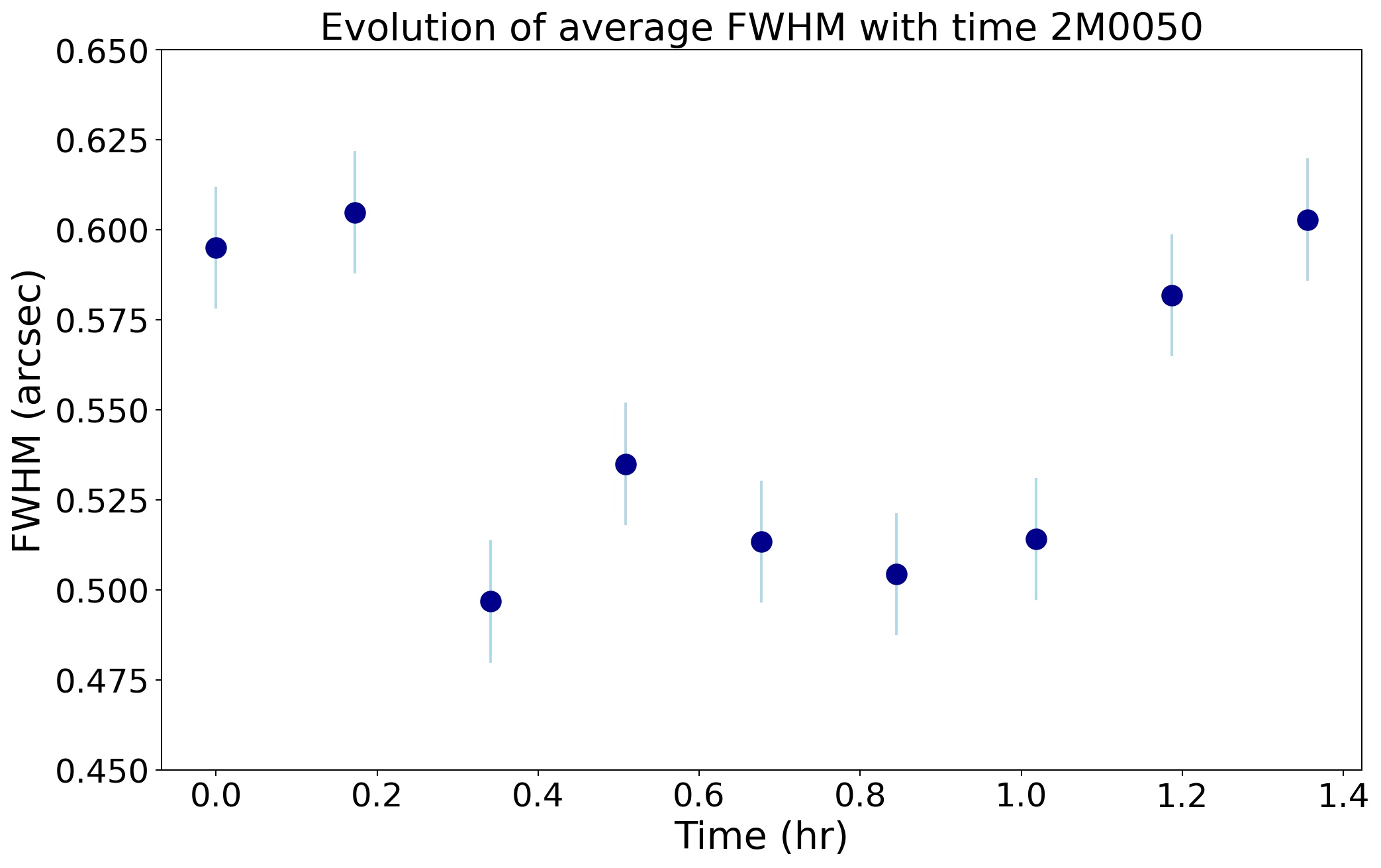}
    \includegraphics[width=0.48\textwidth]{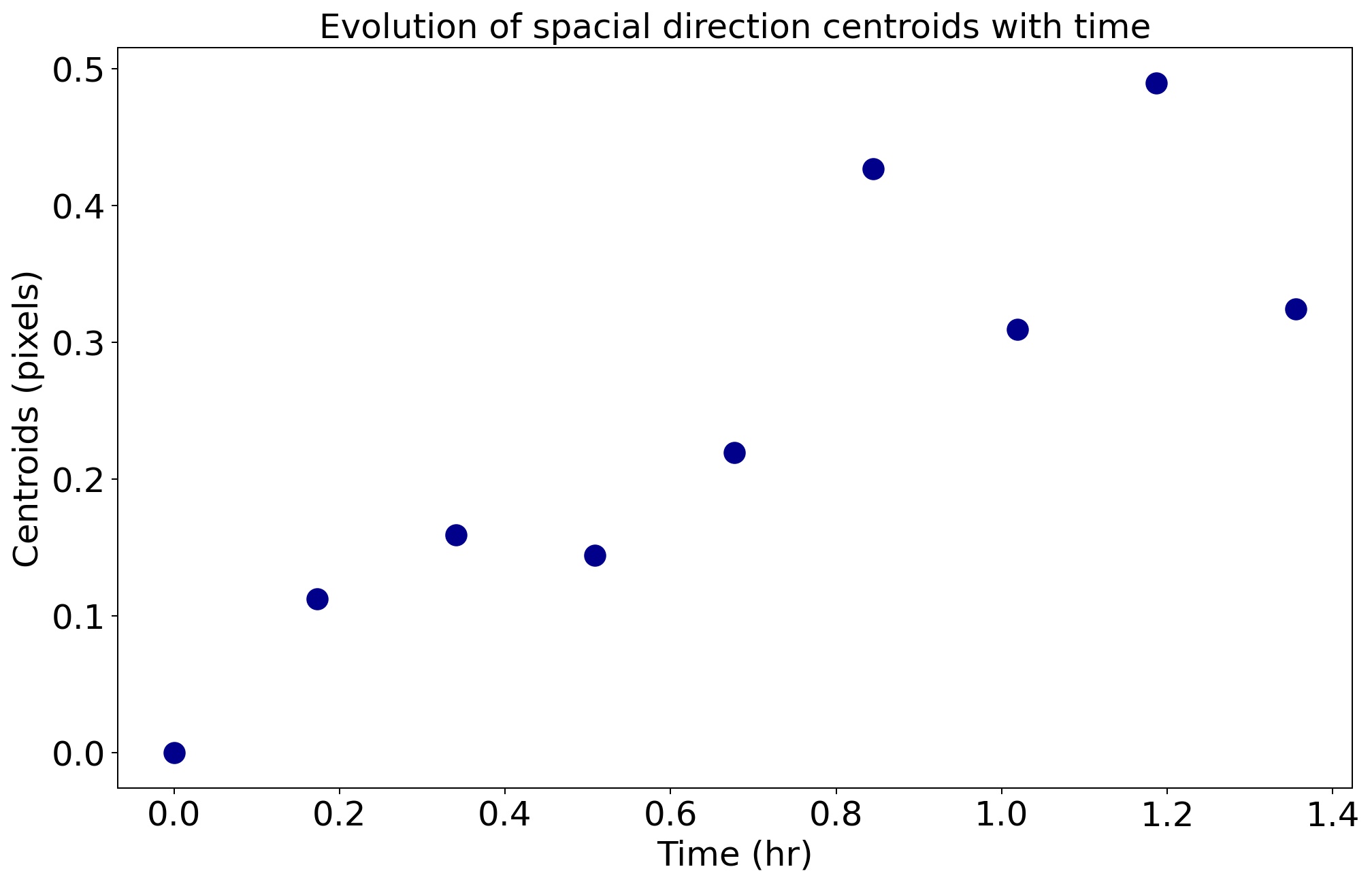}
    \caption{Left: Evolution of the calculated FWHM for the $H$-band 2D spectra with time. Right: Evolution of the centroids of the $H$-band 2D spectra with time.}
    \label{evol_FWHM_centroids}
\end{figure}

\begin{figure}
    \centering
    \includegraphics[width=0.48\textwidth]{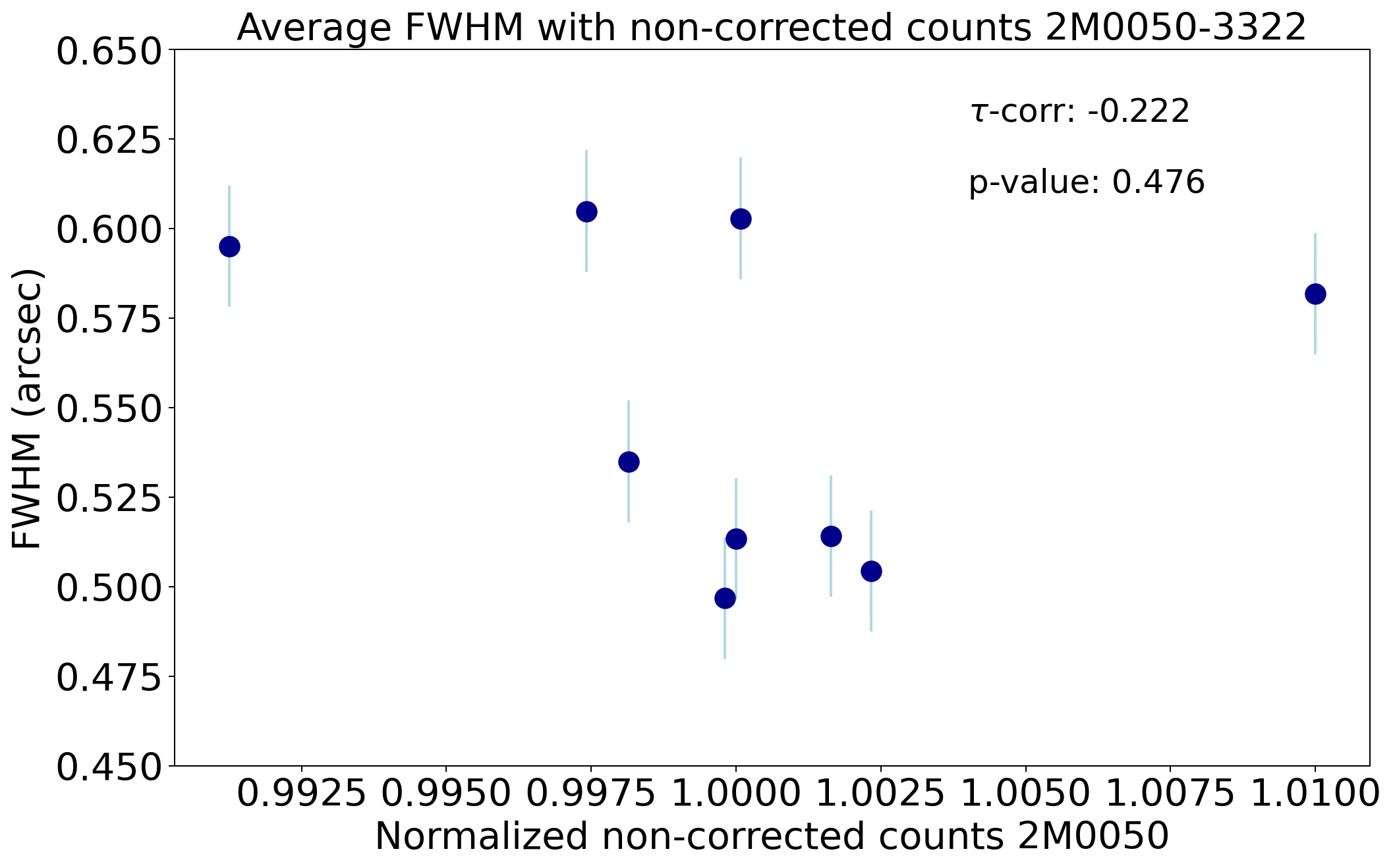}
    \includegraphics[width=0.48\textwidth]{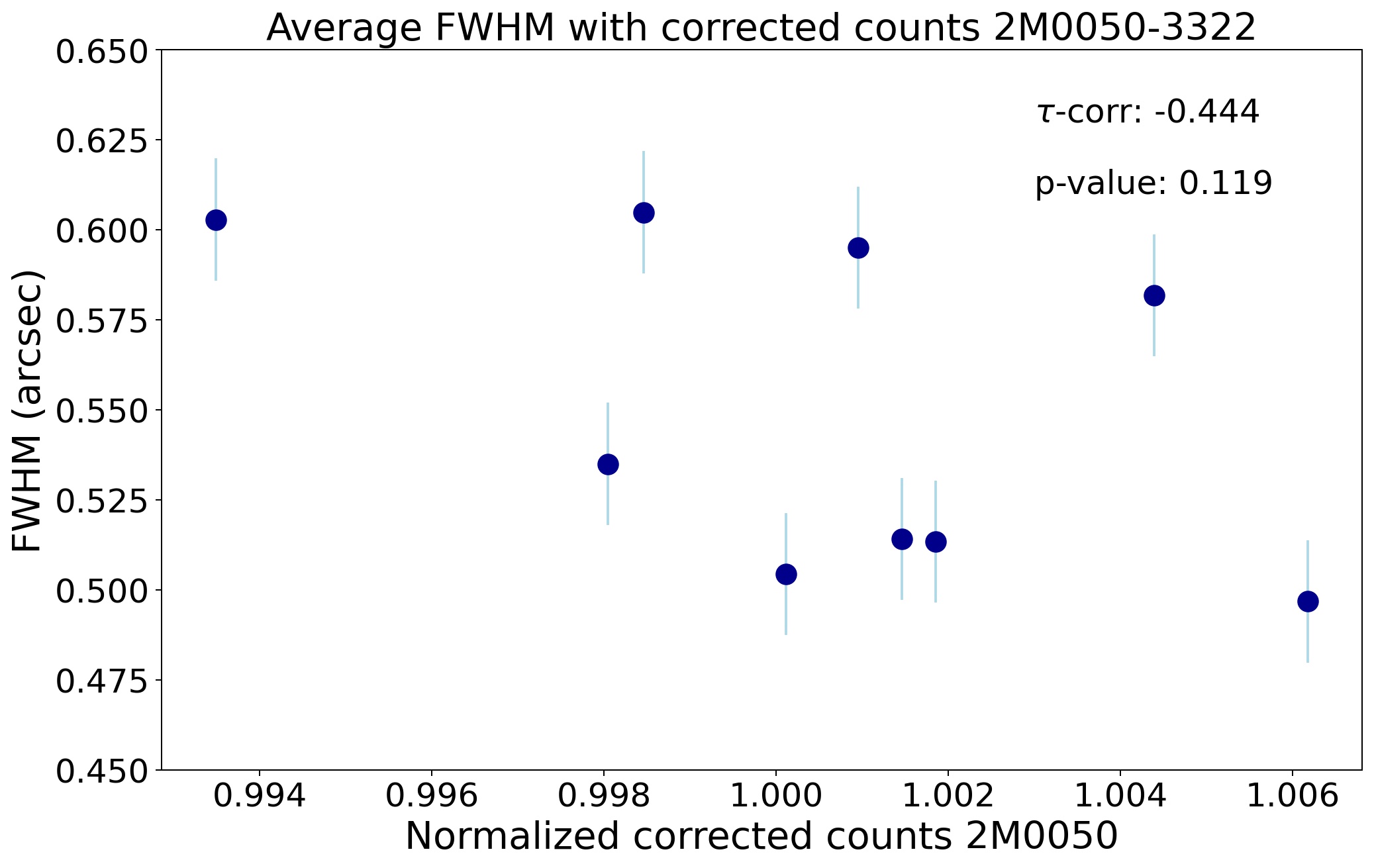}
    \caption{Correlation between the FWHM of the 2D $H$-band spectra and the corrected and non-corrected $H$-band light curve.}
    \label{corr_FWHM}
\end{figure}

\begin{figure}
    \centering
    \includegraphics[width=0.48\textwidth]{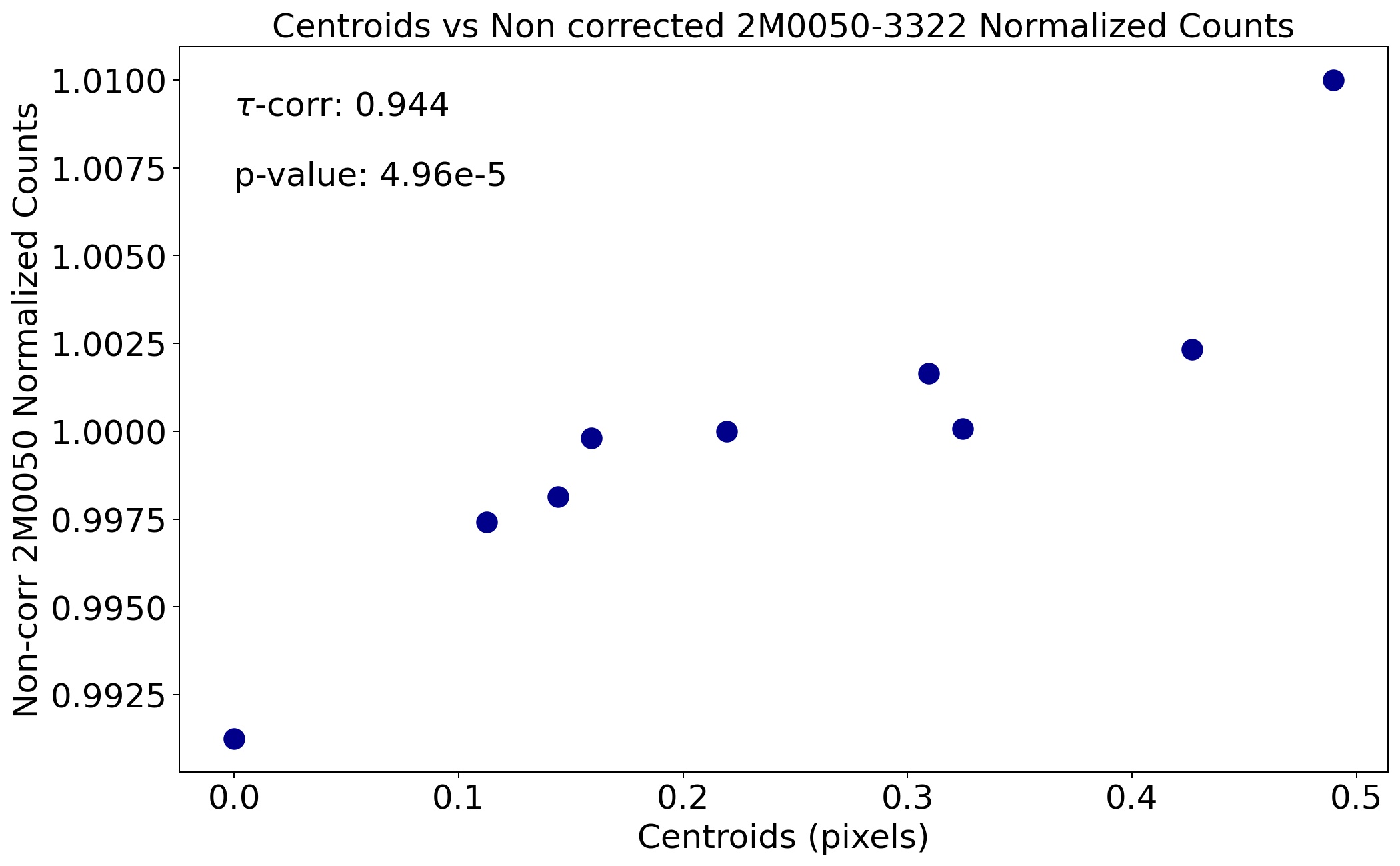}
    \includegraphics[width=0.48\textwidth]{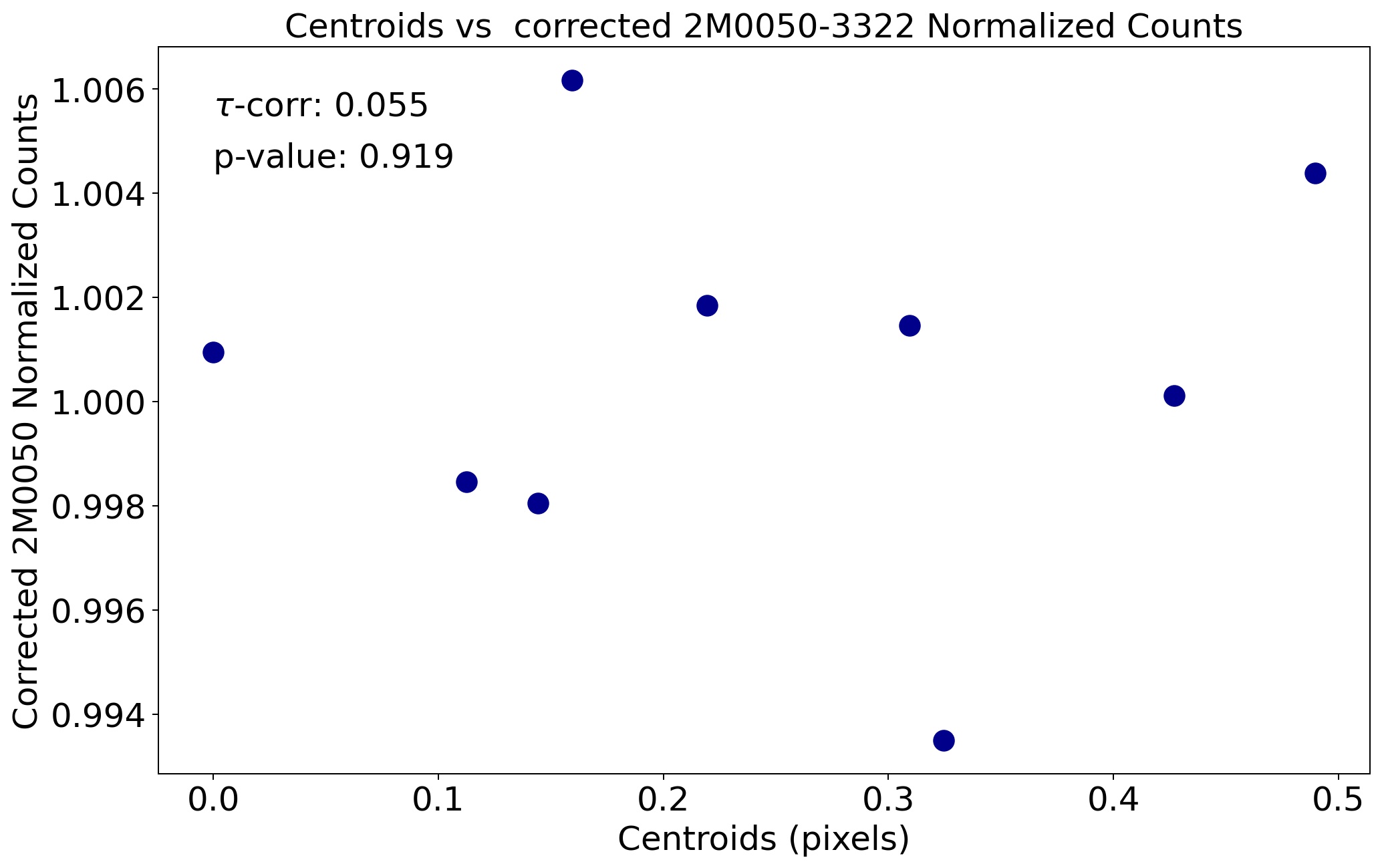}
    \caption{Correlation between the centroids of the 2D $H$-band spectra and the non-corrected $H$-band light curve.}
    \label{corr_pointing}
\end{figure}

\begin{figure}
    \centering
    \includegraphics[width=0.48\textwidth]{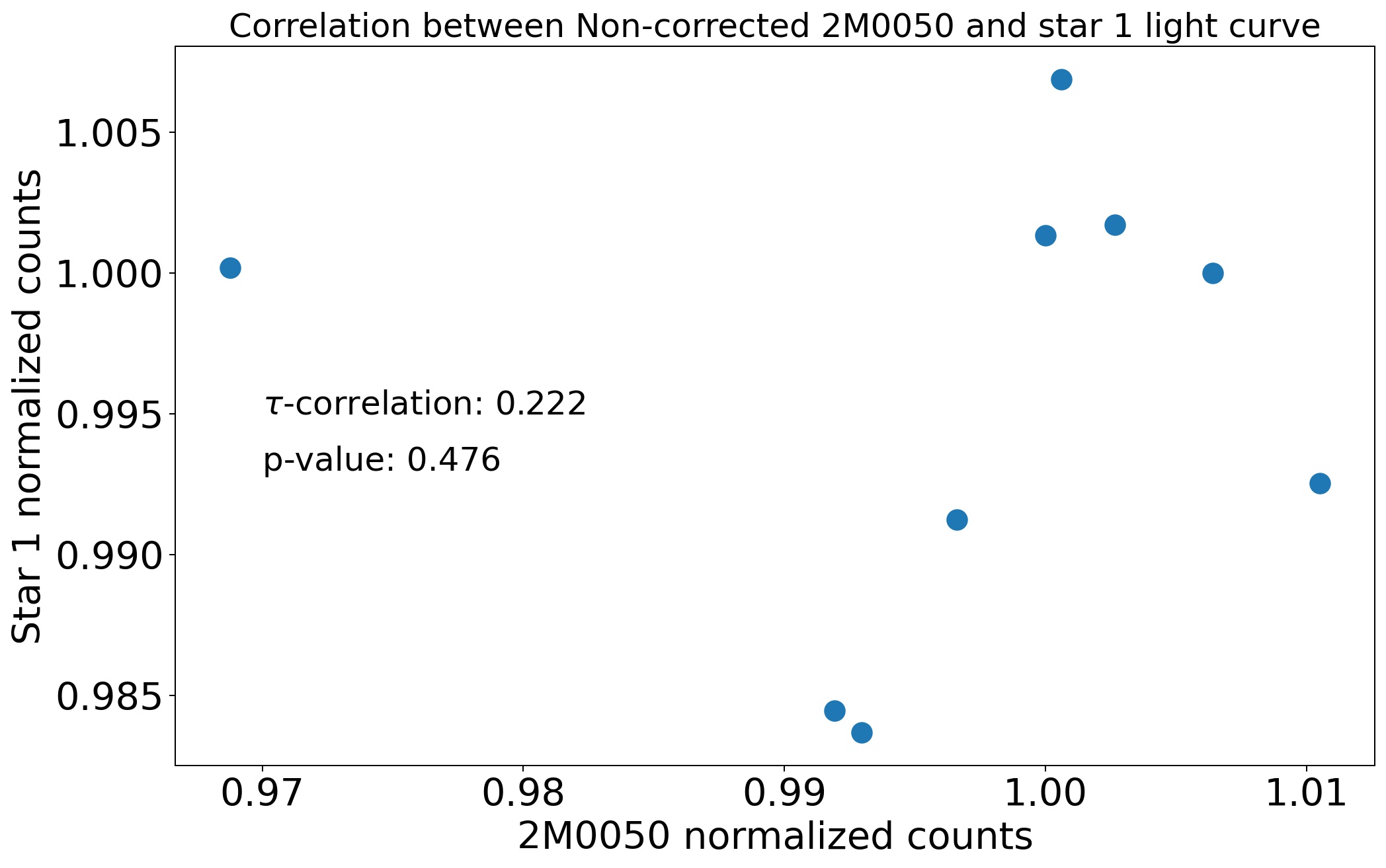}
    \includegraphics[width=0.48\textwidth]{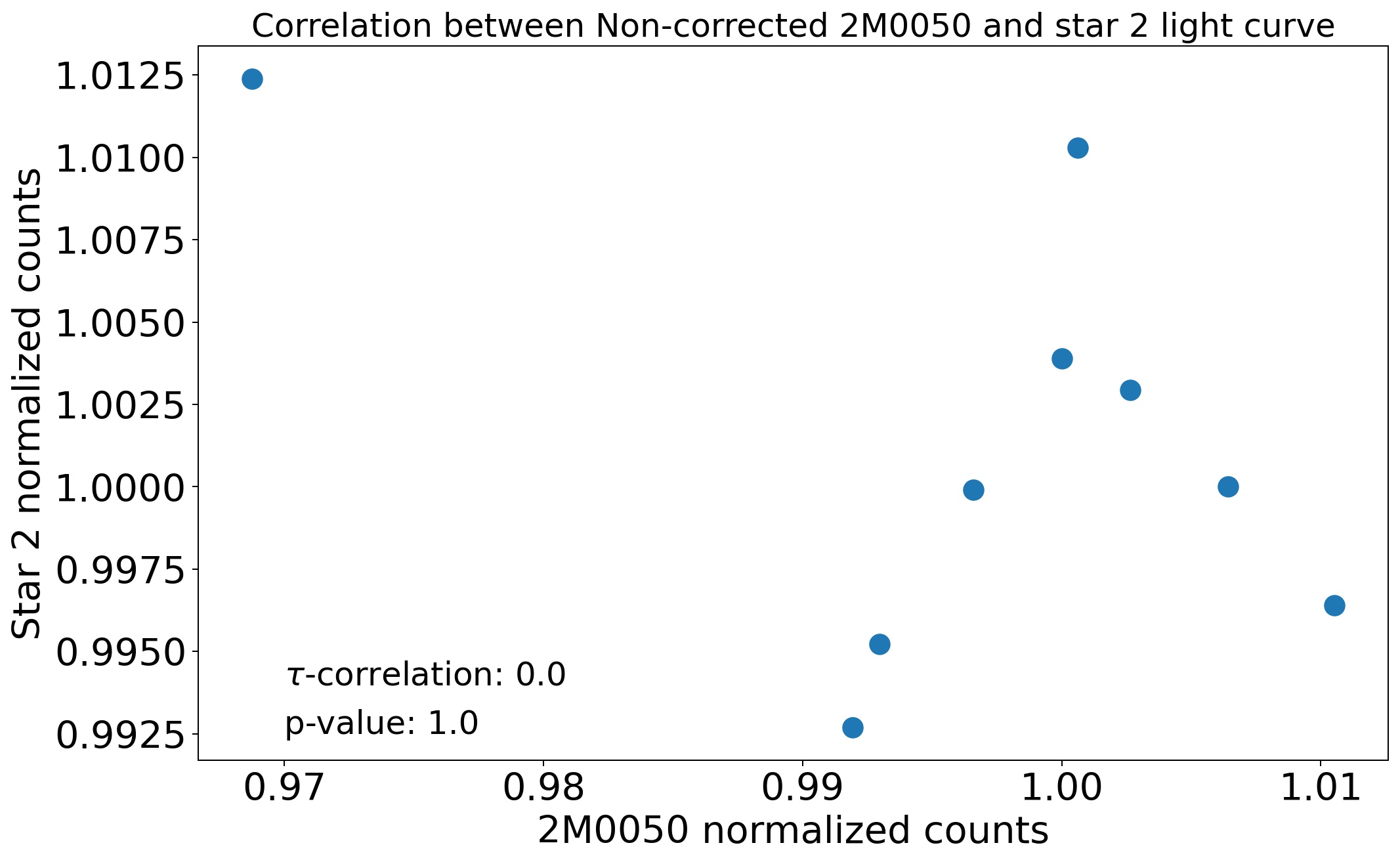}
    \caption{Correlation between the $\mathrm{CH_{4}-H_{2}O}$ target's non-corrected light curve, and the non-corrected calibration stars light curves.}
    \label{corr_noncorr_stars_CH4_H2O_band}
\end{figure}

\begin{figure}
    \centering
    \includegraphics[width=0.48\textwidth]{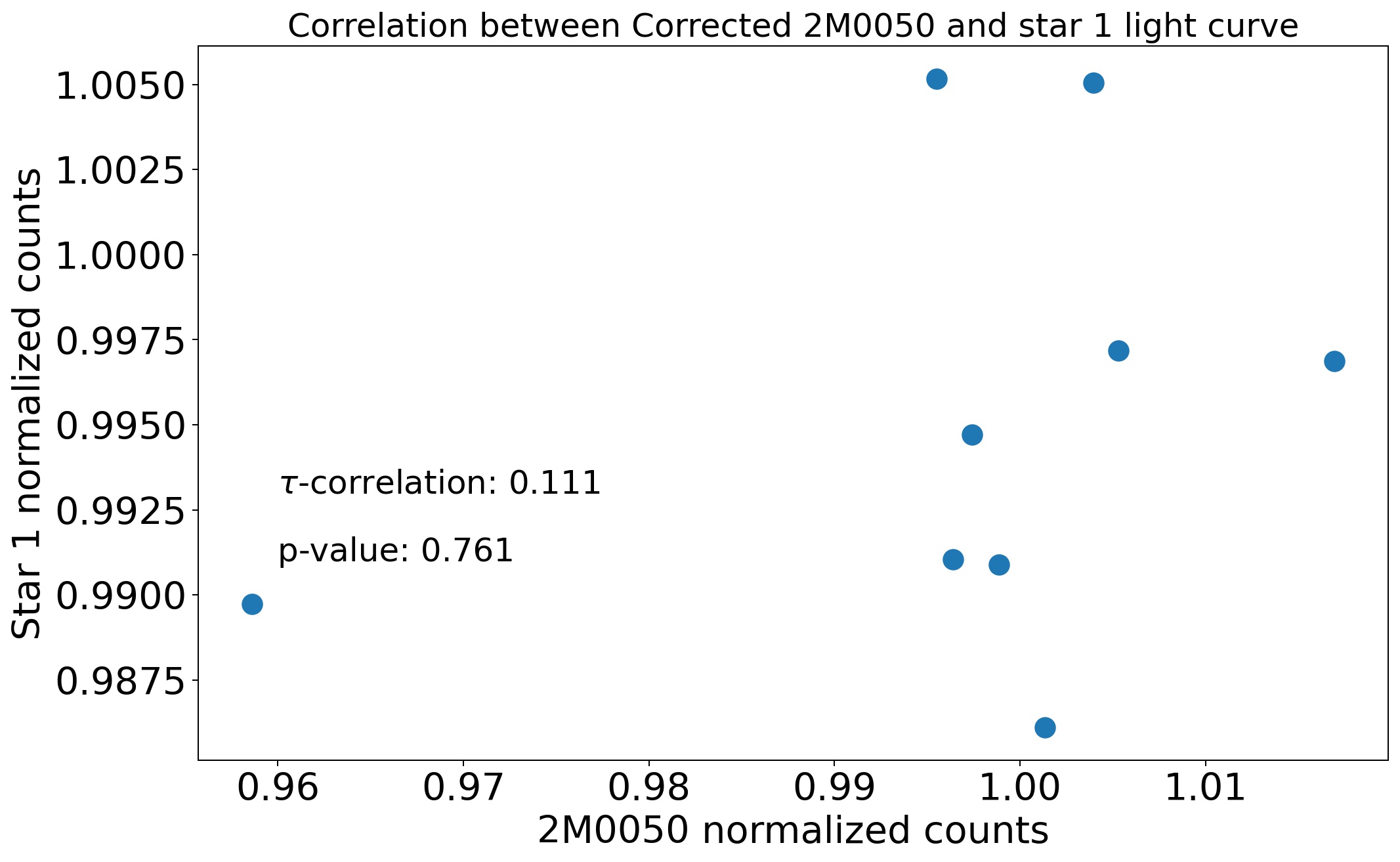}
    \includegraphics[width=0.48\textwidth]{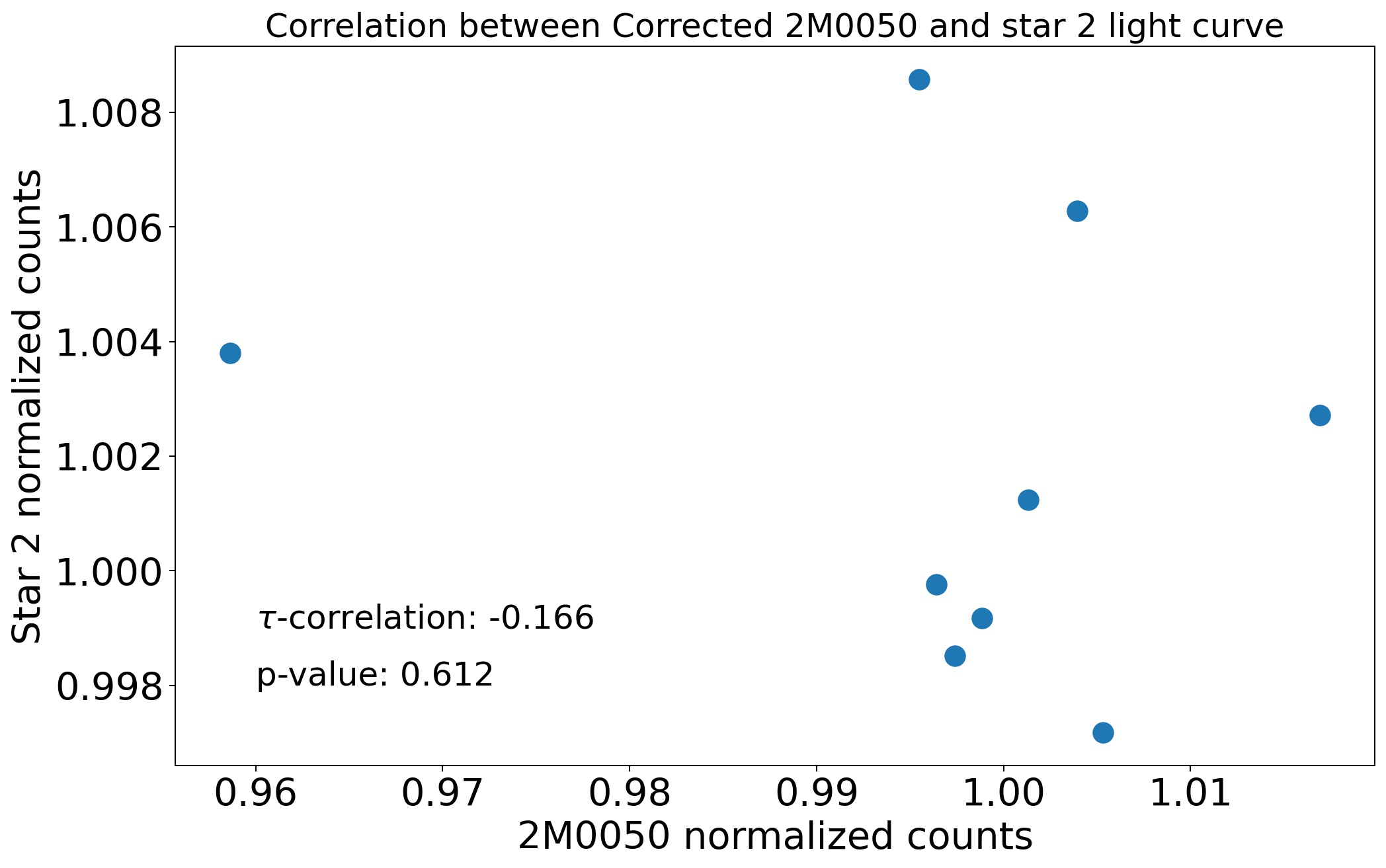}
    \caption{Correlation between the $\mathrm{CH_{4}-H_{2}O}$ target's corrected light curve, and the corrected calibration stars light curves.}
    \label{corr_corr_stars_CH4_H2O_band}
\end{figure}

\begin{figure}
    \centering
    \includegraphics[width=0.48\textwidth]{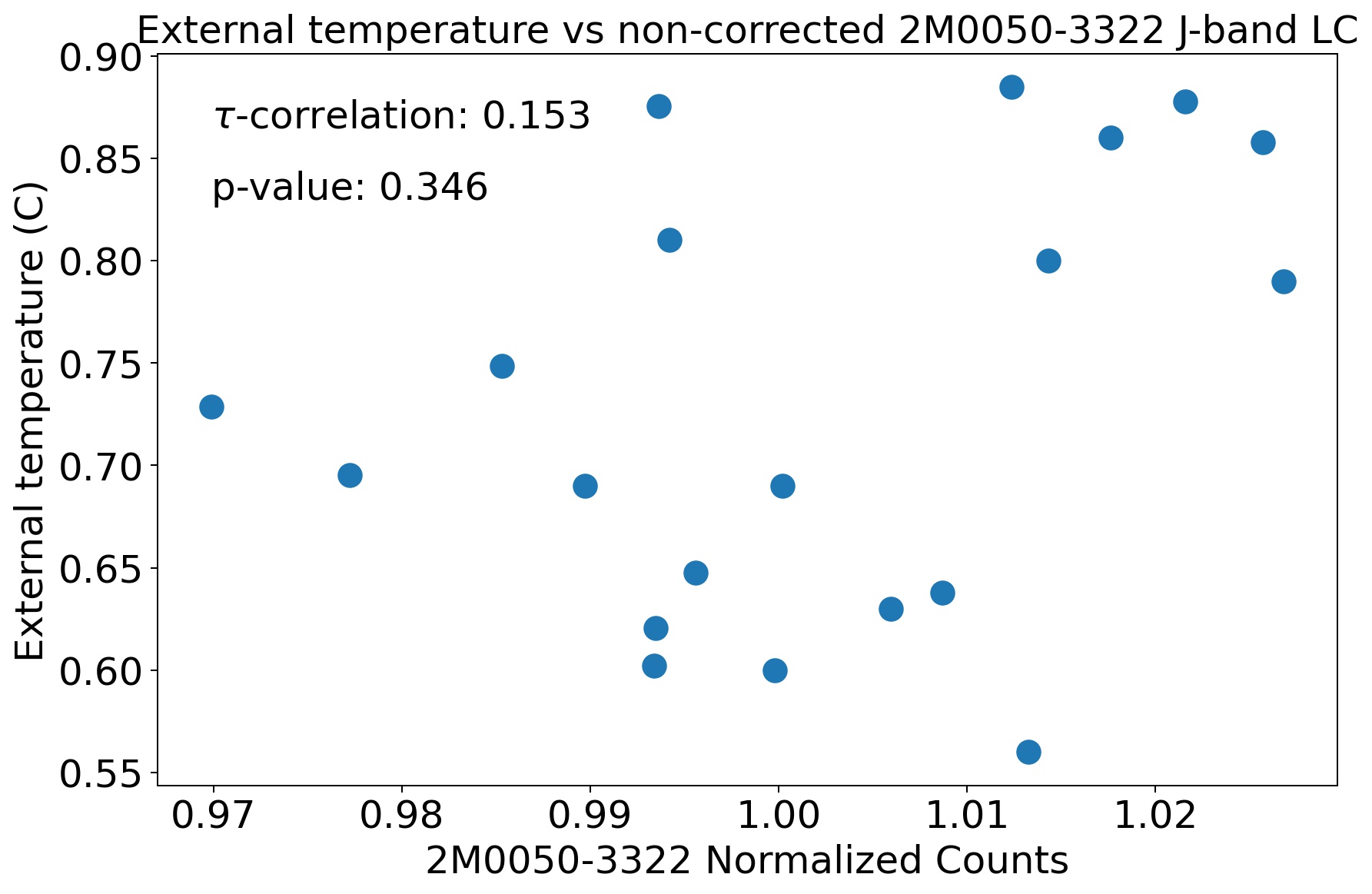}
    \includegraphics[width=0.48\textwidth]{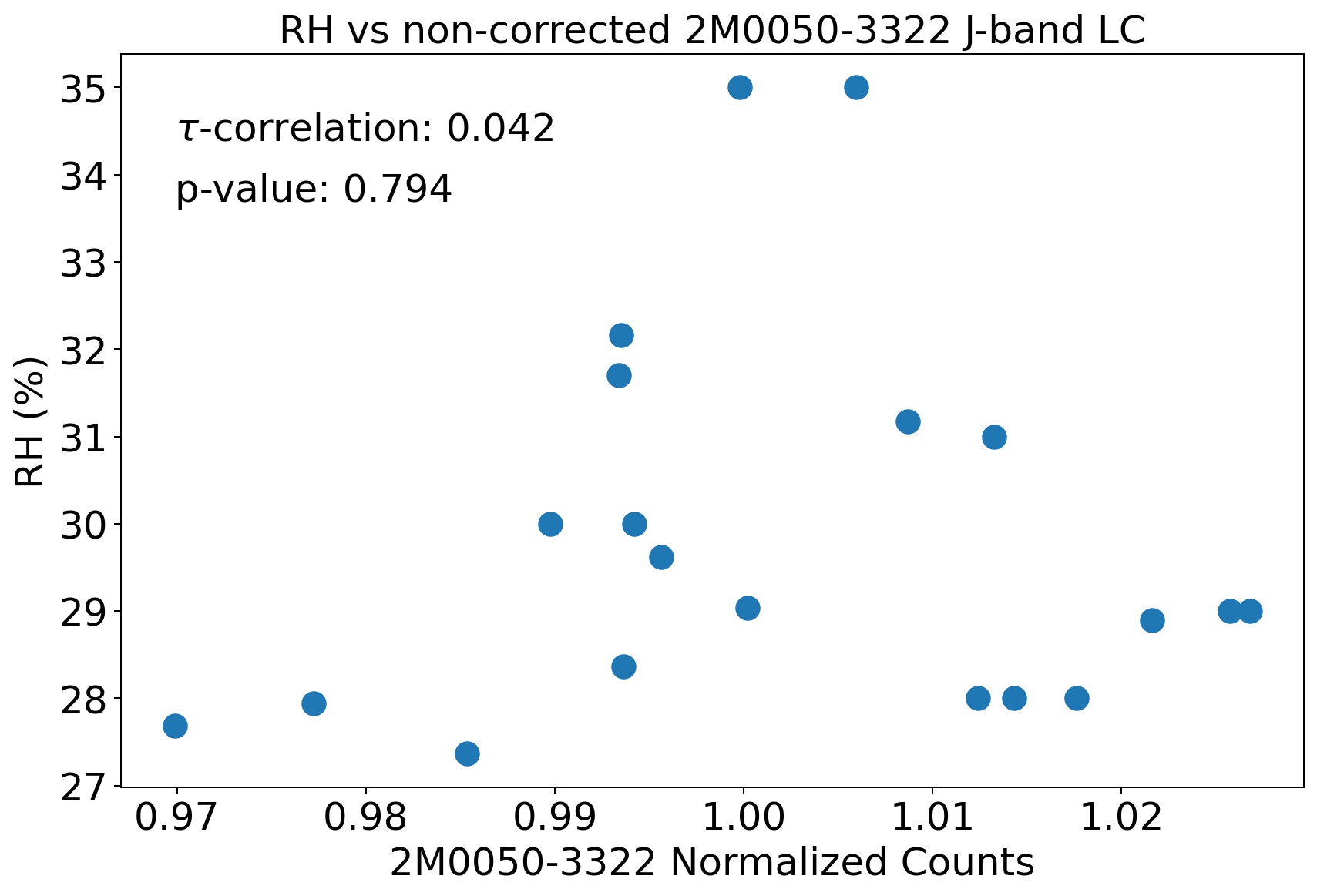}
    \includegraphics[width=0.48\textwidth]{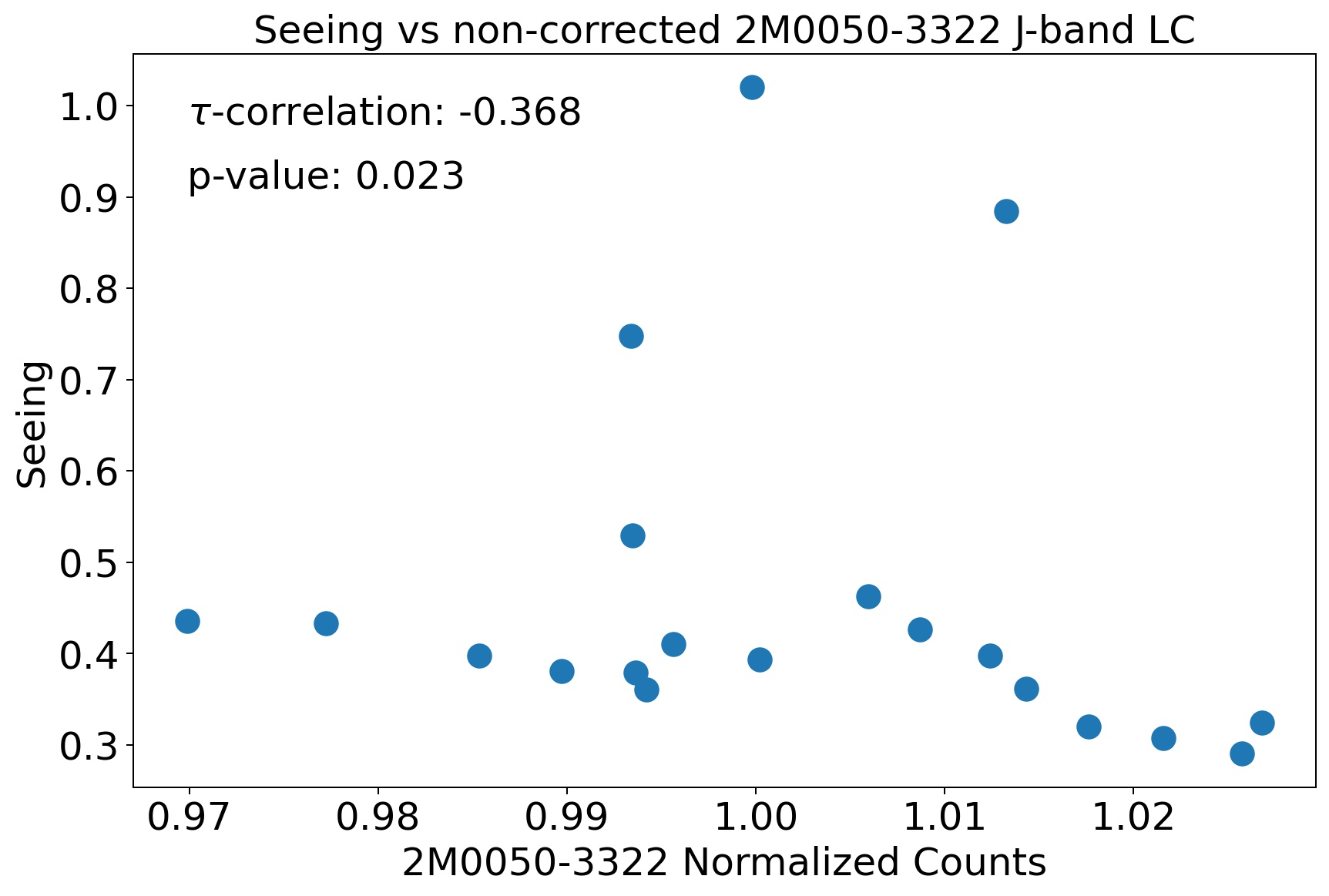}
    \includegraphics[width=0.48\textwidth]{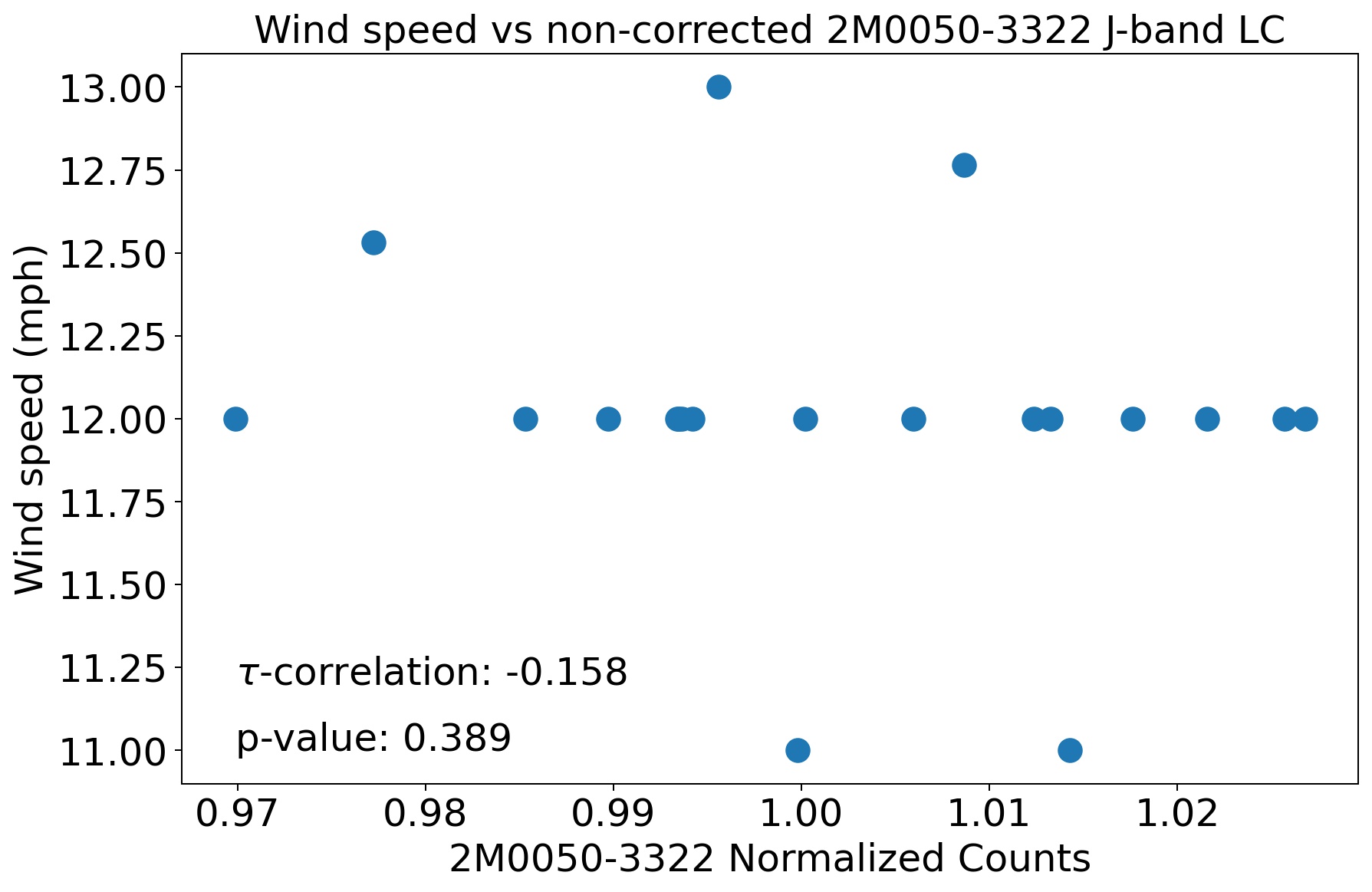}
     \includegraphics[width=0.48\textwidth]{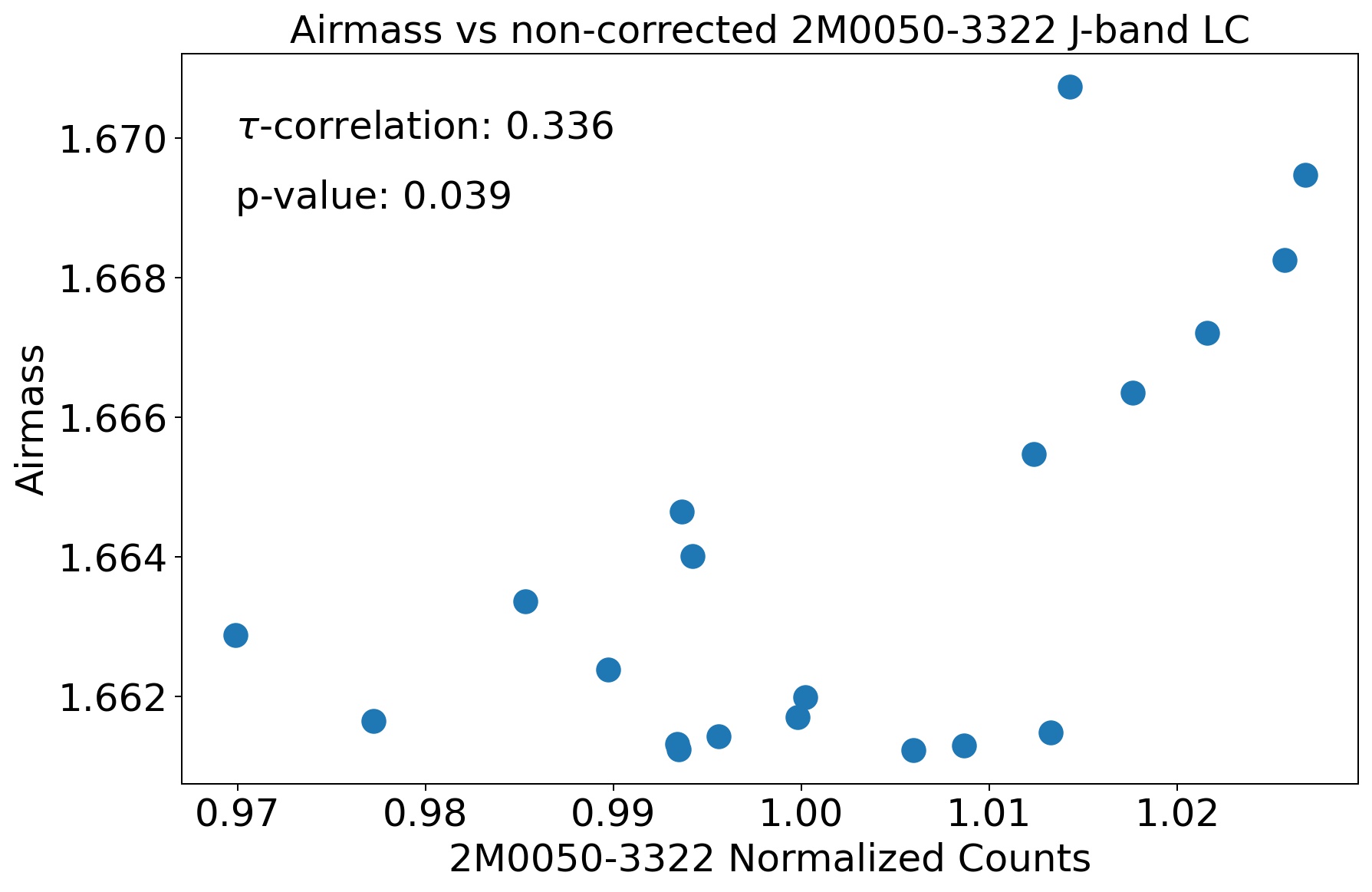}
    \caption{Correlations between the non-corrected $J$-band imaging light curve and different atmospheric parameters.}
    \label{J_band_noncorr}
\end{figure}

\begin{figure}
    \centering
    \includegraphics[width=0.48\textwidth]{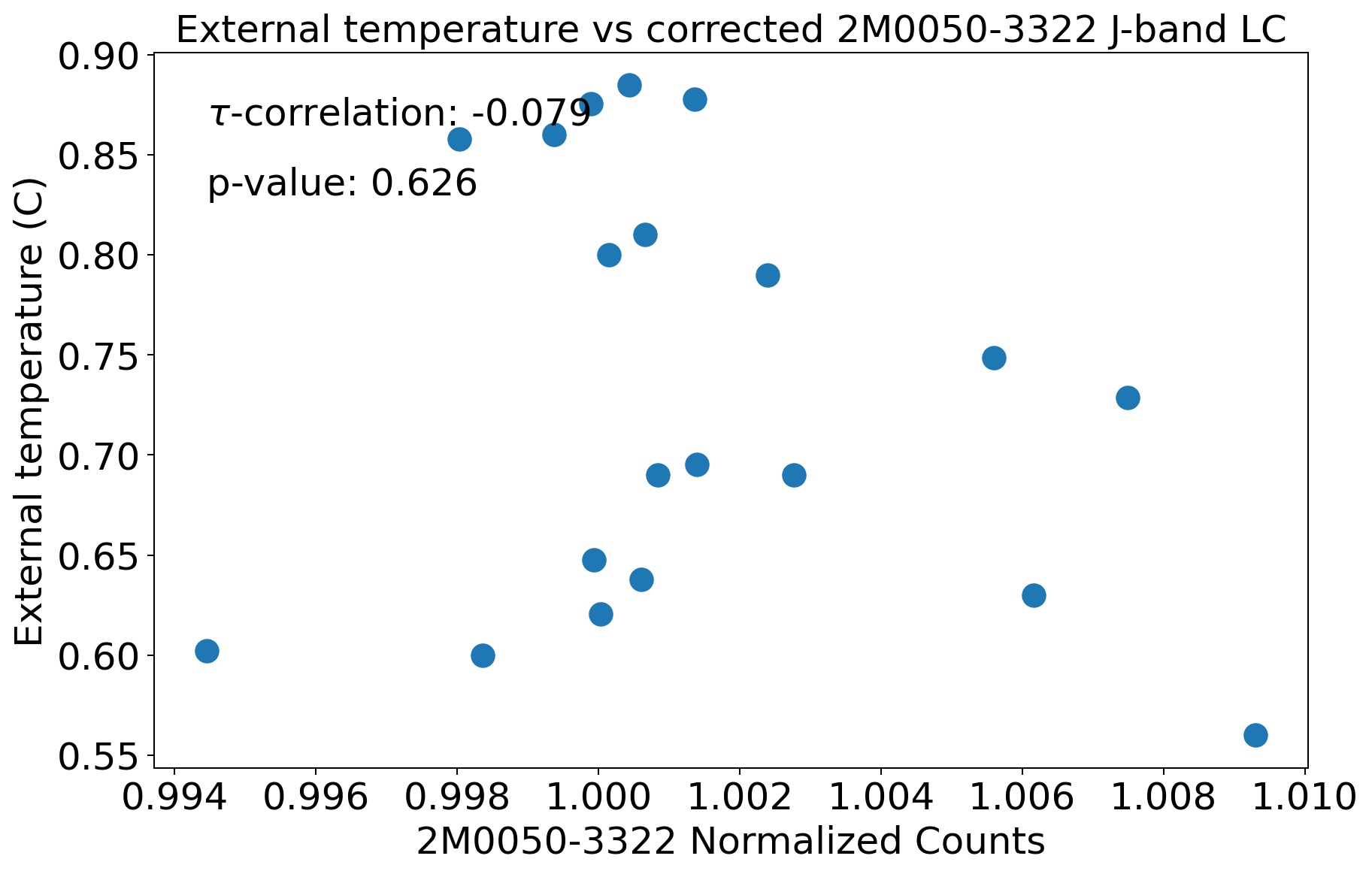}
    \includegraphics[width=0.48\textwidth]{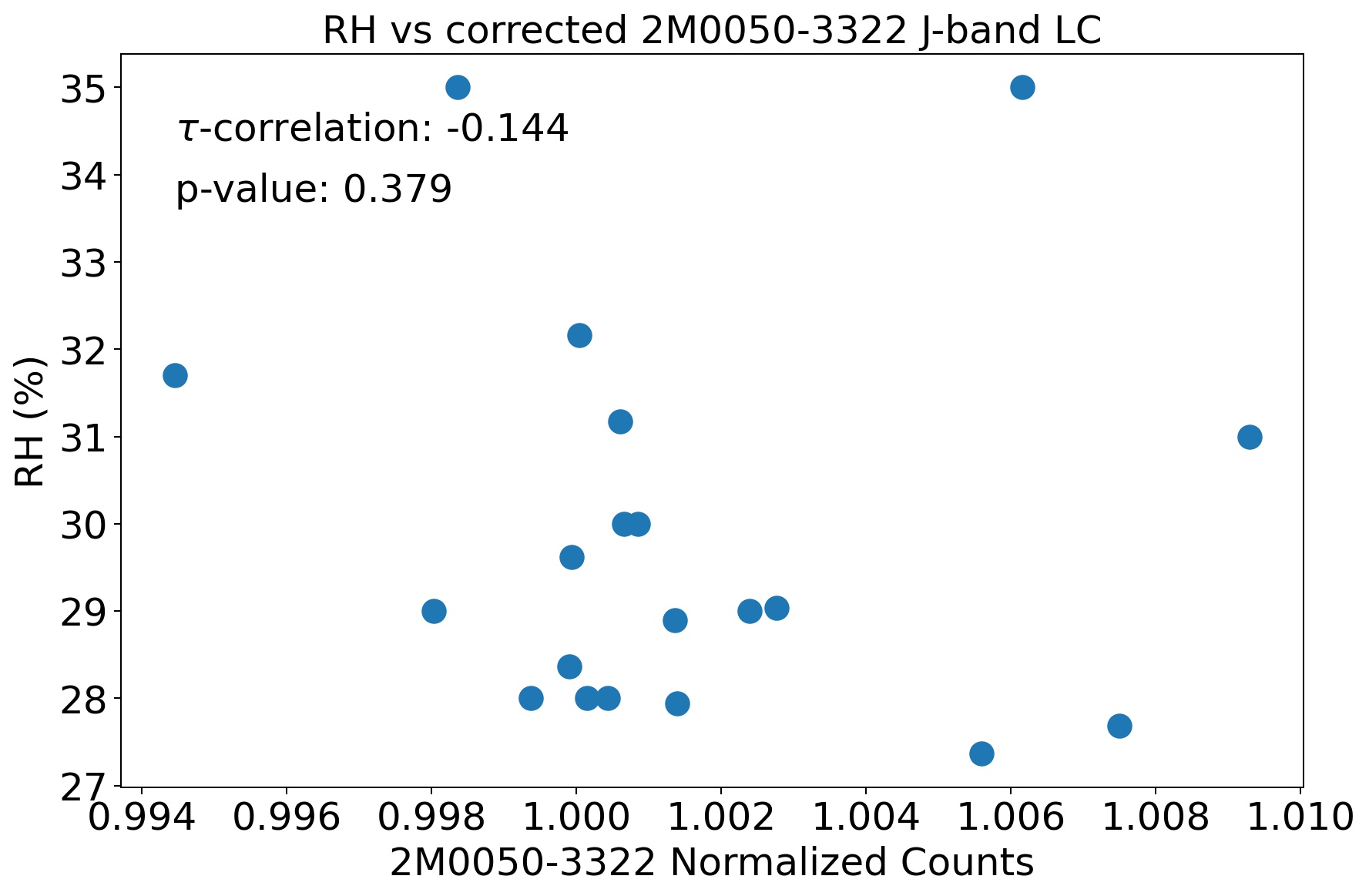}
    \includegraphics[width=0.48\textwidth]{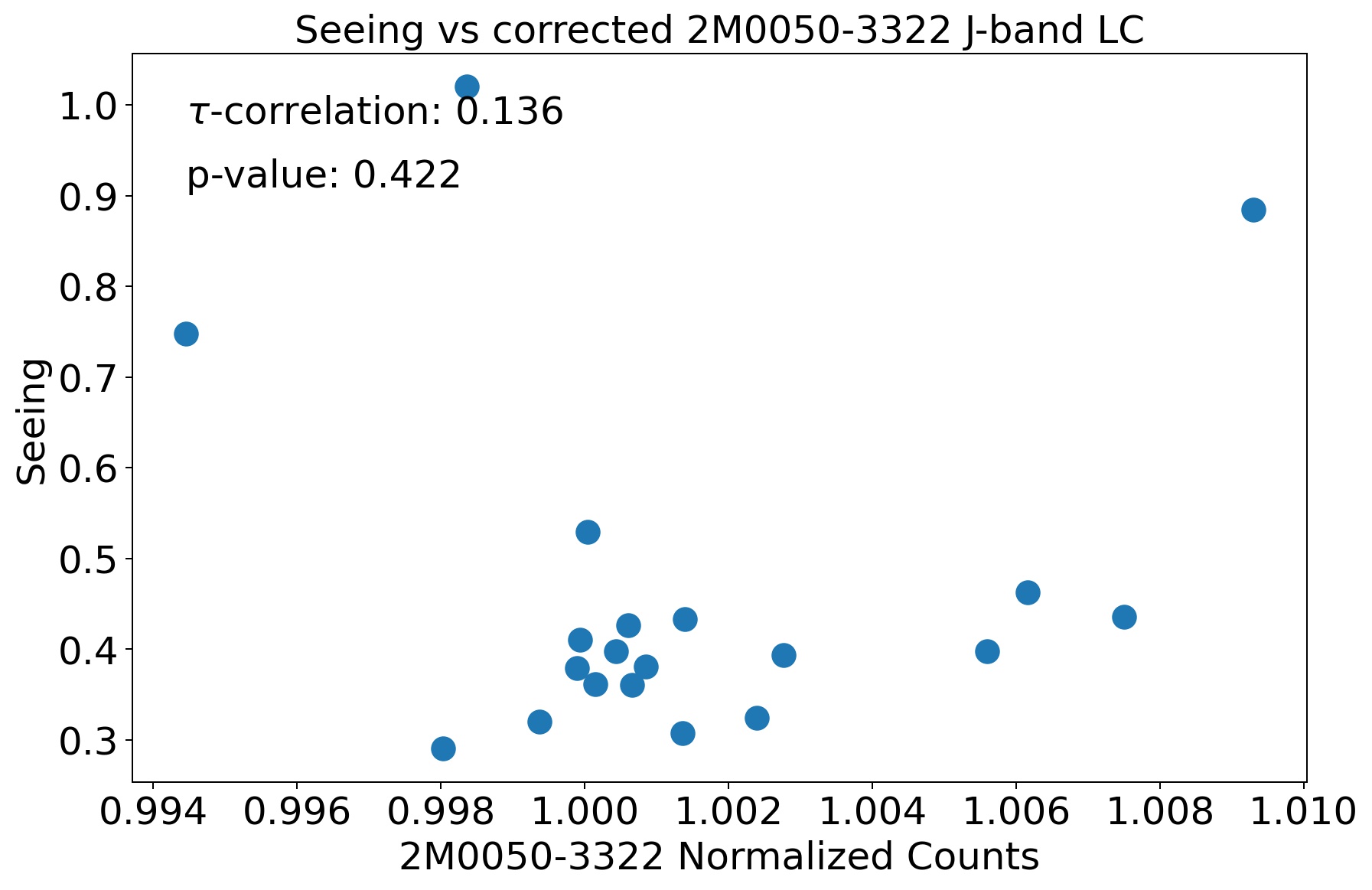}
    \includegraphics[width=0.48\textwidth]{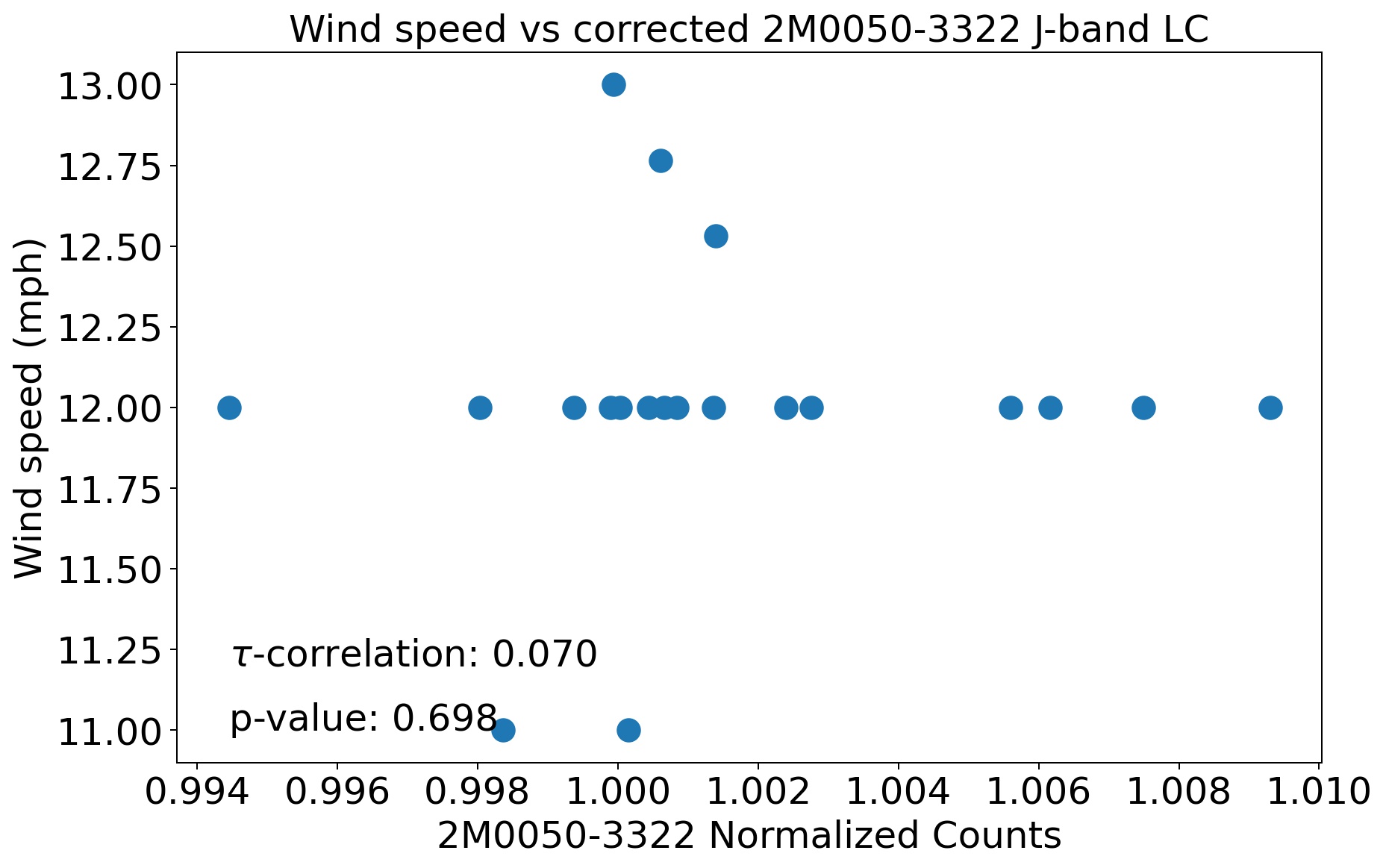}
     \includegraphics[width=0.48\textwidth]{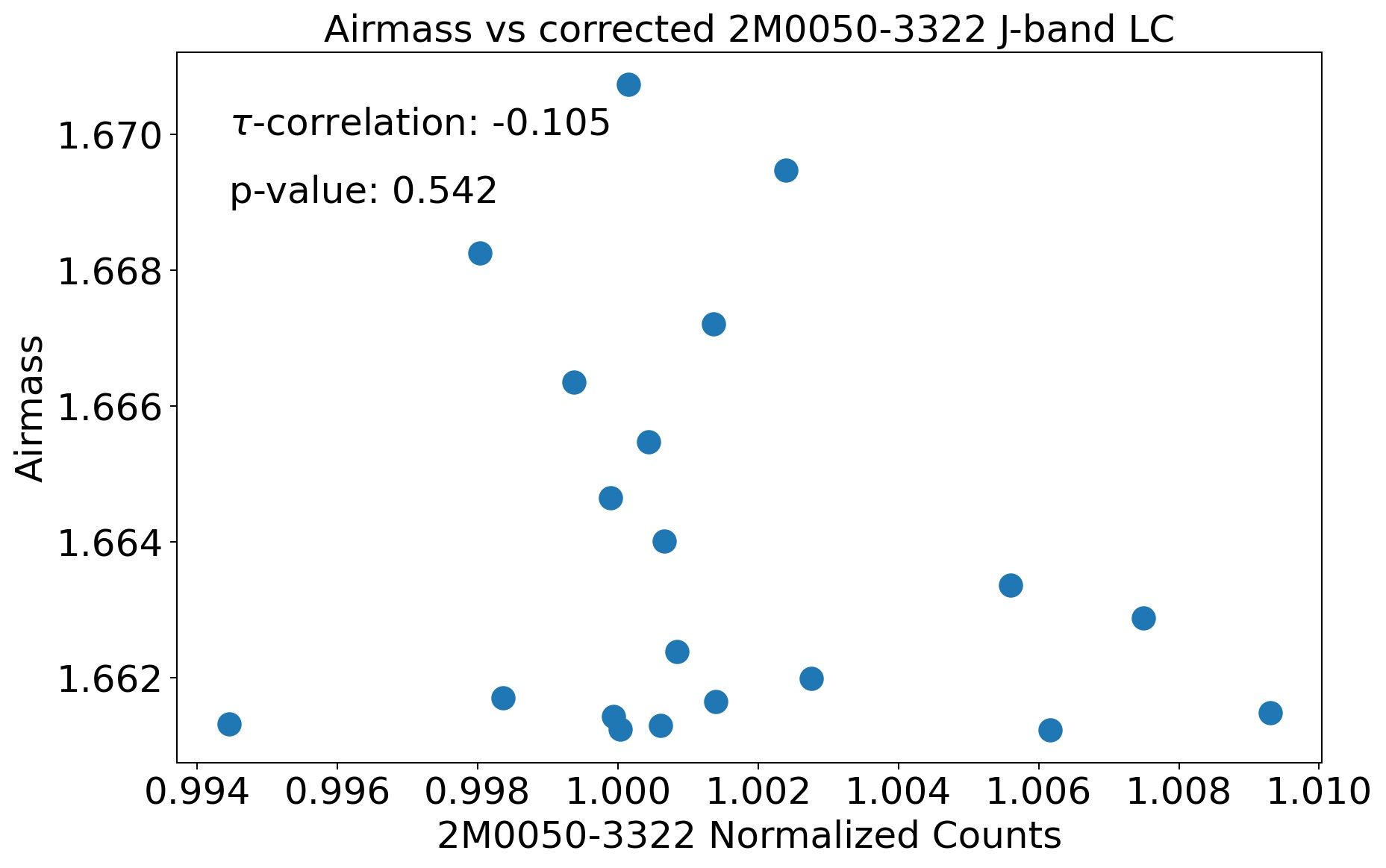}
    \caption{Correlations between the corrected $J$-band imaging light curve and different atmospheric parameters.}
    \label{J_band_corr}
\end{figure}

\begin{figure}
    \centering
   
    \includegraphics[width=0.48\textwidth]{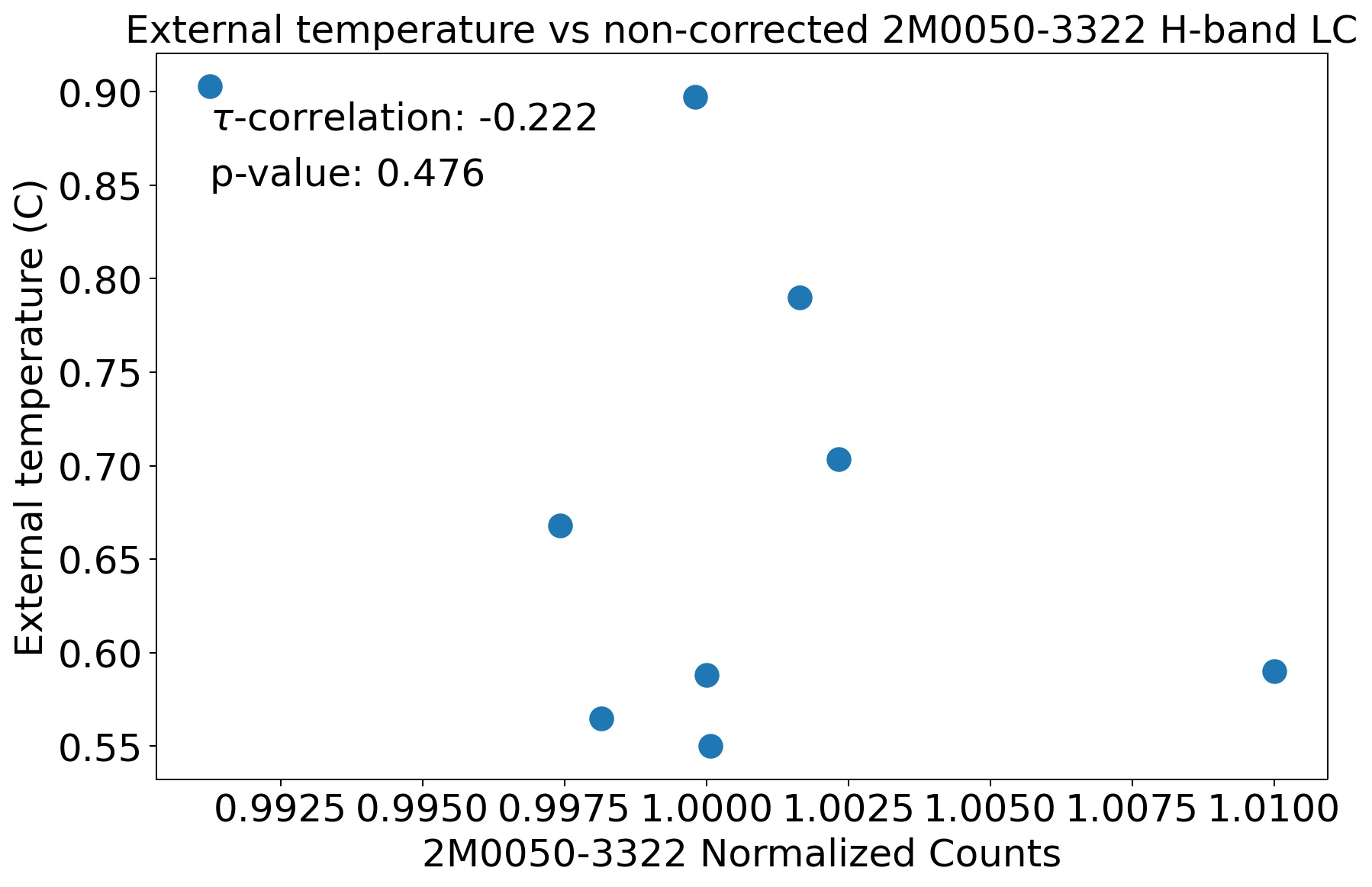}
    \includegraphics[width=0.48\textwidth]{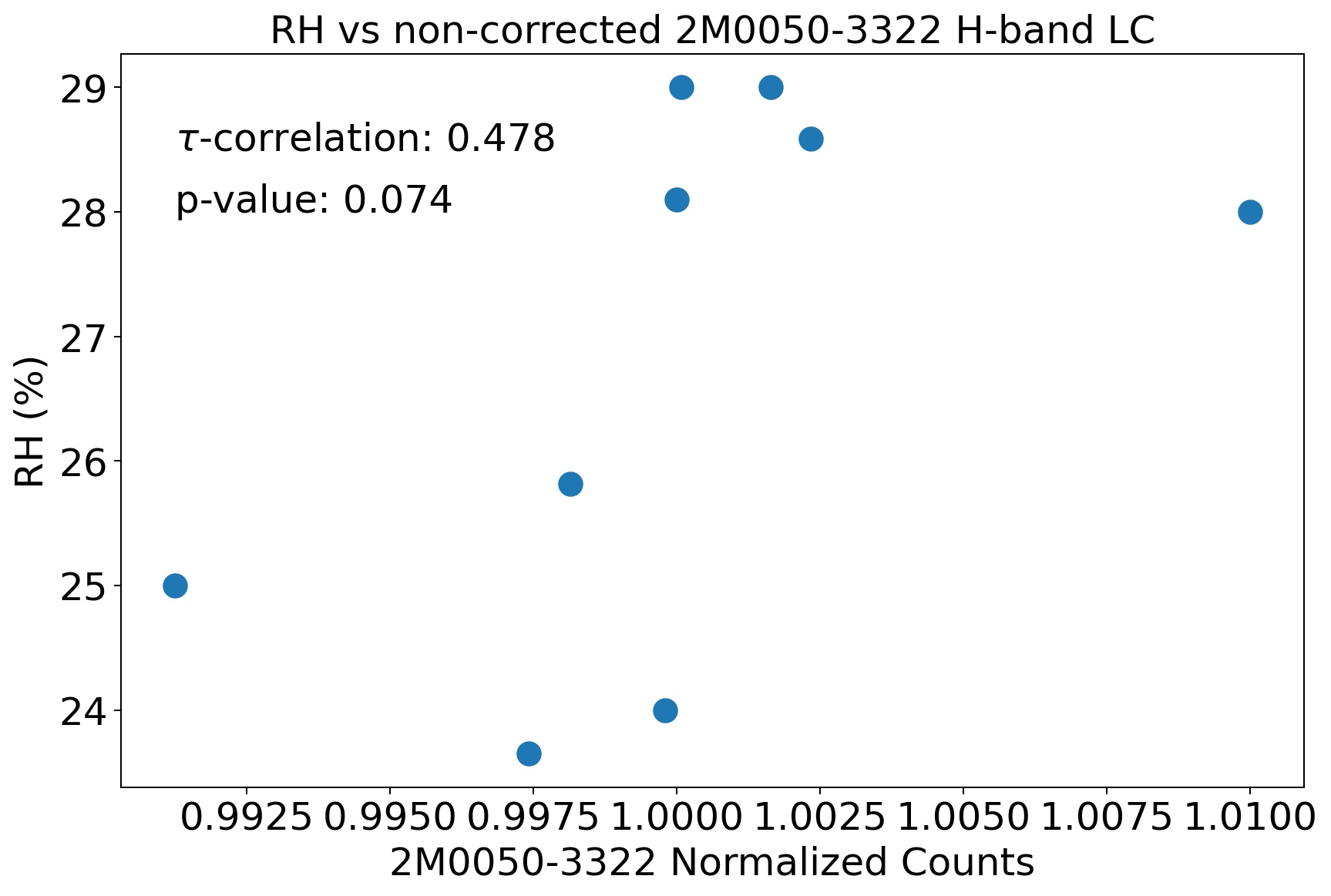}
    \includegraphics[width=0.48\textwidth]{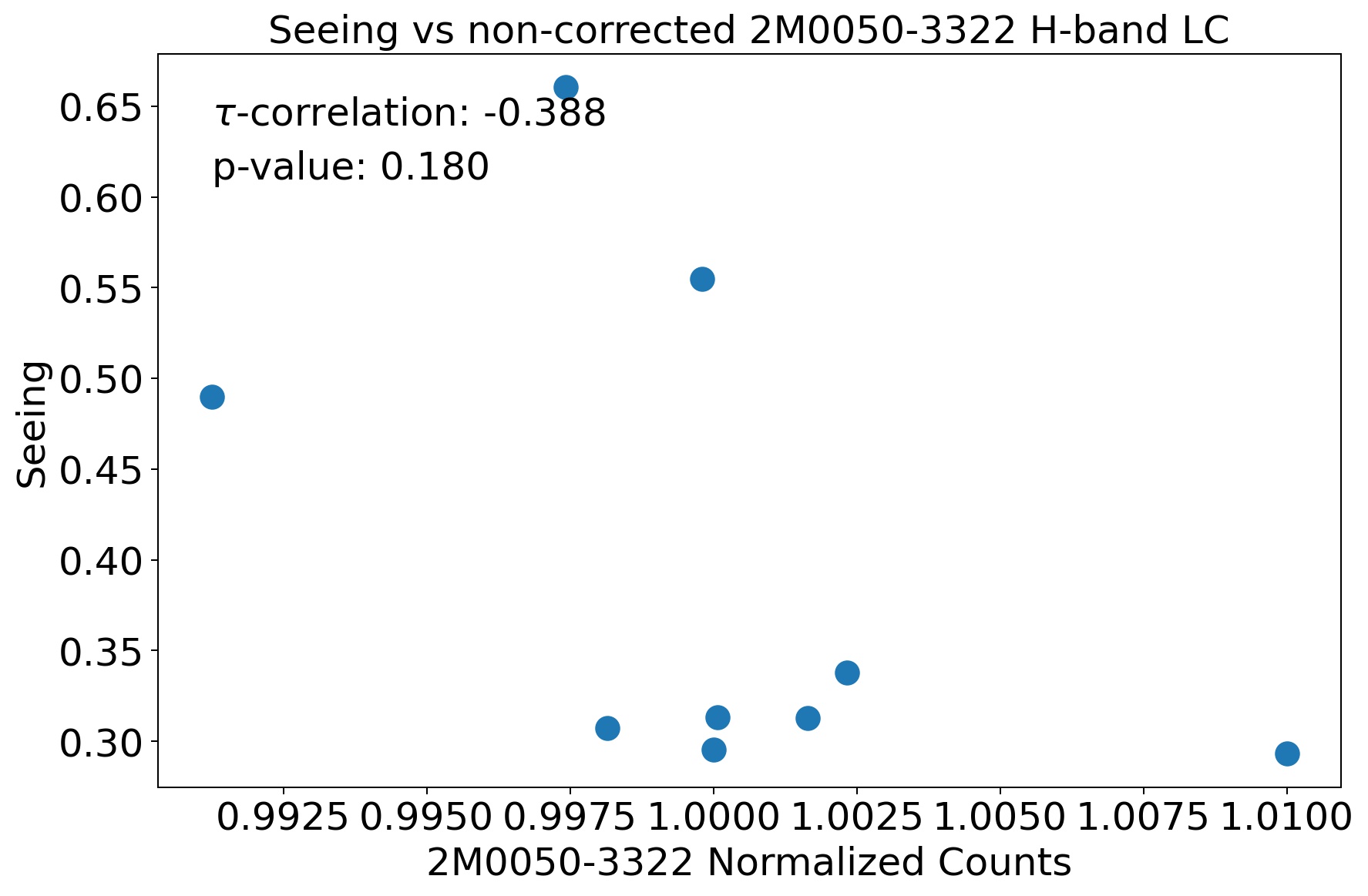}
    \includegraphics[width=0.48\textwidth]{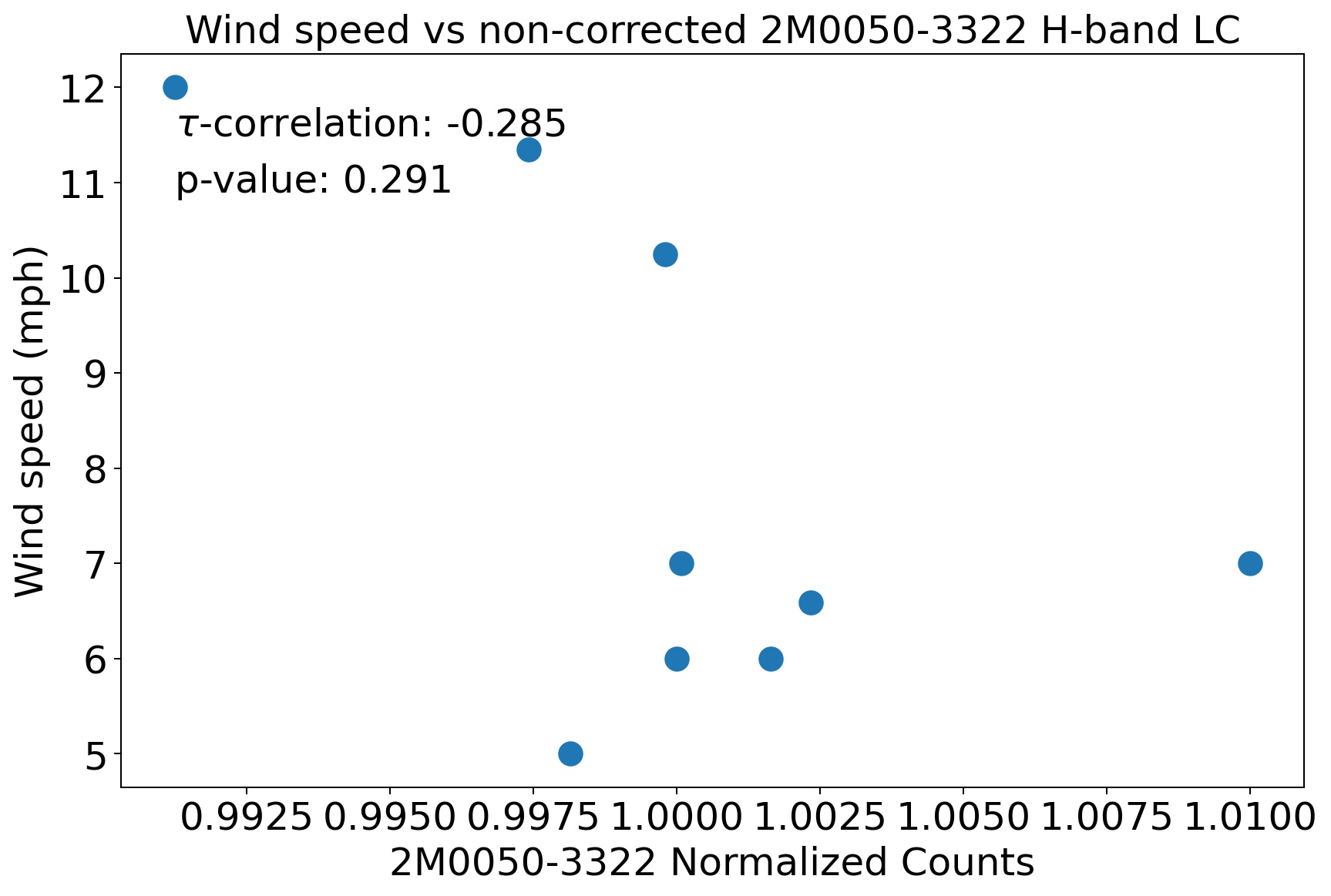}
    \includegraphics[width=0.48\textwidth]{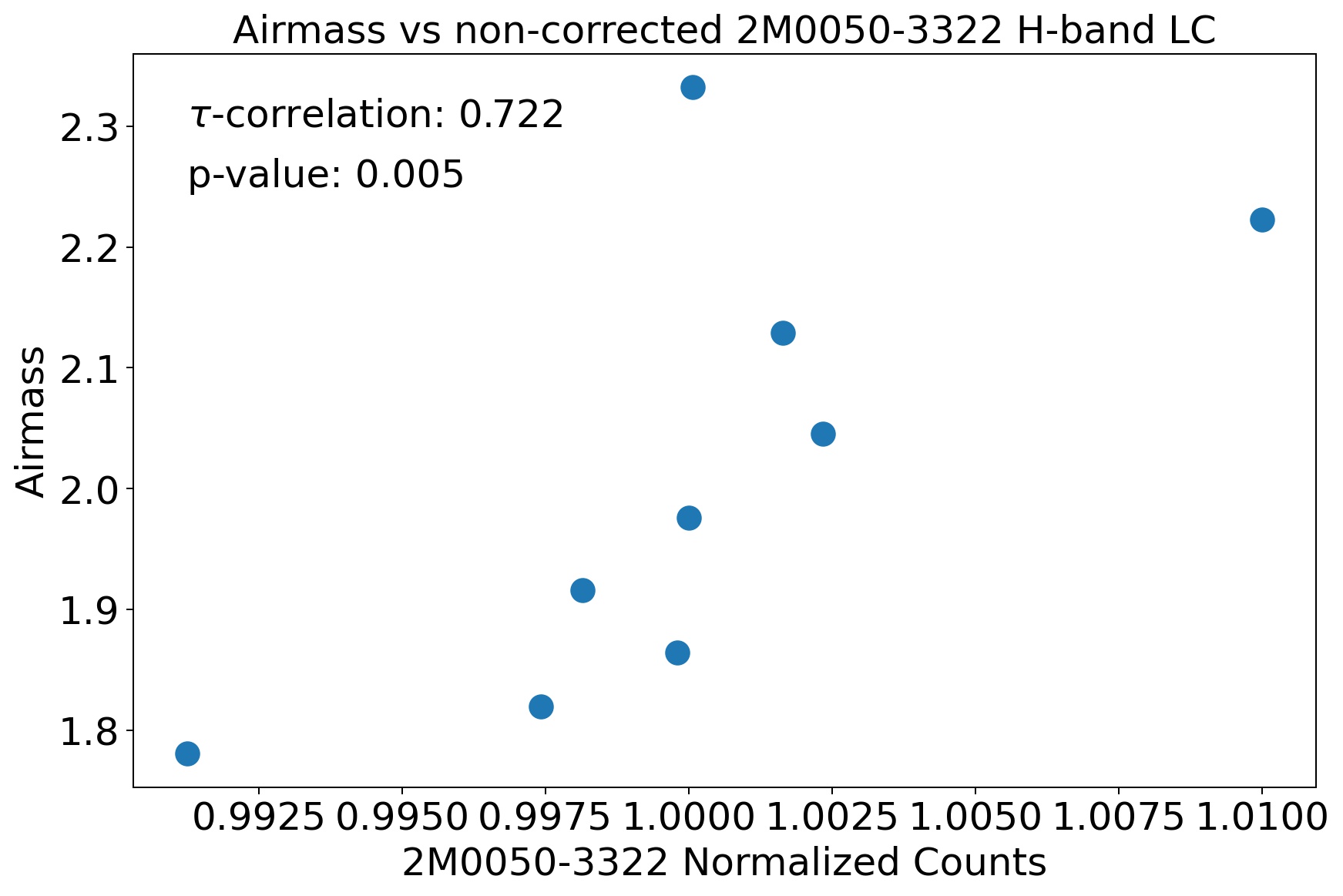}
    \caption{Correlations between the non-corrected $H$-band  light curve and different atmospheric parameters.}
    \label{H_band_noncorr}
\end{figure}

\begin{figure}
    \centering
   
    \includegraphics[width=0.48\textwidth]{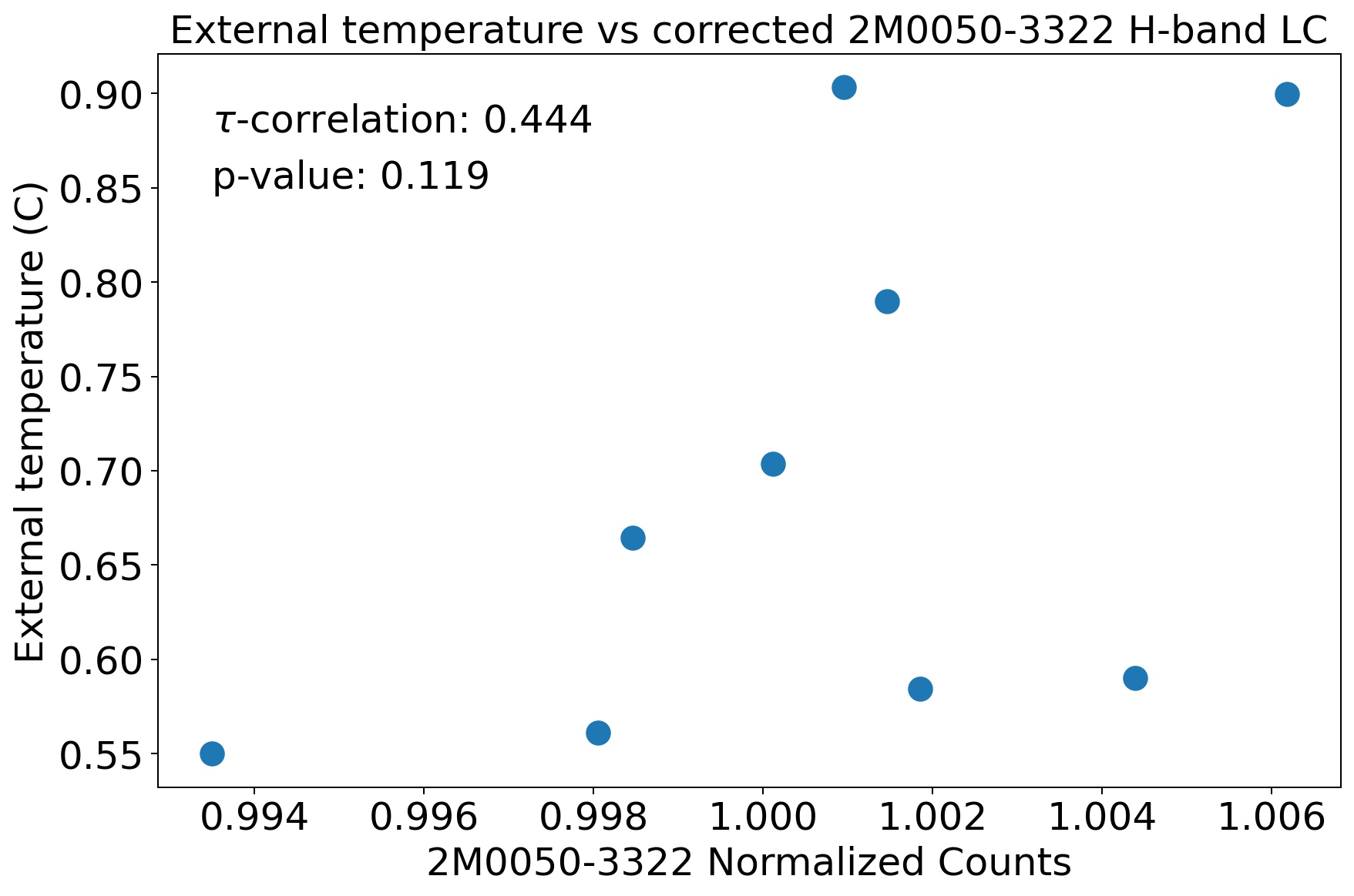}
    \includegraphics[width=0.48\textwidth]{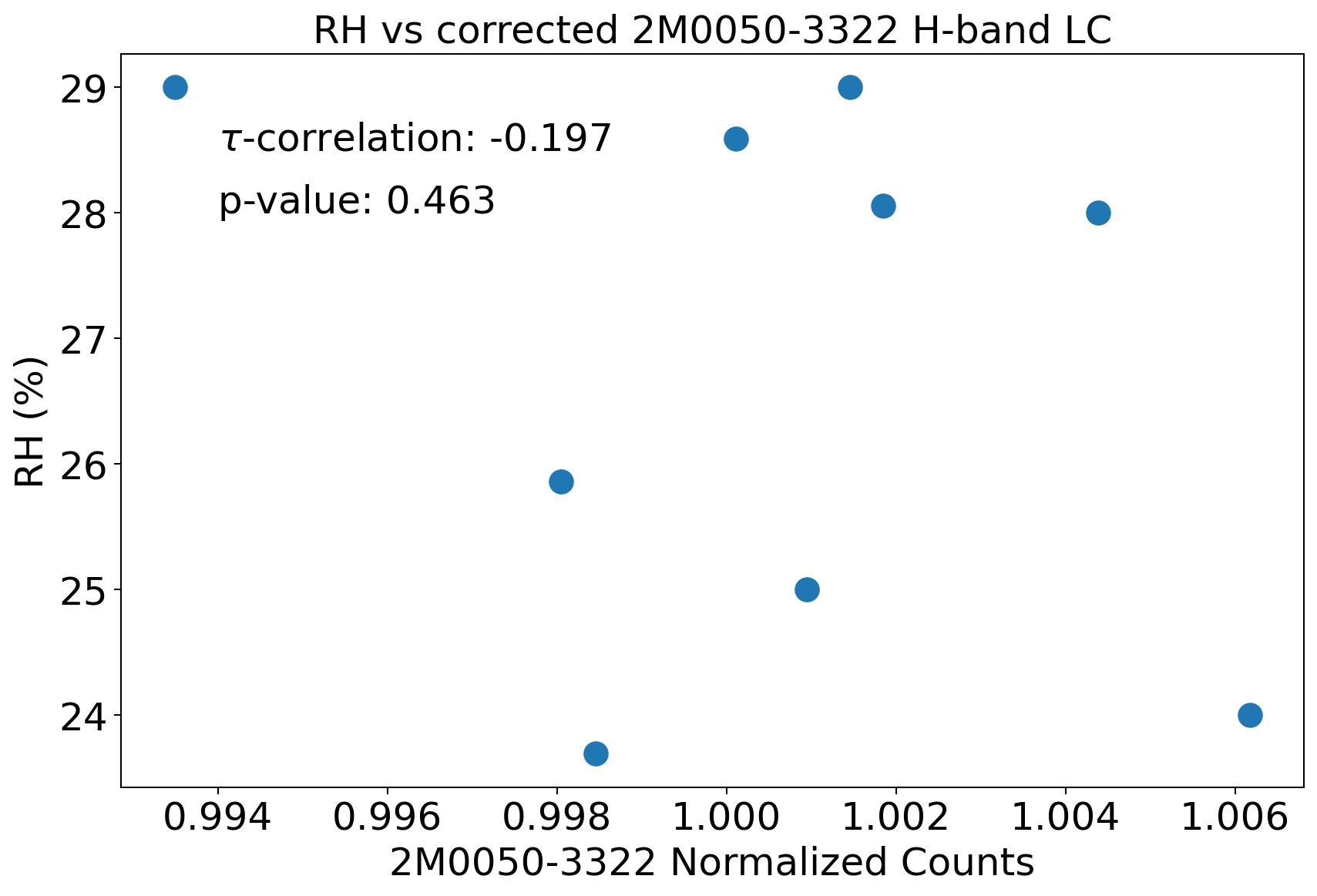}
    \includegraphics[width=0.48\textwidth]{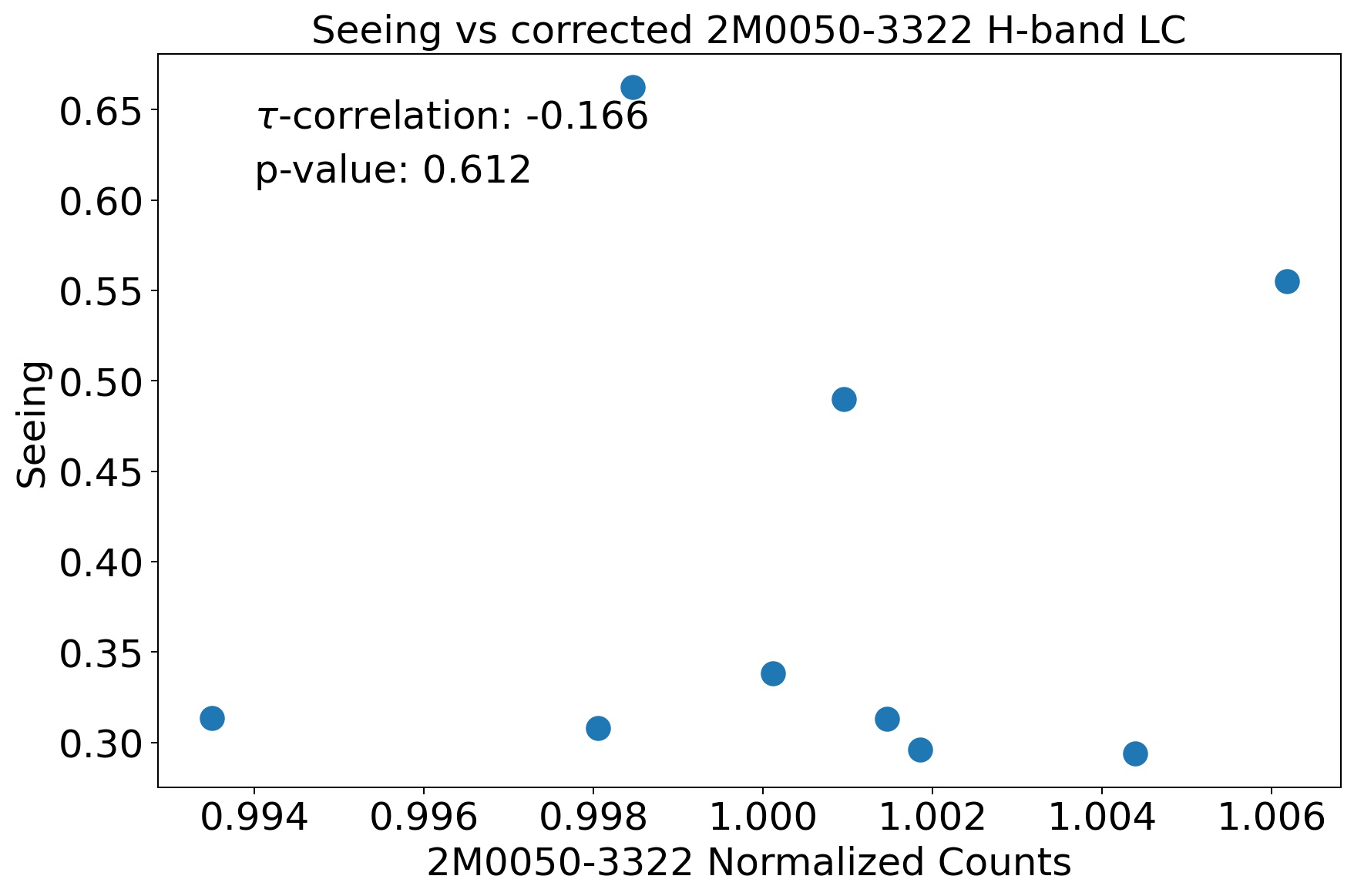}
    \includegraphics[width=0.48\textwidth]{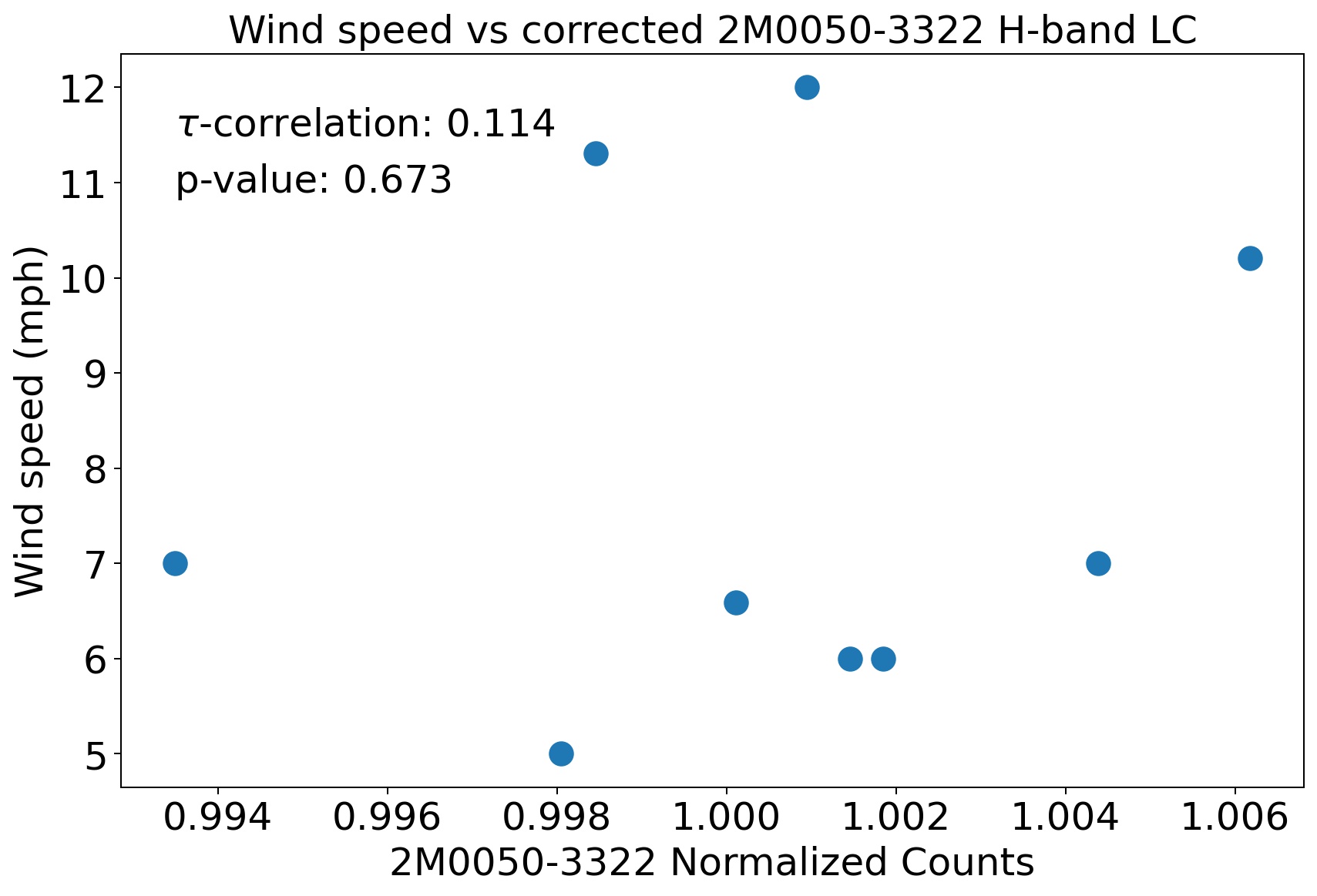}
    \includegraphics[width=0.48\textwidth]{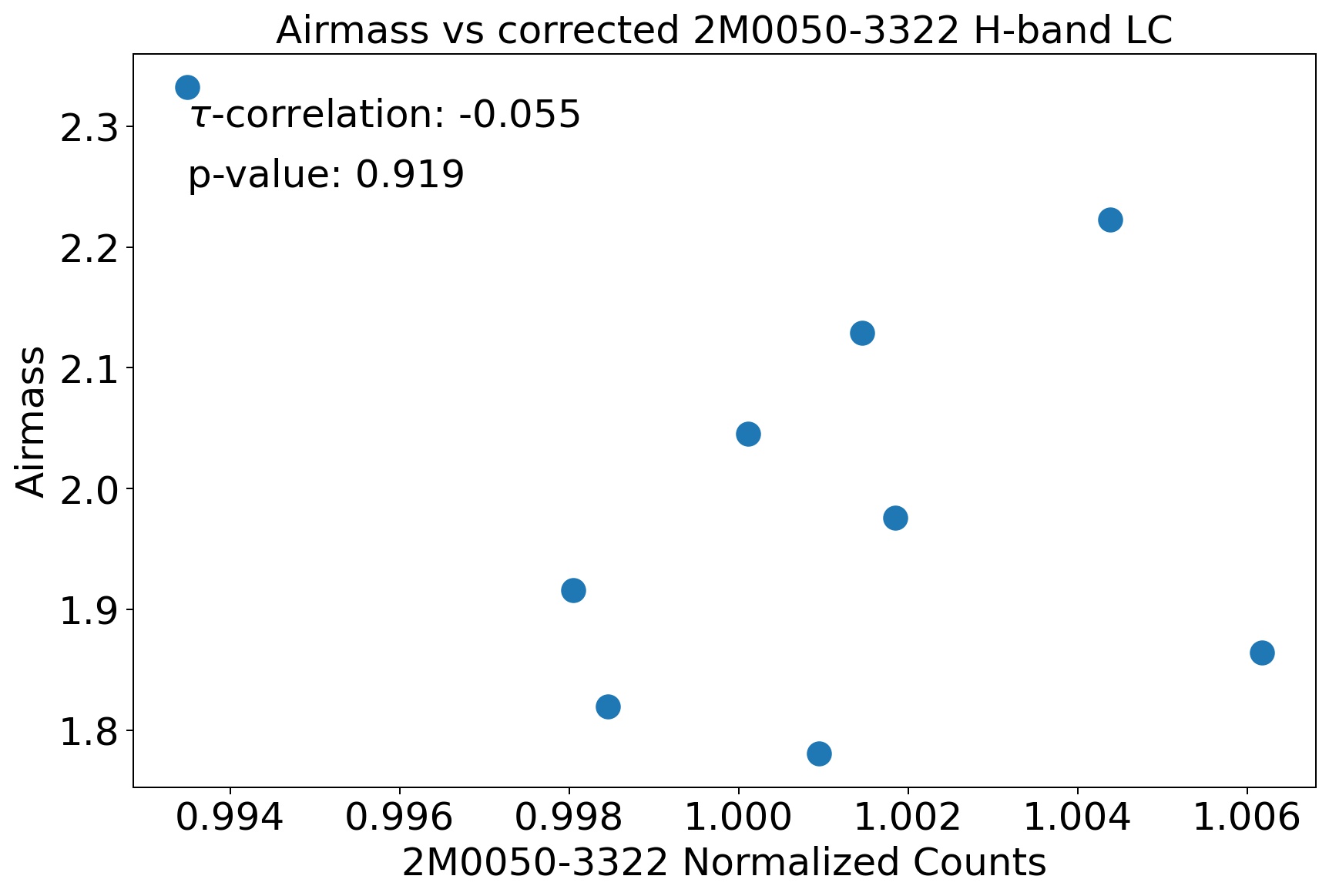}
    \caption{Correlations between the corrected $H$-band  light curve and different atmospheric parameters.}
    \label{H_band_corr}
\end{figure}

\begin{figure}
    \centering
   
    \includegraphics[width=0.48\textwidth]{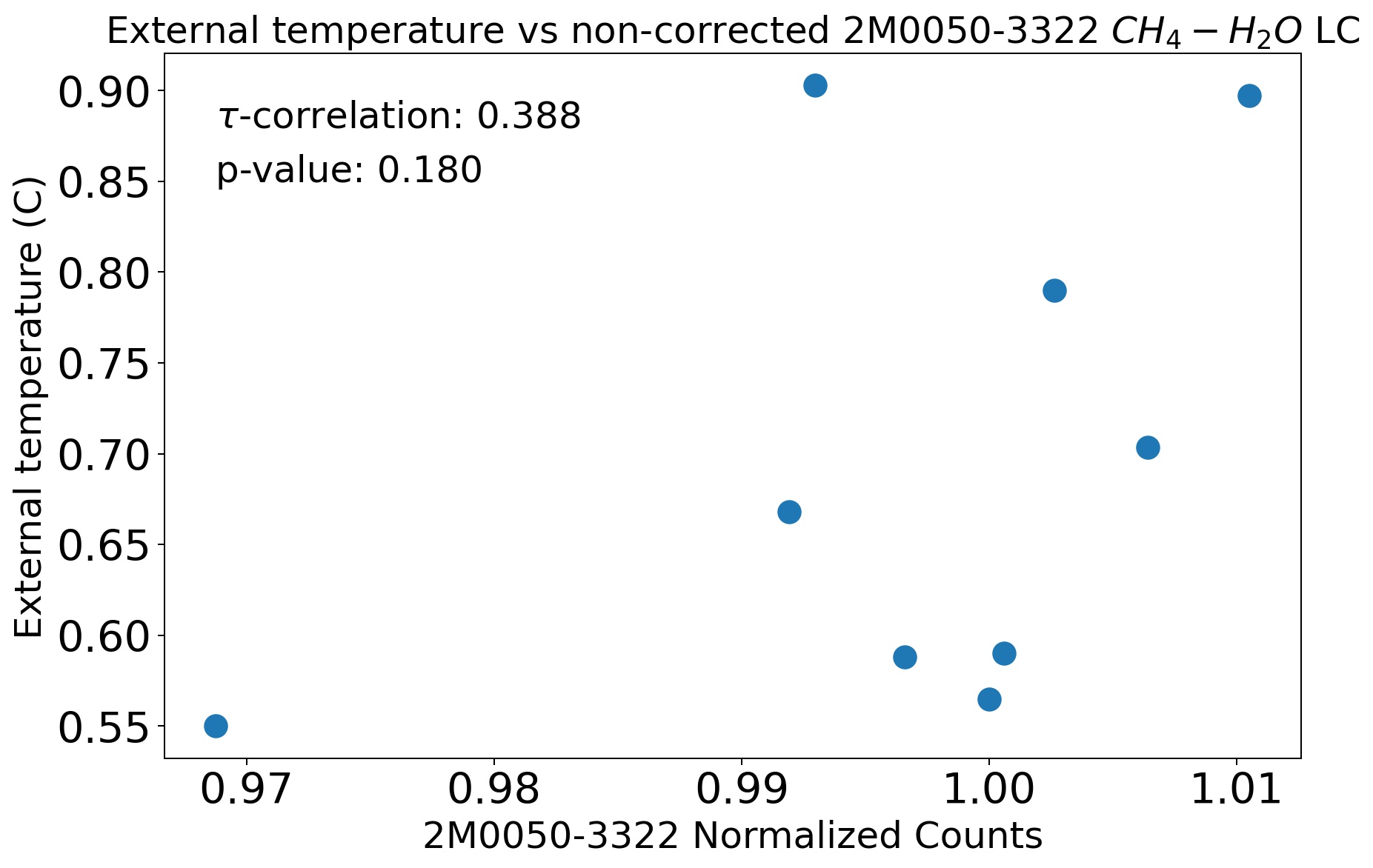}
    \includegraphics[width=0.48\textwidth]{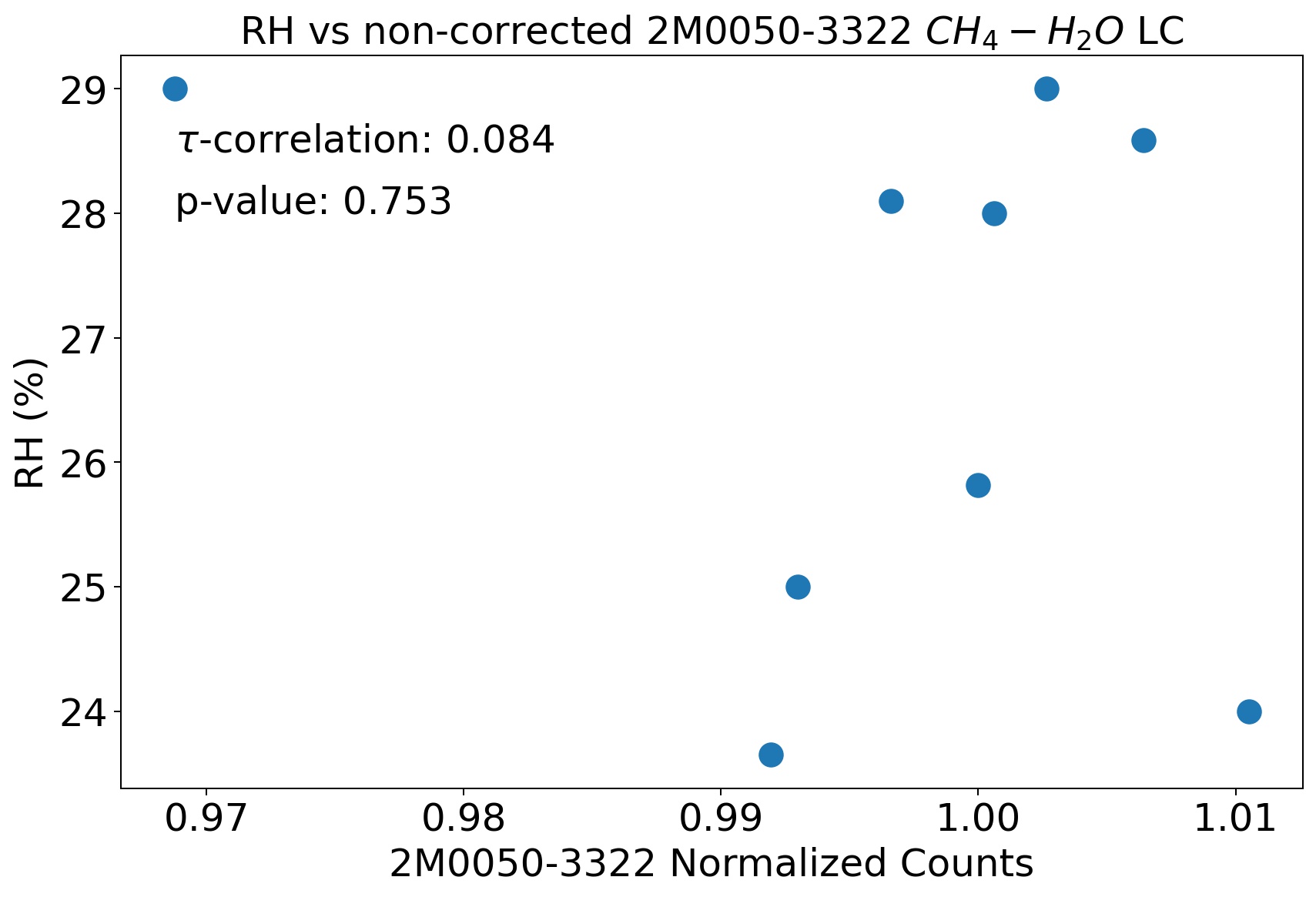}
    \includegraphics[width=0.48\textwidth]{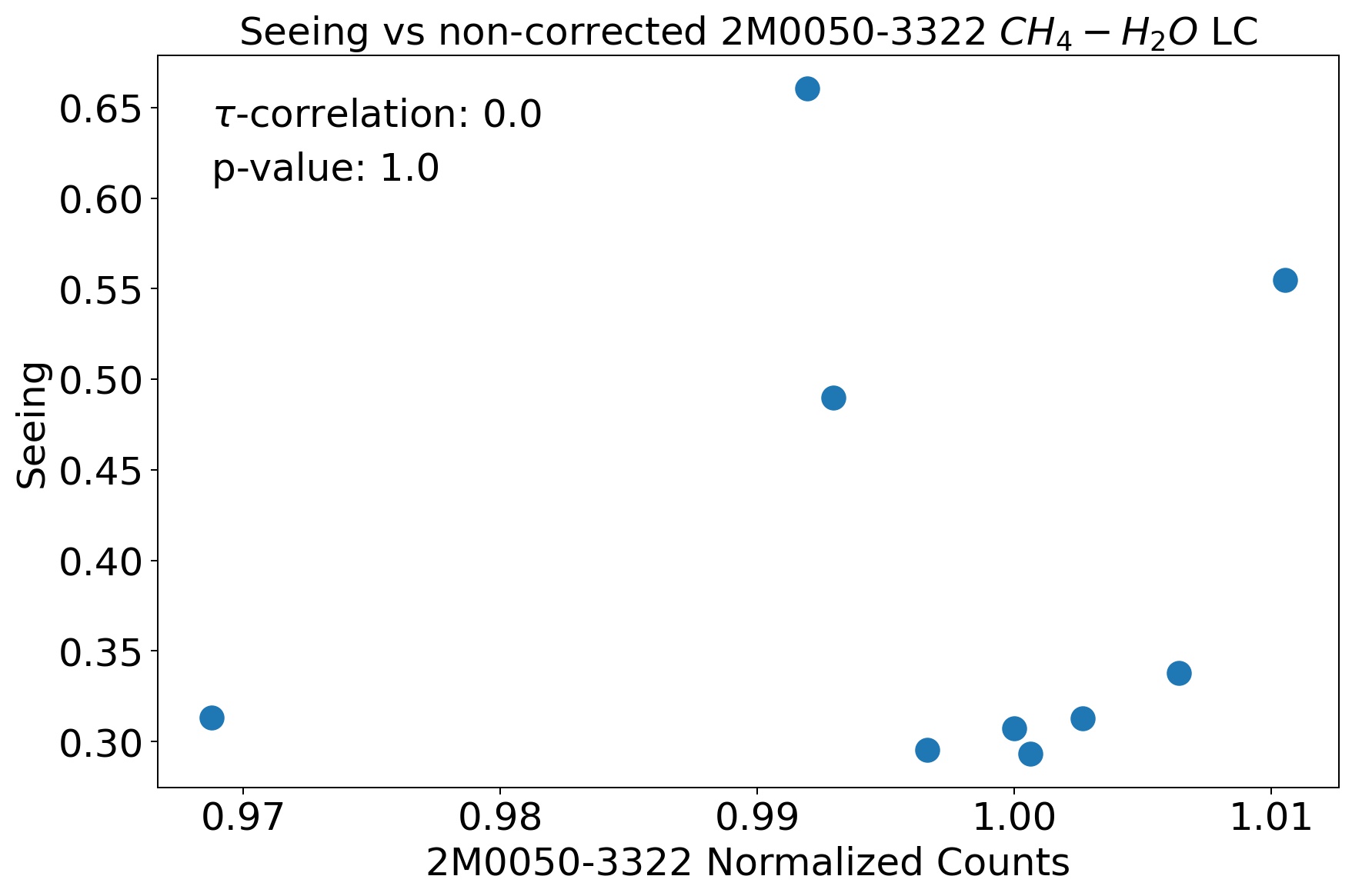}
    \includegraphics[width=0.48\textwidth]{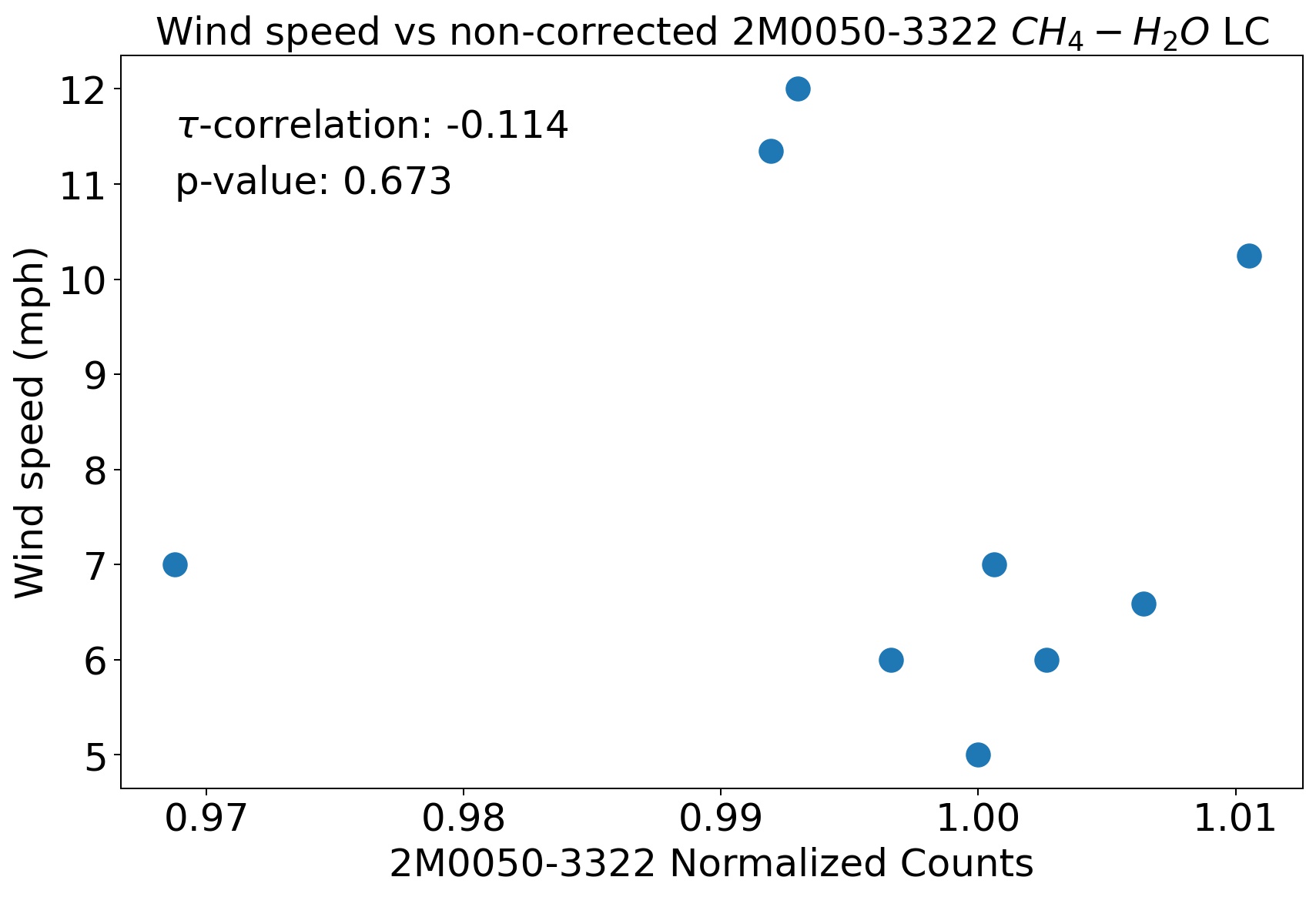}
    \includegraphics[width=0.48\textwidth]{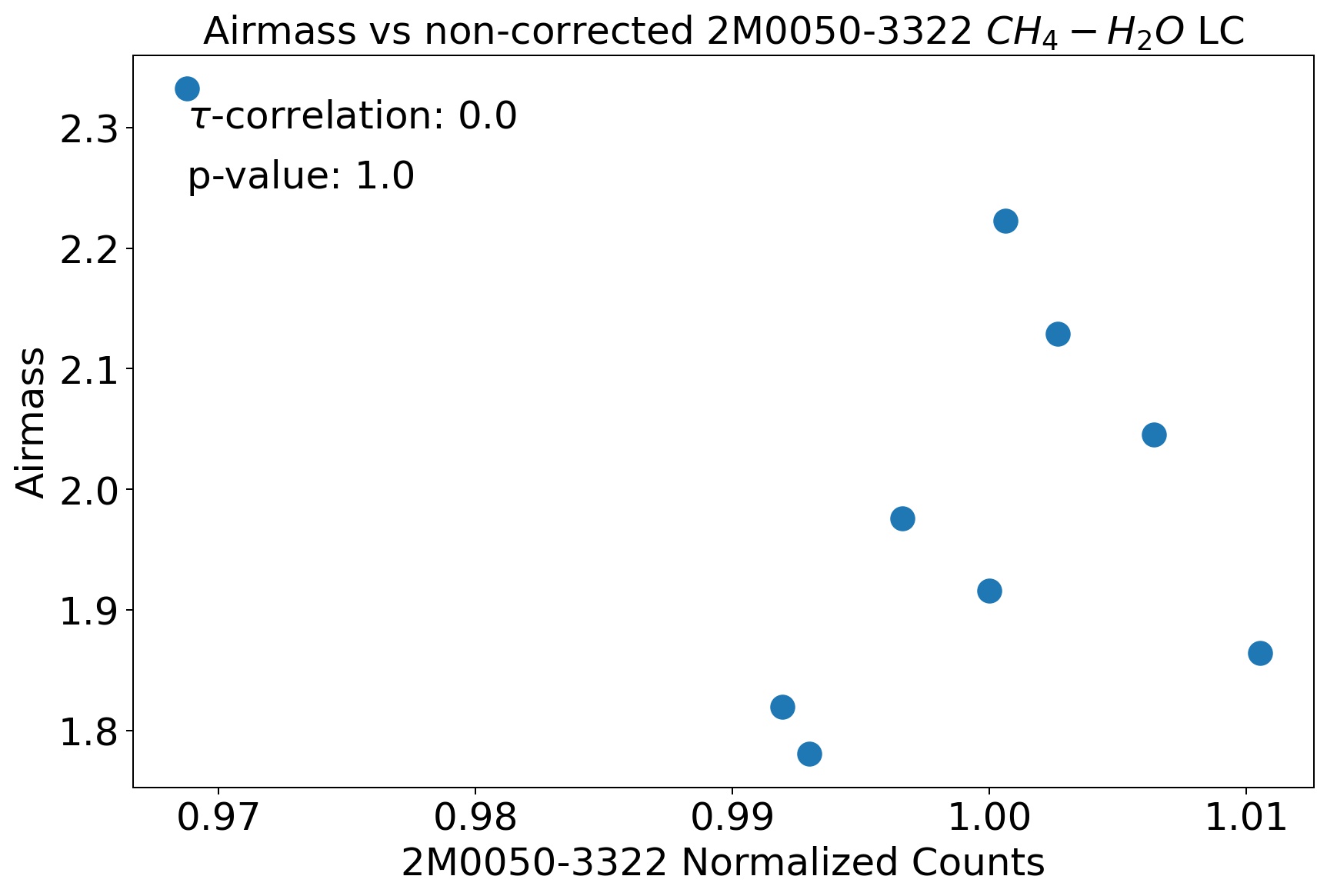}
    \caption{Correlations between the non-corrected $\mathrm{CH_{4}-H_{2}O}$  light curve and different atmospheric parameters.}
    \label{CH4-H2O_noncorr}
\end{figure}

\begin{figure}
    \centering
   
    \includegraphics[width=0.48\textwidth]{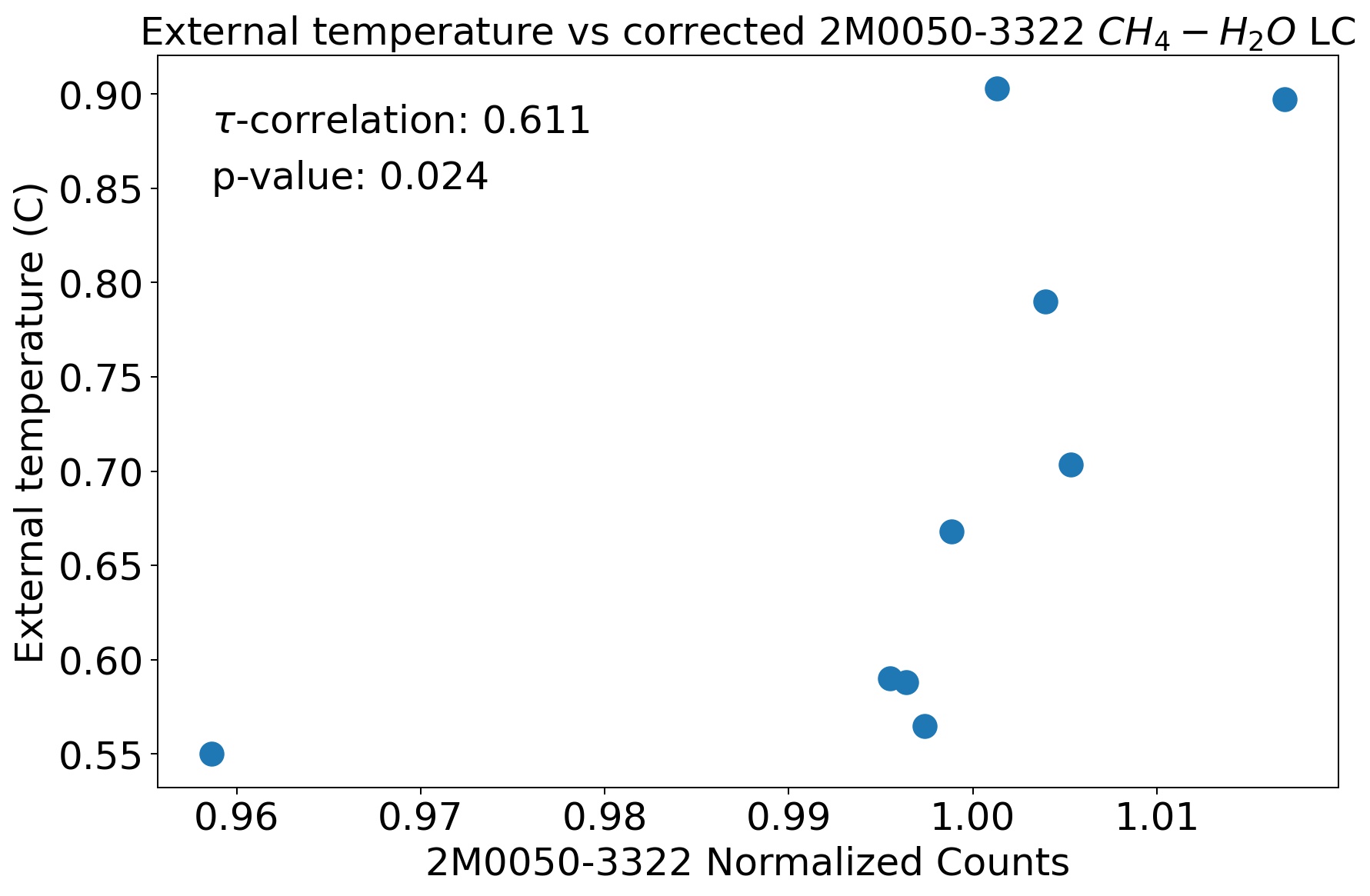}
    \includegraphics[width=0.48\textwidth]{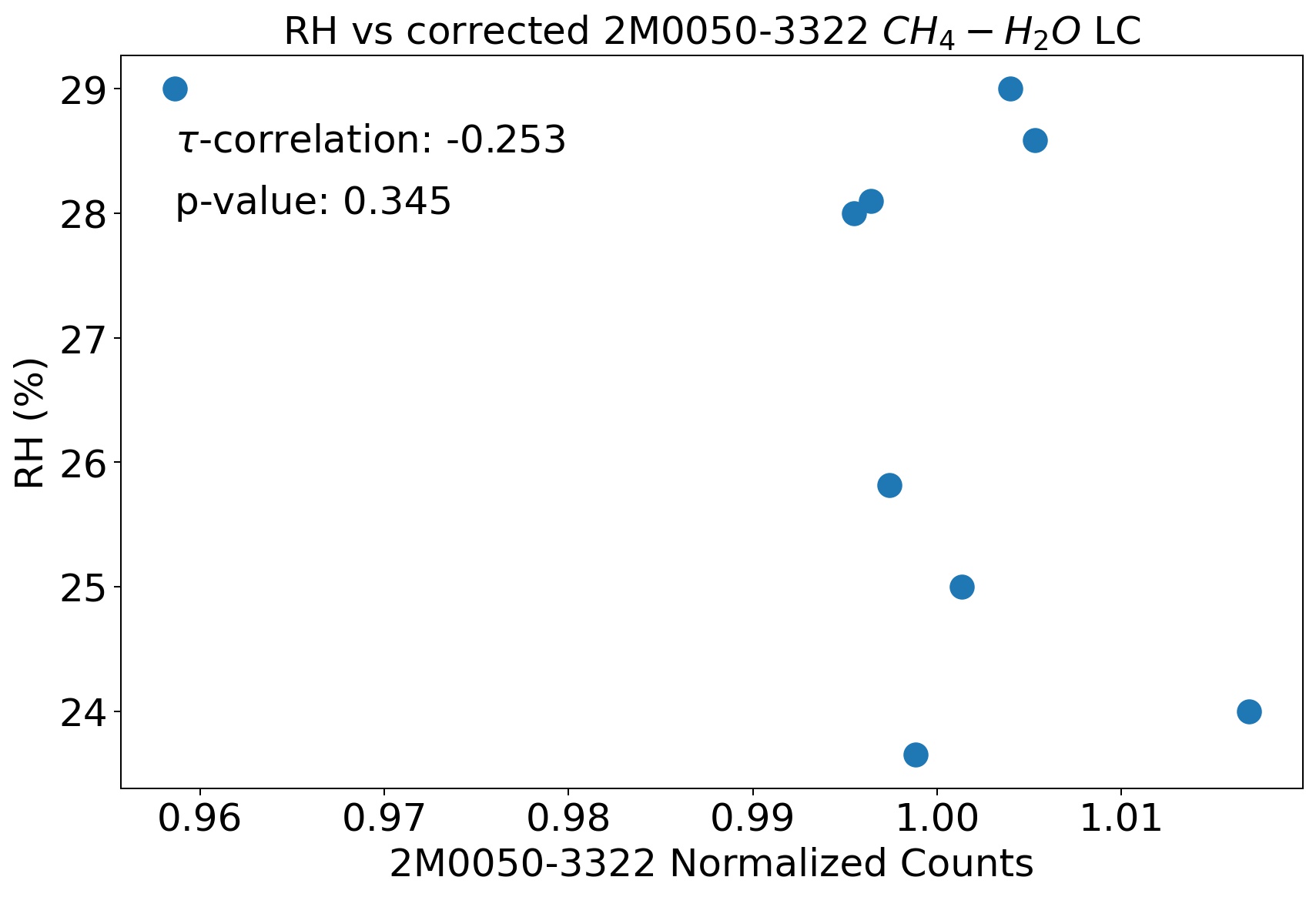}
    \includegraphics[width=0.48\textwidth]{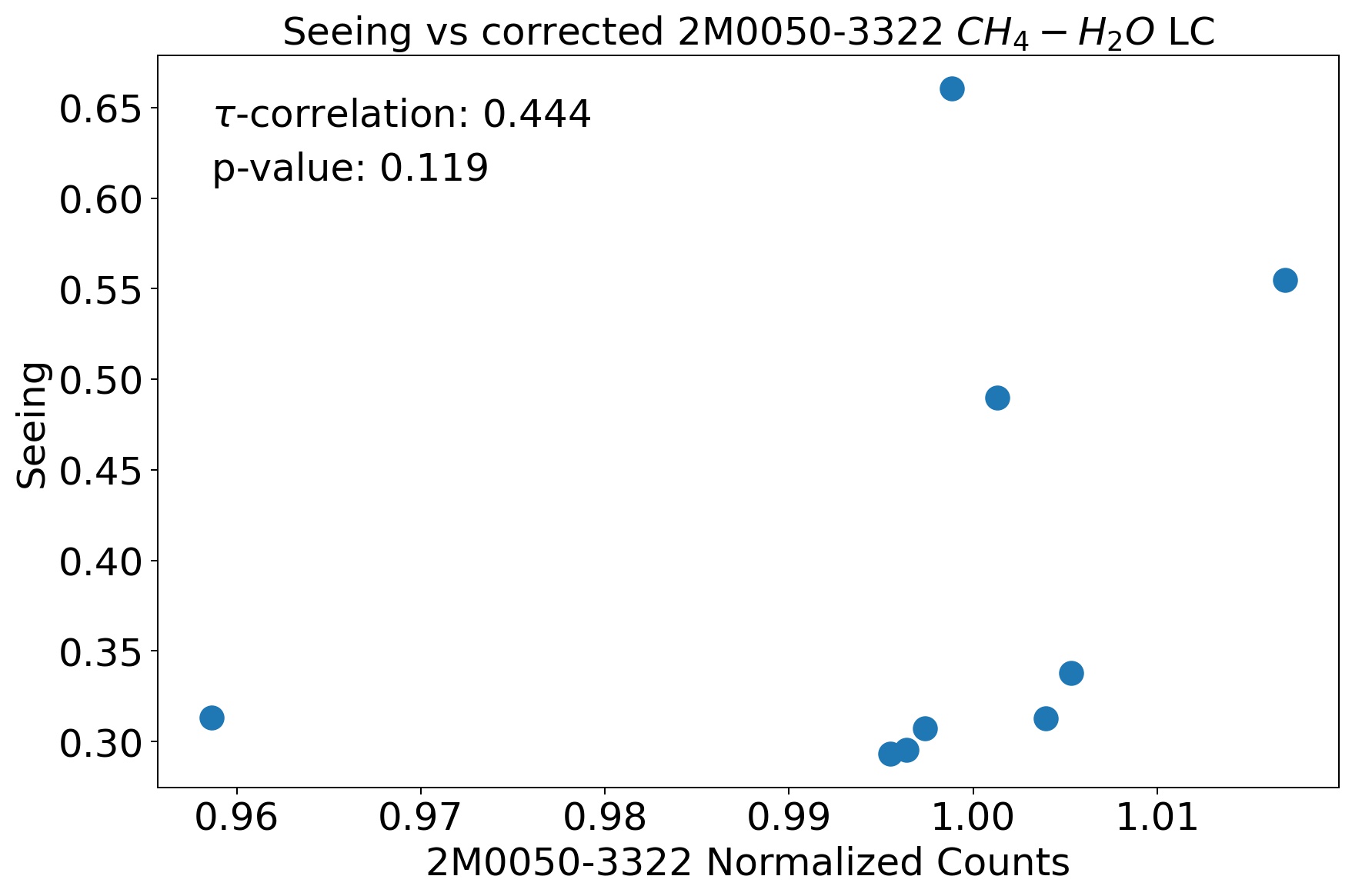}
    \includegraphics[width=0.48\textwidth]{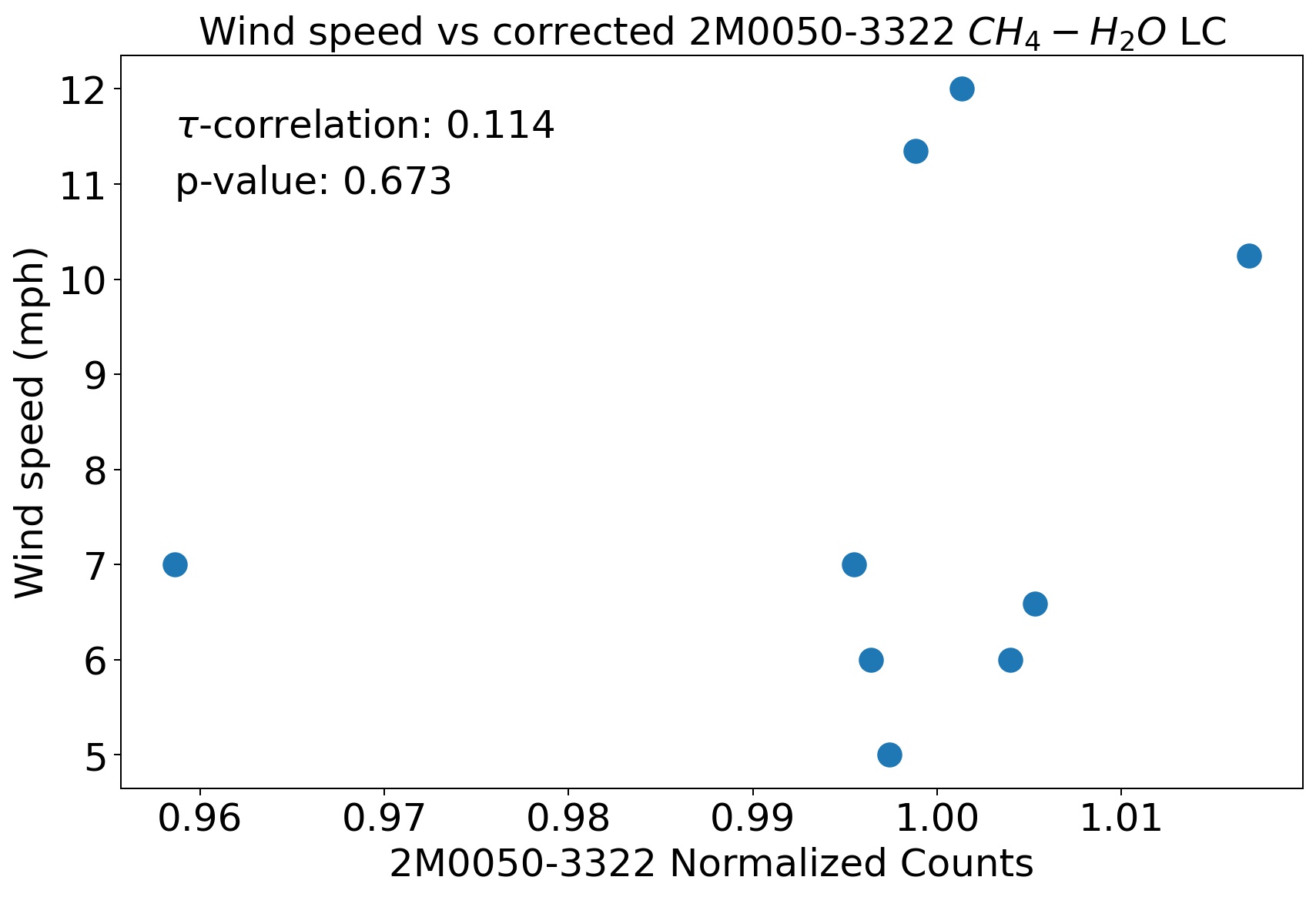}
    \includegraphics[width=0.48\textwidth]{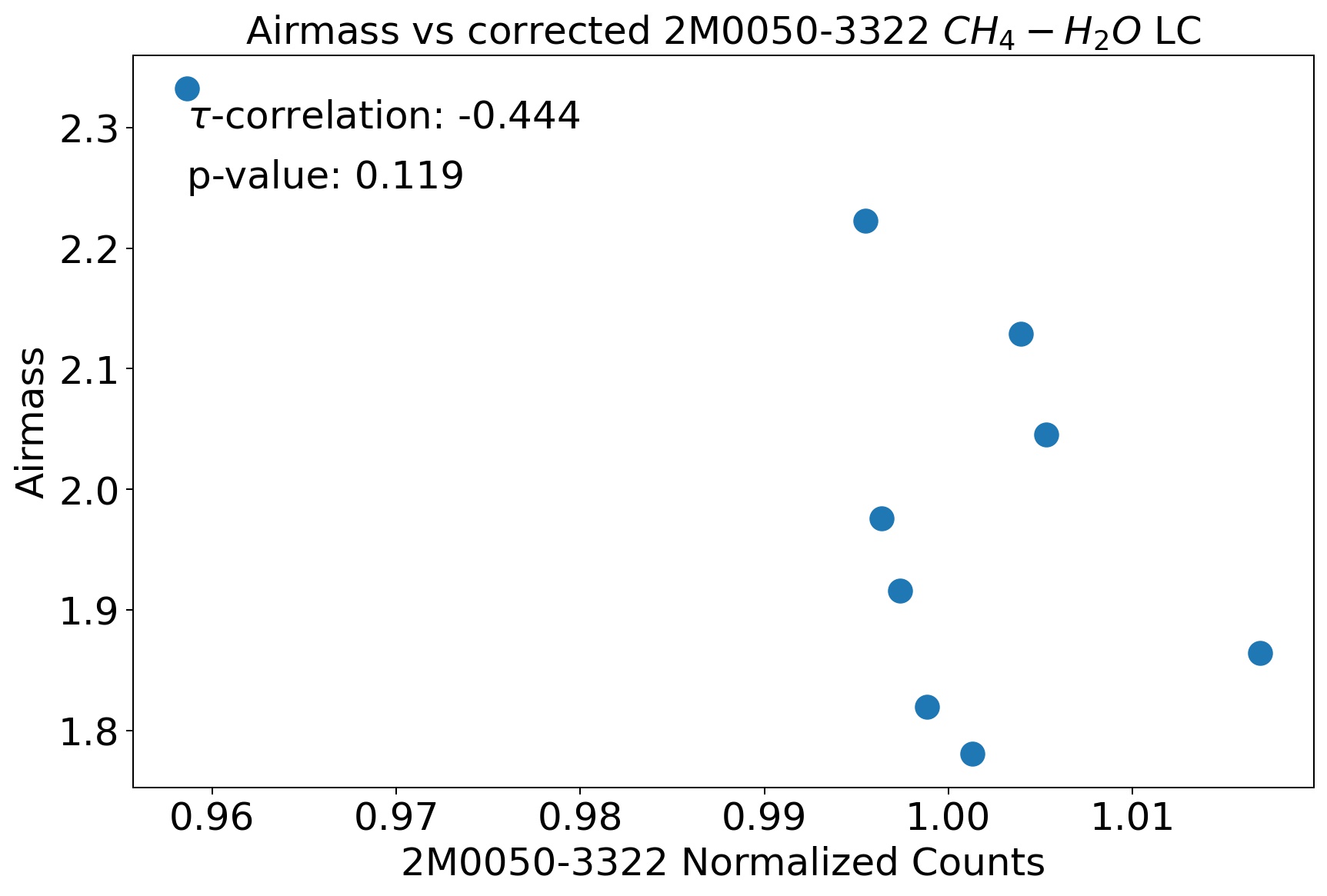}
    \caption{Correlations between the corrected $\mathrm{CH_{4}-H_{2}O}$  light curve and different atmospheric parameters.}
    \label{CH4-H2O_corr}
\end{figure}

%\begin{figure}
%    \centering
% 
%    \includegraphics[width=0.8\textwidth]{Bayesian_LS.png}
%    \caption{Bayesian Generalized Lomb-Scargle periodogram for 2M0055--3322.}
%    \label{LS_periodogram}
%\end{figure}

\section{The General Circulation Model for the late-T dwarf 2M0050--3322}\label{ch.gcm}

We use the general circulation model (GCM) originally developed in \cite{Tan2021} and \cite{Tan2021global} to simulate the global three-dimensional temperature, wind and cloud structures for the atmosphere of the late-T brown dwarf  2M0050–3322. The GCM solves the standard dynamical meteorology equations governing the horizontal angular momentum, hydrostatic balance, mass continuity, thermodynamics and equation of state. A set of tracer equations are solved simultaneously for the dynamical transport of the condensable vapor and cloud components of the atmosphere. A grey (wavelength averaged), two-stream radiative transfer scheme is used to solve for the thermal radiation. The opacity sources include the Rosseland-mean gas opacity of solar metallicity obtained from \cite{Freedman2014}, and the  absorption and scattering due to localtion- and time-dependent cloud particles. The strong   radiative heating/cooling rates due to inhomogeneous cloud structures drive the general circulation, which in turn maintains the large-scale patchy cloud configuration. 

A similar set of GCMs for brown dwarfs Luhman A and B is performed in \cite{Mukherjee2021} in which the major cloud species was assumed to be ${\rm MgSiO_3}$, appropriate for those L/T transition dwarfs. The GCM in this study differs to those used in \cite{Mukherjee2021} in that the major cloud species here are ${\rm MnS}$ and ${\rm Na_2S}$ which are the two most representative species near the photospheres of  late-T dwarfs (e.g., \citealp{morley2012}, see also Figure \ref{vertical_structure}). The temperature-pressure-dependent  condensation curves of ${\rm MnS}$ and ${\rm Na_2S}$ at solar metallicity are given by \cite{visscher2006}. More details of the cloud scheme are referred to \cite{Tan2021,Tan2021global} and \cite{Mukherjee2021}. We assume a fixed cloud particle number per dry air mass $N_c=10^{-9}\;{\rm kg^{-1}}$ throughout the atmosphere. When clouds form, the characteristic  particle size is determined by the amount of local cloud mass mixing ratio and $N_c$. For solar abundance of condensable gases, clouds {particles} are sub-micron in our model. 

The GCM uses a horizontal resolution equivalent to 256 and 512 grid points in the latitudinal and longitudinal directions. There are 45 vertical layers in the model, which are evenly discretized in the log-pressure space in between 0.005 bar and 100 bars. A relatively weak frictional drag with a drag timescale of $10^7$ s is applied at deep layers with pressures larger than 50 bars, representing angular momentum mixing between the modeled layer and the deep interior. The model has been integrated over 1200 Earth days and has reached statistical equilibrium. 

\bibliography{2M0050_variability}{}
\bibliographystyle{aasjournal}



\end{document}